\documentclass[twocolumn,pra,superscriptaddress, aps, 10pt]{revtex4-2}
\usepackage{graphicx,subfigure}          % For the figures and subfigures
\usepackage{amsfonts,amsmath,amssymb}    % For various math symbols
\usepackage{dsfont}                      % For the identity operator
\usepackage[dvipsnames]{xcolor}          % To use colors
\usepackage[
colorlinks=true,
linkcolor=blue,
citecolor=blue,
urlcolor=blue]{hyperref}                % For the use of hyperlinks
\usepackage{comment}                    % Comment package for MacBook
\usepackage{physics}                    % To use Dirac Notation
\usepackage[normalem]{ulem}
\usepackage{soul}

\newcommand{\remove}[1]{{\sout{#1}}}
\definecolor{fg}{rgb}{0.13, 0.55, 0.13}

\begin{document}
	%%%%%%%%%%%%%%%%%%%%%%%%%%%%%%%%%%%%%%%%%%%%%%%%%%%%%%%%%%%%%%%%%%%%%%%%%%%%%%%%%%%%%%%%
	
	\title{Deterministic generation of hybrid entangled states using quantum walks}

	\author{Jaskaran Singh}
	\email{jaskaran@iitmandi.ac.in}
	\affiliation{Center for Quantum Science and Technology, Indian Institute of Technology, Mandi, Himachal Pradesh 175075, India}
	\affiliation{Department of Physics and Center for Quantum Frontiers of Research \&
		Technology (QFort), National Cheng Kung University, Tainan 701, Taiwan}

	\author{Vikash Mittal}
	\email{vikashmittal.iiser@gmail.com}
	\affiliation{Department of Physics, National Tsing Hua University, Hsinchu 300044, Taiwan}
	
	\author{Soumyakanti Bose}
	\email{soumyakanti.bose09@gmail.com}
	\affiliation{Department of Physics \& Astronomy, Seoul National University, Gwanak-ro 1, Gwanak-gu, Seoul 08826, Korea}
	
	%%%%%%%%%%%%%%%%%%%%%%%%%%%%%%%%%%%%%%%%%%%%%%%%%%%%%%%%%%%%%%%%%%%%%%%%%%%%%%%%%%%%%%%%
	
	\begin{abstract}
		In recent times, hybrid-entanglement (HE) between a qubit and a coherent state has demonstrated superior performance in various quantum information processing tasks, particularly in quantum key distribution.
		Despite its theoretical advantages, efficient generation of such states in the laboratory has been a challenge.
		Here, we introduce a deterministic and efficient approach for generating HE states using quantum walks. 
		Our method achieves a remarkable fidelity of $99.9\%$ with just $20$ time steps in a one-dimensional split-step quantum walk. 
		This represents a significant improvement over prior approaches for probabilistic generation of HE states with fidelity as low as $80\%$.
		Our scheme not only provides a robust solution to the generation of HE states but also highlights a unique advantage of quantum walks, thereby contributing to the advancement of this burgeoning field. Moreover, our scheme is experimentally feasible with the current technology.
	\end{abstract}
	
	%%%%%%%%%%%%%%%%%%%%%%%%%%%%%%%%%%%%%%%%%%%%%%%%%%%%%%%%%%%%%%%%%%%%%%%%%%%%%%%%%%%%%%%%
	
	\maketitle

	%%%%%%%%%%%%%%%%%%%%%%%%%%%%%%%%%%%%%%%%%%%%%%%%%%%%%%%%%%%%%%%%%%%%%%%%%%%%%%%%%%%%%%%%
	
	\section{Introduction}
	
	Entanglement is a crucial resource in many quantum information processing tasks and is one of the defining key properties of quantum states that differentiates it from classical resources~\cite{HHH09}. 
	It has been generally studied in multipartite systems that are entangled in either all discrete variable (DV) degrees of freedom~\cite{HHH96}, including qudit-qudit entanglement or all continuous variable (CV) degrees of freedom~\cite{S00, DGC00}, which generally include two-mode squeezed coherent states of light. 
	Both CV and DV entangled states are of paramount importance in several information processing tasks, including quantum key distribution~\cite{E91, NDN22, ZVR22, LZZ22, JSA21, JBA17, SB21, KSB19}, quantum computation~\cite{J05, CSR12, SS99, SB02}, teleportation~\cite{BVK98, BBC93, MHS12} as well as tests of Bell nonlocality and contextuality~\cite{CHS69, XSS23, SC23, TJA18, SS21, SF20,JRA23, DJK19, DJK23}. 
	
	However, a third class of bipartite entangled systems has been recently studied in which one of the subsystems is in a qubit state while the other is in a coherent state. 
	Such states are known as hybrid entangled (HE) states~\cite{CHZ02, PLJ12, KJ13}. 
	These strongly correlated HE states incorporate advantages from both the CV and DV subsystems and have been proven to be extremely useful in quantum computation, communication, EPR steering, and certification of Bell nonlocality~\cite{LJ13, ANV15, OTJ20, OTL21, BJ22, HM22, KJ13, Sheng2013, YLim2016, Laurat2018}. 
	Moreover, they are known to outperform both CV and DV entangled states in the case of long-distance quantum key distribution~\cite{BSC24} and also allow deterministic teleportation of qubits~\cite{TMF13}. 
	Moreover, the development of networks in which both CV and DV systems can be interfaced together or combined and interchanged also motivates the research into such states and similar resources~\cite{ANV15, V11}. 
	
	Given their extensive applications and usefulness in quantum information, it is imperative that HE states be efficiently generated in the lab. 
	Till now, several experimental setups have managed to generate such states with limited success and fidelity~\cite{JZK14, MHL14, USP17, SUT18, DAG23}. 
	These experimental techniques rely on costly single-photon resources and photon-subtraction techniques to generate HE states with low probability and low fidelity.

	In principle, a HE state can be produced by introducing weak cross-Kerr nonlinear interactions between a qubit state and a coherent state with appropriate displacement operations~\cite{LJ13, NM04, MNS05}. 
	These approaches, while theoretically straightforward, can be extremely challenging to implement experimentally, requiring either very long optical fibers ($\approx 3000$ km) or very high coherent amplitudes. Moreover, the resultant HE states are highly decohered. Afterward, different schemes were proposed where it was only possible to probabilistically generate a HE state with a high coherent state amplitude and having a fidelity less than $80 \%$~\cite{HJM19, MHL14}. As such, these schemes are based on photon addition or subtraction and are relatively more involved than previous approaches. In summary, previous theoretical schemes were simple and straightforward but experimentally infeasible and extremely challenging, while currently accepted schemes yield HE states probabilistically with low fidelity.
	
	The aim of this article is to provide a deterministic scheme to generate a large class of HE states using single particle quantum walk~\cite{Aharonov1993,Ambainis2001,Kempe2003} with high fidelity.
	The idea is to entangle the coin, which is a qubit system, and the lattice on which we generate a coherent state.
	This coin-lattice system has been physically realized using trapped atoms and trapped ions~\cite{Milburn2002, Schmitz2009, Roos2010, Karski2009}, waveguide arrays~\cite{Regensburger2012},
	photonic setups~\cite{Schreiber2010, Schreiber2011, Regensburger2012, Broome2010, Zhang2007, Sephton2019}, nuclear magnetic resonance~\cite{Du2003, Laflamme2005}, Bose-Einstein condensates~\cite{Sandro2017} orbital angular momentum states of light~\cite{Sephton2019} and photons~\cite{Galton1998}.
	
	In this paper, we propose an efficient scheme that can deterministically produce HE states with fidelity higher than $98 \%$, which is a marked departure from earlier schemes, where it was only possible to do so for a limited class of HE states~\cite{JZK14}. 
	While prior schemes relied on single-photon sources and single-photon detectors, our scheme, instead, only requires an implementation of a quantum walk of a coin on a lattice, where the initial state of the walker is in a superposition of lattice sites. 
	For our purposes, it is possible to consider a one-dimensional lattice formed by the energy eigenstates of a harmonic oscillator, while the coin can be considered a two-dimensional system formed by the electronic states of an ion. Such a system has already been studied in the context of quantum random walks~\cite{Schmitz2009}. 
	However, we provide a generalized treatment, which can be accommodated for any physical system on which coherent states can be prepared. 
	
	Quantum walks are comparatively easier to handle as they are just an iteration over simple unitary operations. 
	Moreover, using our current scheme, we are able to efficiently prepare a larger class of HE states than earlier schemes. 
	We provide a characterization of the different HE states that we are able to generate with our scheme and provide the necessary ingredients to achieve the same (with high fidelity) in the lab.
	
	Our scheme offers an extremely good fidelity of preparation of such states, which decreases with the number of steps taken in the quantum walk. 
	However, even with as high as $60$ time steps (which, to the best of our knowledge, is hard to manage), we are still able to achieve a fidelity greater than $98 \%$. 
	For as low as $20$ time steps (which is much more practical~\cite{QWStep23,QWStep23a}), we are able to generate HE states with fidelity greater than $99\%$.
	This makes our scheme experimentally viable with the current technology and allows for better efficiency, fidelity and rate of generation than earlier proposals. 
	
	Furthermore, our scheme not only showcases a practical application of quantum walks but also underscores its unique quantum advantage. 
	In particular, current research in quantum walks has been focused on the generation of several different classes of quantum states~\cite{Innocenti2017,Majury2018,Giordani2019,Fabio2023}. 
	Specifically, the generation of highly entangled states using quantum walks is a topic of ongoing research~\cite{Gustavo2013, Chandru2022}. 
	In this context, and given the increasing significance of HE states in quantum communication, our approach holds the promise of becoming a pioneering contribution to this burgeoning field, further advancing its overall development.
	
	We organize the paper as follows. In Sec.~\ref{sec:back}, we provide the necessary background details on HE states and quantum walks, which will be used throughout the paper. 
	In Sec.~\ref{sec:he_quant_walk}, we detail our scheme to generate HE states using quantum walks and provide a method to characterize the resultant state. 
	In Sec.~\ref{sec:fidel}, we evaluate the fidelity of the resultant HE state. 
	Since it is only possible to numerically evaluate the fidelity given the initial state of the coin-lattice system, we provide complete details on how it can be done for arbitrary scenarios. 
	In Sec.~\ref{sec:conc}, we conclude our results by discussing its advantages and disadvantages, provide a comparison with prior implementations, and also lay out the groundwork for future prospects of our scheme.

	%%%%%%%%%%%%%%%%%%%%%%%%%%%%%%%%%%%%%%%%%%%%%%%%%%%%%%%%%%%%%%%%%%%%%%%%%%%%%%%%%%%%%%%%
	
	\section{Background}
	\label{sec:back}
	
	In this section, we provide the necessary background information on HE states and quantum walks that are relevant to our work. We also fix several notations that we use throughout the paper.
	
	%%%%%%%%%%%%%%%%%%%%%%%%%%%%%%%%%%%%%%%%%%%%%%%%%%%%%%%%%%%%%%%%%%%%%%%%%%%%%%%%%%%%%%%%
	\subsection{Hybrid entangled states}
	\label{sec:he}
	
	We start by giving a brief description of the HE states, which have been generated so far on optical systems. 
	Bipartite HE states are composed of two subsystems, where one of the subsystems corresponds to a qubit state, and the other corresponds to a coherent state. 
	Generally, the qubit subsystem is defined using the basis $\lbrace \ket{0}, \ket{1}\rbrace$, while the subsystem corresponding to a coherent state is defined by $\lbrace \ket{\alpha_1}, \ket{\alpha_2}\rbrace $, where $\alpha_1$ and $\alpha_2$ are the coherent amplitudes. 
	A hybrid entangled (HE) state is an entangled pair where the entanglement is between qubit states and coherent states. 
	A general HE state can be mathematically written as~\cite{KP12, JZK14}
	\begin{equation}
		\ket{\psi}_{a_1 a_2}=\frac{1}{\sqrt{2}}\left( \ket{0}_{a_1}\ket{\alpha_1}_{a_2} + \ket{1}_{a_1}\ket{\alpha_2}_{a_2} \right),
		\label{eq:he_state}
	\end{equation}
	where $a_1$ and $a_2$ are the two subsystems pertaining to the qubit and coherent subsystem, respectively. 
	For $\alpha_1 = -\alpha_2 = \alpha$, we obtain a symmetric form of a HE state~\cite{KP12}, which can be written as
	\begin{equation}
		\ket{\psi}_{a_1 a_2}=\frac{1}{\sqrt{2}}\left( \ket{0}_{a_1}\ket{\alpha}_{a_2} + \ket{1}_{a_1}\ket{-\alpha}_{a_2} \right).
		\label{eq:symm_he_state}
	\end{equation}
	
	Such states with small coherent amplitudes, $\alpha\lessapprox 1$ are experimentally available~\cite{JZK14}. 
	They have been generated experimentally using photon addition on a coherent state as well as photon subtraction on two-mode squeezed states.
	These experimental techniques produce HE states with non-unit probability and very low fidelity ($\lessapprox 80 \%$) for low coherent amplitude $0.4\lessapprox \alpha \leq 1$~\cite{JZK14}.
	
	A coherent state is mathematically represented as a superposition of energy eigenstates $\lbrace \ket{j} \rbrace$ of a harmonic oscillator as~\cite{SZ97}
	\begin{equation}
		\ket{\alpha} = e^{-\frac{\abs{\alpha}^2}{2}} \sum_{j = 0}^{\infty} \frac{\alpha^j}{\sqrt{j!}} \ket{j}.
		\label{eq:coherent}
	\end{equation}
	
	Coherent states were originally studied by Schr{\"o}dinger for a single particle having some mass confined in a harmonic oscillator potential energy well. 
	However, such states are now widely studied in (quantum) optics, where the energy eigenstates can be replaced by the number states of photons (Fock states). 
	A crucial point to note here is that the energy eigenstates (number states) $\lbrace \ket{j}\rbrace$ are discrete states which belong to a countably infinite-dimensional Hilbert space, while the coherent amplitude $\alpha$ can take a continuous range of values.

	%%%%%%%%%%%%%%%%%%%%%%%%%%%%%%%%%%%%%%%%%%%%%%%%%%%%%%%%%%%%%%%%%%%%%%%%%%%%%%%%%%%%%%%%
	\subsection{Quantum walks}
	\label{sec:quant_walk}
	
	\begin{figure}
		\centering
		\includegraphics[scale = 0.4]{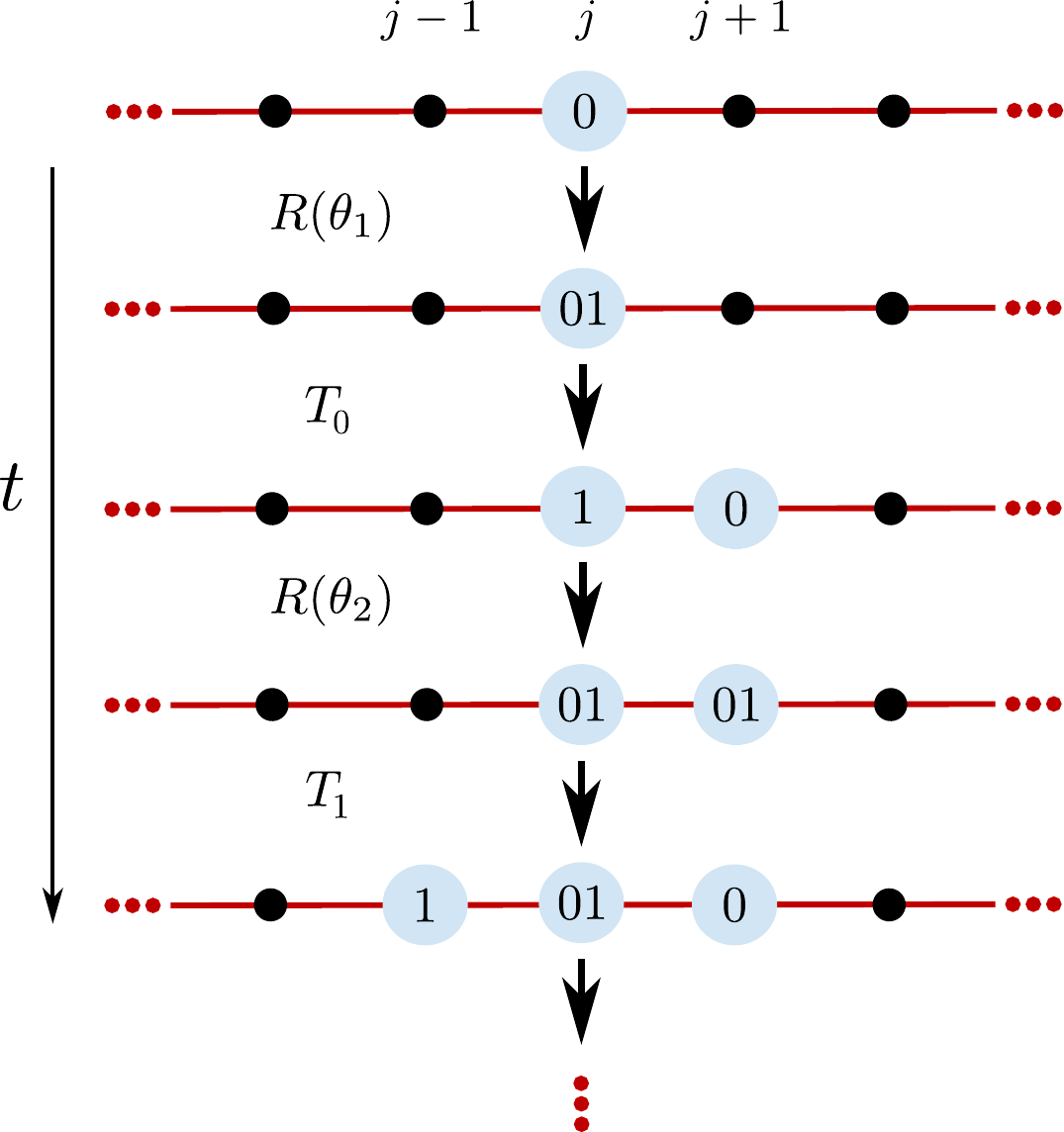}
		\caption{A schematic representation of a $1$-D SSQW consisting of two shift operators and two coin operators. The black dots represent lattice points, and the blue circles represent the walker whose internal degrees of freedom are represented by either $0$, $1$, or $01$ (superposition of $0$ and $1$).}
		\label{fig:fig1}
	\end{figure}
	
	Quantum walks are a quantum counterpart of classical random walks~\cite{Aharonov1993, Kempe2003}, where the variance of the position of the walker grows quadratically in time, unlike linearly in classical random walks~\cite{VenegasAndraca2012}. 
	They have been instrumental in the development of  
	quantum algorithms~\cite{Ambainis2003, Childs2004, Shenvi2003, Agliari2010}, quantum simulations~\cite{Nicola2014}, universal quantum computation~\cite{Childs2009, Childs2013, Lovett2010}, implementation of generalized measurements~\cite{QWMeasurement2013,QWStep23a} and in the study of topological phases~\cite{Kitagawa2010,Kitagawa2012,Asboth2012,Mittal2021}. 
	They are also of great interest in comprehending several non-classical features, such as quantum correlations, entanglement, and the phenomenon of decoherence~\cite{QWEntang2005, QWDeco2007, QWDeco2014}. 
	
	Quantum walks are broadly categorized into discrete-time and continuous-time quantum walks. 
	The dynamics of the walker in the latter category can be defined in terms of the lattice only~\cite{ContinousQW1998}, while in the former category, we require an additional coin operator~\cite{Ambainis2001}. 
	Quantum walks with a coin have been shown to be faster (for search algorithms) as compared to other protocols of quantum walks without the coin degree of freedom ~\cite{Ambainis2005}. 
	For the remainder of this paper, we will only focus on discrete-time quantum walks (DTQW), which use a coin operator to define the dynamics of the walker on a lattice.
	
	A DTQW is defined on a lattice with a quantum coin (also termed as a walker). 
	We define a $d$-dimensional lattice as a set of $N$ discrete points on which a walker can hop to, where $N \in \mathbb{Z}$ and $d$ is the real dimension of the lattice. 
	Corresponding to a $d$-dimensional lattice, we generally take a $2^d$ level system as a quantum coin. 
	The state of the coin determines in which direction the walker will move on the lattice. 
	For the remainder of this paper, we will focus on a $1$-D lattice, where each lattice point corresponds to a discrete quantum state $\ket{j} \in \mathcal{H}_L$ ($j \in \lbrace 0, 1,\ldots, N\rbrace$) where $\mathcal{H}_L$ is the Hilbert space of the lattice, and a two-level quantum coin with discrete quantum states denoted by $\lbrace \ket{0}, \ket{1}\rbrace \in \mathcal{H}_C$, where $\mathcal{H}_C$ is the Hilbert space of the coin. 
	We define the Hilbert space of the composite system of the coin and lattice as $\mathcal{H} = \mathcal{H_C} \otimes \mathcal{H_L}$. 
	Depending on the state of the quantum coin, the walker can either move to the right or the left of its current position. 
	
	The simplest quantum walk on a $1$-D lattice consists of a single coin operator $R(\theta)$ and a translation operator $T$, acting on the quantum coin and the walker, respectively. 
	The operation of the latter depends on the state of the quantum coin. 
	For example, the translation operator shifts the walker either to the right or to the left when the state of the coin is either $\ket{0}$ or $\ket{1}$, respectively. 
	The coin and the translation operators can be written as
	\begin{equation}
		R(\theta) = e^{-i \vb{n} \vdot \boldsymbol{\sigma} \theta/2} \otimes \mathds{1}_L,
		\label{eq:1D-rotation}
	\end{equation}
	and
	\begin{equation}
		T = \sum_{j = 1}^{N - 1} \dyad{0} \otimes \dyad{j+1}{j}  + \dyad{1} \otimes \dyad{j-1}{j},
		\label{eq:1D-Translation}
	\end{equation}
	
	where $\vb{n}$ is a unit vector, $\boldsymbol{\sigma}$ is a vector of Pauli matrices, $\theta \in \left[-2\pi, 2\pi\right]$ and $\mathds{1}_L$ is the identity operator acting on the lattice. Note that we do not impose any boundary conditions on the translation operators, which would also be the case in a practical scenario. Due to this, the translation operators are no longer unitary. While this does not affect our technique (or the quantum walk in general), the translation operators can be approximated to unitary transformations when the number of lattice points $N$ are taken to be large
	As a consequence, the summation in Eq.~\eqref{eq:1D-Translation} starts from $j = 1$ and not $j = 0$ because, in the absence of any boundary conditions, there is no place for the walker to hop to below $j = 0$. 
	Therefore, the translation operator has no physical meaning for the lattice point $j = 0$ (other than the interpretation that the particle is annihilated). 
	Additionally, there is also an upper bound, $j = N - 1$, in the summation. We impose this upper bound solely for the purpose of numerical simulation. The number of lattice points will depend on the physical architecture being used.
	This leads to the operator governing the time evolution of the walker for a single time step of the quantum walk, which is written as 
	\begin{equation}
		U(\theta) = T R(\theta).
	\end{equation}
	
	DTQW proceeds in discrete time steps, where at each step, the state of the coin is entangled with the position of the walker on the lattice. 
	The coin operation is applied first, modifying the state of the coin, and then a shift operation is applied, shifting the walker on the lattice according to the modified state of the coin. 
	By iterating over these coin and shift operations over multiple time steps, denoted by $t$, the state of the walker and, hence, the quantum walk evolves. 
	This leads to a spread of the walker's probability distribution of its position across the lattice, whose variance increases linearly with $t$ and with a much higher slope than for classical random walks (For more details, refer to Appendix~\ref{appendixA}). 
	
	A more generalized version of the $1$-D DTQW is defined by splitting the shift operator into two, separated by an additional coin operator. 
	This generalization is known as $1$-D split step quantum walk ($1$-D SSQW)~\cite{Kitagawa2010}. 
	The $1$-D SSQW consists of two shift operators namely, $T_{0}$ and $T_{1}$, which act on the walker with coin state $\ket{0}$ and $\ket{1}$ separately and are given as 
	\begin{equation}
		\begin{aligned}
			T_{0} &= \sum_{j = 1}^{N - 1} \dyad{0} \otimes \dyad{j+1}{j}  + \dyad{1}  \otimes \mathds{1}_L \\
			T_{1} &= \sum_{j = 1}^{N - 1} \dyad{0}  \otimes \mathds{1}_L  + \dyad{1} \otimes \dyad{j-1}{j},
		\end{aligned}
		\label{eq:translation}
	\end{equation}
	
	The operator governing the time evolution of the $1$-D SSQW for a single time step then becomes~\cite{Kitagawa2010}
	\begin{equation} 
		\label{eq:SSQW-Unitary}
		U_{_{\text{SS}}}(\theta_1, \theta_2) = T_{1} R(\theta_2) T_{0} R(\theta_1).
	\end{equation}
	with $\theta_1, \theta_2 \in [-2\pi, 2\pi]$. As can be seen, for $\theta_2 = 0$, the resultant $1$-D SSQW reduces to a conventional $1$-D DTQW. 
	
	We note that quantum walk is a powerful technique that can be implemented on many experimental platforms. 
	For an experimental realization, the coin-lattice system can be appropriately chosen to closely resemble the properties that are theoretically required.

	%%%%%%%%%%%%%%%%%%%%%%%%%%%%%%%%%%%%%%%%%%%%%%%%%%%%%%%%%%%%%%%%%%%%%%%%%%%%%%%%%%%%%%%%
	\section{Generating hybrid entangled states using quantum walks}
	\label{sec:he_quant_walk}
	
	\begin{figure*}
		\centering
		\subfigure[]{\includegraphics[width=0.23\textwidth]{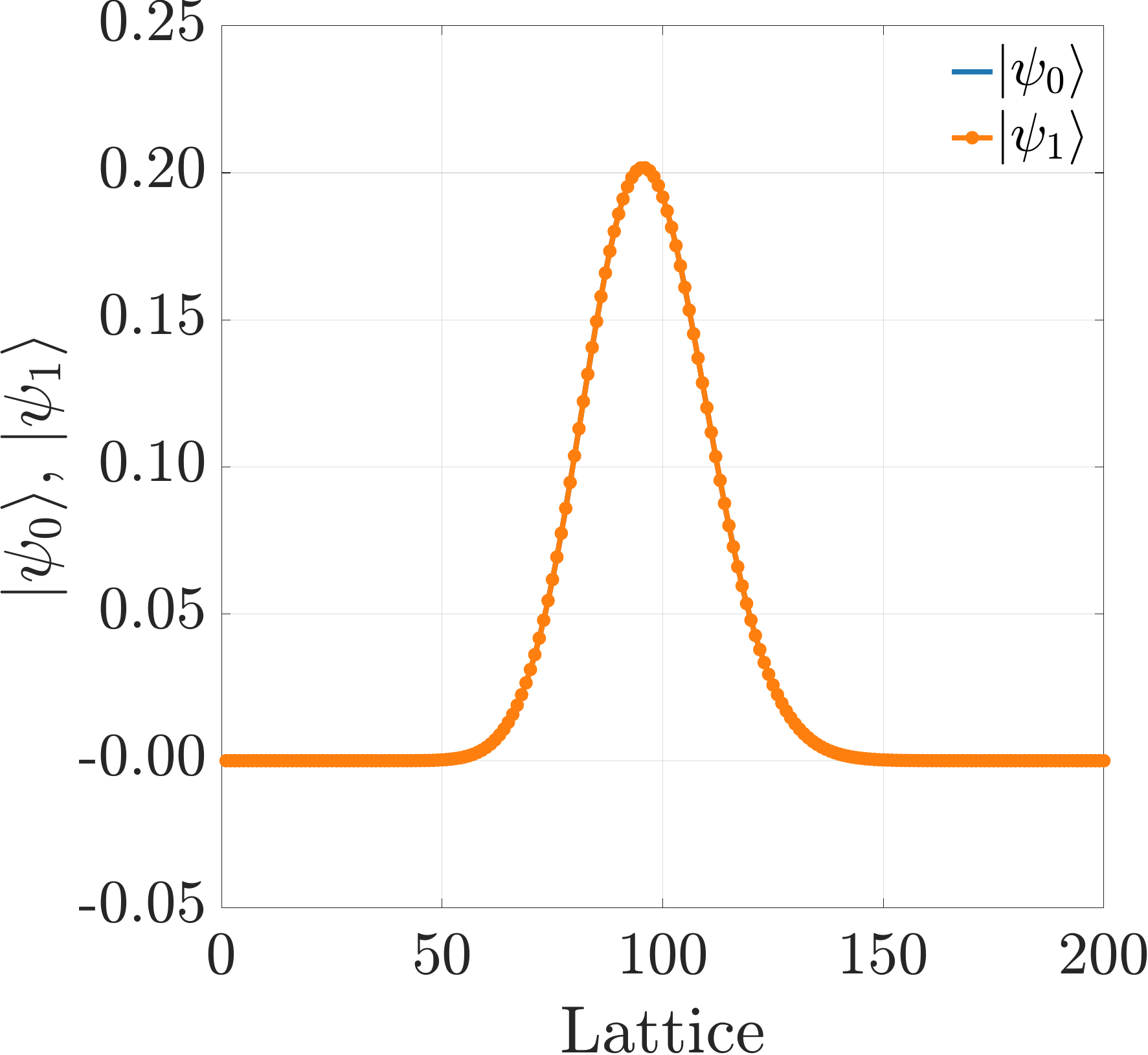}
			\label{fig:step0}}
		\subfigure[]{\includegraphics[width=0.23\textwidth]{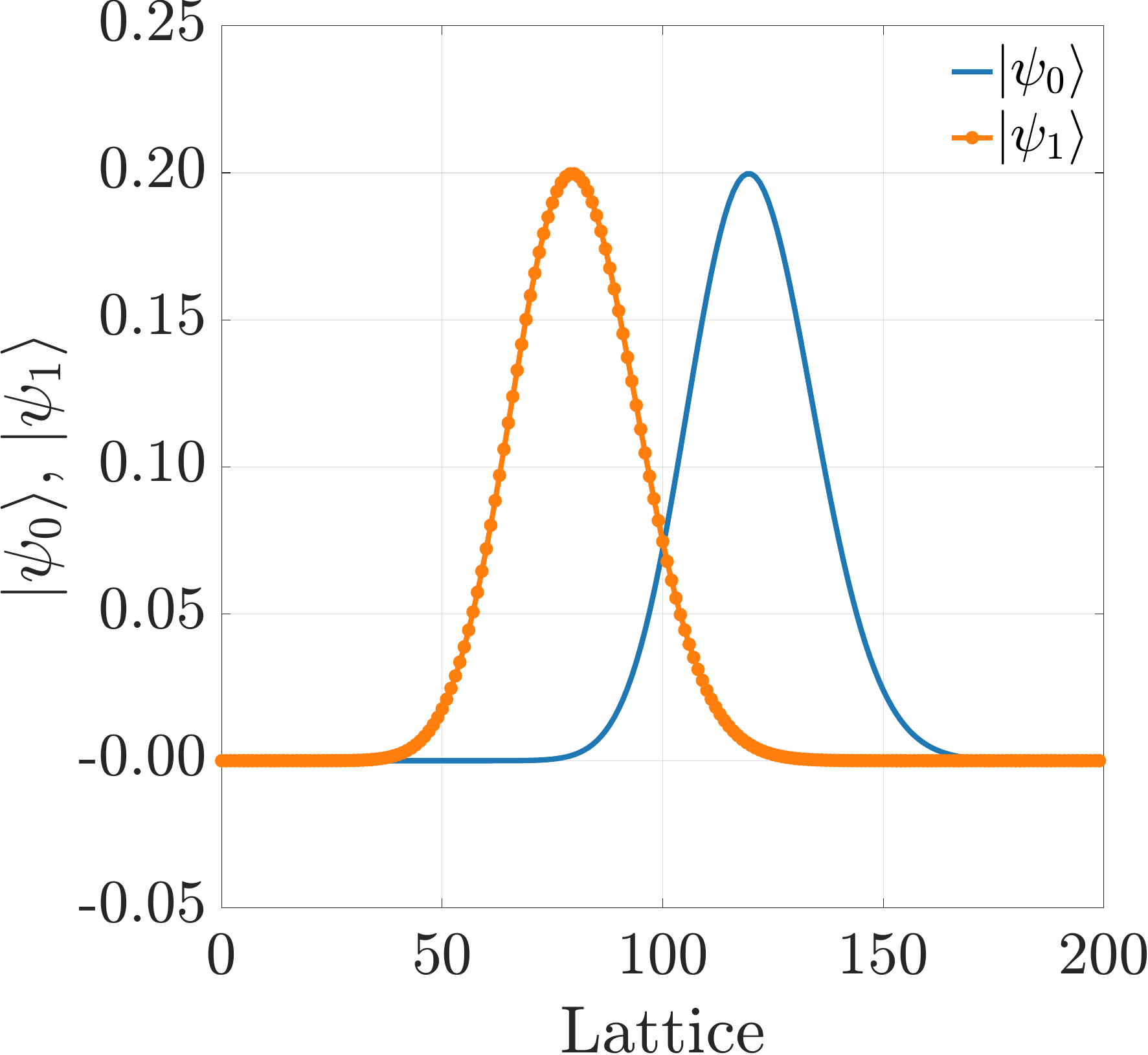}
			\label{fig:step20a}}
		\subfigure[]{\includegraphics[width=0.23\textwidth]{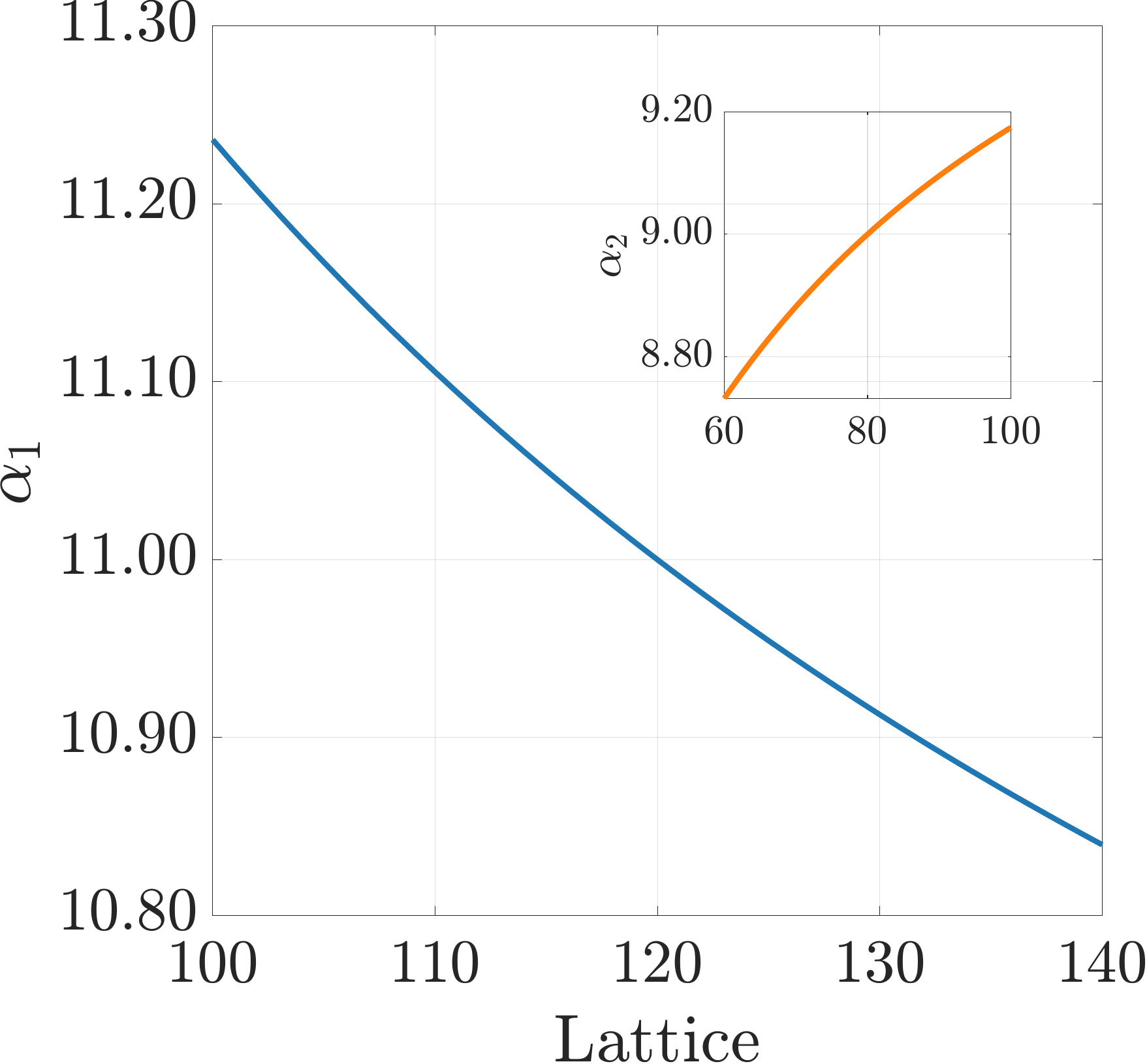}
			\label{fig:step20b}}
		
		\subfigure[]{\includegraphics[width=0.23\textwidth]{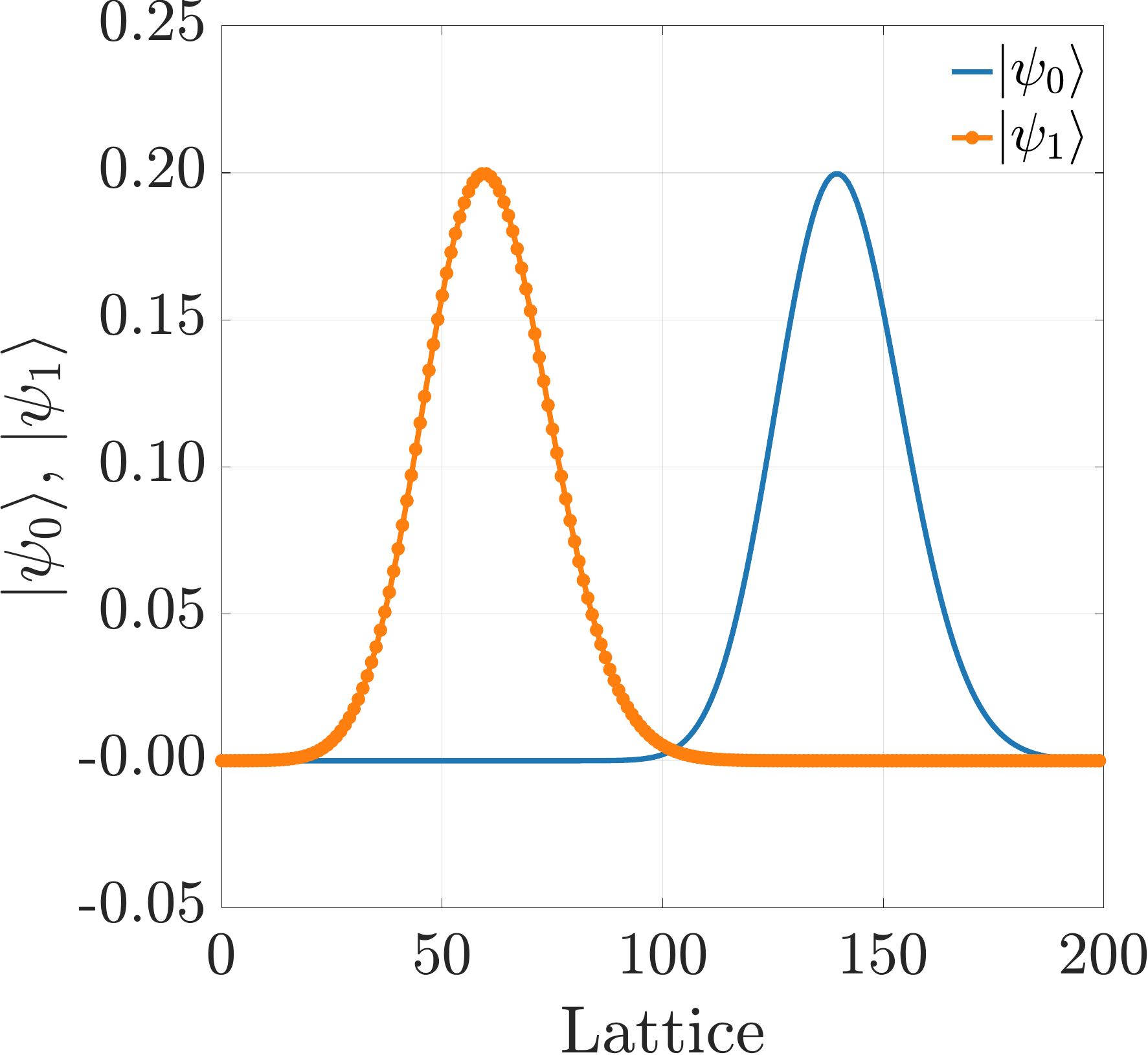}
			\label{fig:step60a}}
		\subfigure[]{\includegraphics[width=0.23\textwidth]{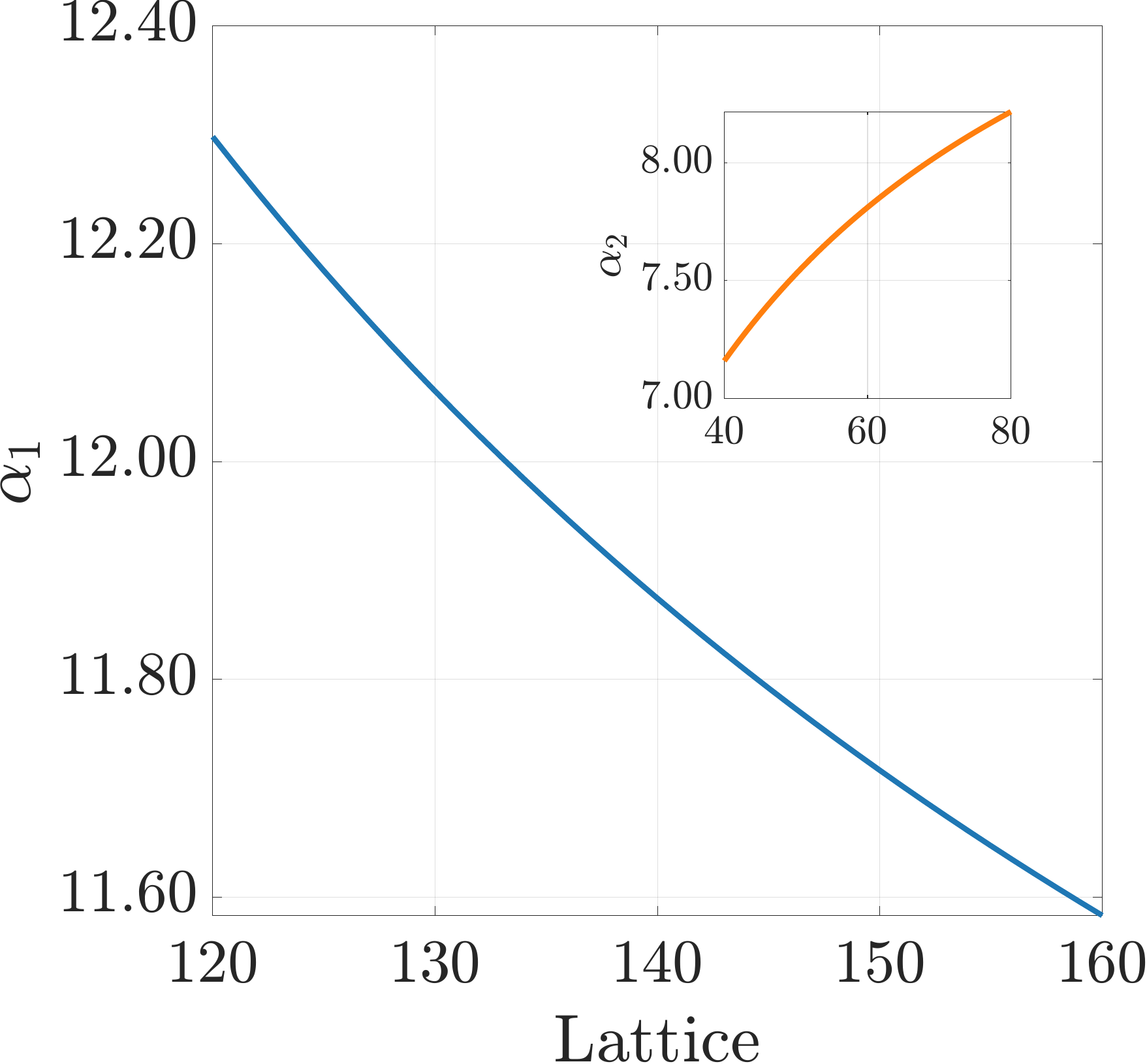}
			\label{fig:step60b}}
		\caption{The evolution of the lattice state after a quantum walk with lattice size $N = 200$ using coin parameters $\theta_1 = \theta_2 = 0$ and $\delta = 0$ for various time steps, with the initial coherent amplitude $\alpha_0 = 10$ and the initial state of the composite system is as defined in Eq.~\eqref{eq:intial_lattice_coin}.
			In (a), we plot the initial state of the lattice ($t = 0$ time steps), while in (b) we plot the final state of the lattice after $t = 20$ time steps. 
			In (c) we plot the variation of $\alpha^j_1$ and $\alpha^j_2$ by only considering a subset of all lattice points around $j^*$, as explained in Sec.~\ref{sec:he_quant_walk} and obtain $\Bar{\alpha}_1 = 11.0131$ and $\Bar{\alpha}_2 = 8.9839$. 
			In (d) we plot the final state of the lattice after $t = 40$ time steps while keeping all other parameters the same as before. 
			In (e) we plot the variation of $\alpha^j_1$ and $\alpha^j_2$, again by considering a subset of all lattice points around $j^*$ and obtain $\Bar{\alpha}_1 = 11.8973$ and $\Bar{\alpha}_2 = 7.7682$.}
		\label{fig:alpha10}
	\end{figure*}
	
	In this section we describe our scheme to generate HE states using quantum walks on a coin-lattice system. 
	However, since our scheme requires the creation of coherent states, the coin-lattice system cannot be chosen arbitrarily. 
	For our purposes, we choose the lattice to correspond to the energy eigenstates of a single particle confined in a harmonic potential well. 
	Additionally, the coin system can be chosen as some two-dimensional internal degrees of freedom of the single particle, say the long-lived electronic levels of an ion. We provide a brief outline of such a scheme in Sec.~\ref{sec:conc}.
	
	In the current paper, we identify the $1$-D lattice points $\lbrace \ket{j}\rbrace$ as energy eigenstates of a single particle confined in a harmonic potential energy well.
	However, for the remainder of the paper, we still keep the term lattice points.

	Moreover, it should be noted that the Hilbert space dimension of a $1$-D lattice is equal to the number of its lattice points, \textit{i.e.} $\text{dim}\mathcal{H}_L = N + 1$.
	Furthermore, the probability of hopping from one lattice point $j$ to either $j + 1$ or $j - 1$ is independent of $j$, as can be seen in Eq.~\eqref{eq:translation}. 
	Note the same is true for the rotation parameters, which are also independent of $j$, which is clear from Eq.~\eqref{eq:1D-rotation}. 
	Such quantum walks are known as homogeneous quantum walks. 
	
	Next, before performing the $1$-D SSQW, we first initialize the state of the lattice and the coin as~\cite{Su2019}
	\begin{equation}
		\begin{aligned}
			\ket{\psi(t = 0)}_L &= e^{-\frac{|\alpha_0|^2}{2}} \sum_{j = 0}^{N} \dfrac{\alpha_0^j}{\sqrt{j !}} \ket{j}\\
			\ket{\psi(t = 0)}_C &= \frac{1}{\sqrt{2}}\left( \ket{0} + e^{i \delta}\ket{1}\right),
		\end{aligned}
		\label{eq:intial_lattice_coin}
	\end{equation}
	where $\ket{\psi(t)}_L$ and $\ket{\psi(t)}_C$ are the states of the lattice and coin after $t$ time steps respectively, $\delta \in \left[0, \frac{\pi}{2}\right]$ is the relative phase factor and $\alpha_0$ is a constant parameter corresponding to the coherent amplitude of coherent states. 
	Here, we note three points:
	\begin{enumerate}
		\item While the number of terms corresponding to $\ket{j}$ appearing in Eq.~\eqref{eq:coherent} are countably infinite, we only employ a finite (but large) number of terms that correspond to the total number of lattice points in Eq.~\eqref{eq:intial_lattice_coin} to perform our computations. 
		The reason for this is because for large values of $j$, the corresponding coefficients satisfy $e^{-\frac{\alpha_0^2}{2}} \frac{|\alpha_0|^j}{\sqrt{j!}} \approx 0$. 
		For our purposes, we can safely take $N = 200$ without any significant repercussions (a lattice size of this order has also been achieved experimentally~\cite{Regensburger2011,DiColandrea23}). Additionally, for a large number of accessible lattice points, the conditional shift operators are almost unitary.
		
		\item In order to preserve the symmetry of the final HE states, we only consider $\delta = \lbrace 0, \frac{\pi}{2}\rbrace$. As can be seen, this ensures that the state of the coin is symmetric, which consequently allows us to retain the symmetry in the generated HE states. We note that other values of $\delta$ are certainly permissible, but they do not yield the desired results.
		
		\item It should be noted that the position of the walker on the lattice is distinct from the real space position of the particle in the harmonic oscillator potential. 
		Remember that the lattice is identified as the energy levels of a harmonic oscillator; and the position of the walker on the lattice corresponds to the energy of the particle. 
	\end{enumerate}

	For the set of initial states in Eq.~\eqref{eq:intial_lattice_coin}, we perform a $1$-D SSQW on the lattice. 
	For simplicity, we choose the two coin operators with parameters
	\remove{$\delta = 0$ and} $\theta_1 = \theta_2 = 0$ as this yields a simpler version of the quantum walk with no local coin operations. Additionally, in the Appendix~\ref{appendixB} we present our findings for 
	$\delta = \pi/2$, $\theta_1 = \pi$ with $\vb{n}_1 = (\hat{x} + \hat{z})/\sqrt{2}$ and $\theta_2 = -\frac{\pi}{2}$ with $\vb{n}_2 = \hat{y}$ respectively. We choose this particular set of values of the parameters $\theta_1$ and $\theta_2$ in order to maximize the fidelity of the HE state produced by the quantum walk with a theoretical HE state. To determine these optimal values, we perform a full parameter scan of the fidelity as a function of $\theta_1$ and $\theta_2$, and choose the parameters that yielded the highest fidelity (we provide more details in Sec.~\ref{sec:fidel}).
	
	However, we also note that arbitrary values of $\theta_1$ and $\theta_2$ do not yield the desired HE states, and they have to be carefully tailored.
	We represent the resultant composite state of the coin and lattice after $t$ time steps of a quantum walk by $\ket{\Psi (t)}_{CL}$.

	We find that the resultant state of the coin-lattice system is very close to the HE state of the form in Eq.~\eqref{eq:he_state} (as shown in Fig.~\ref{fig:alpha10}). 
	The resultant HE state of the coin-lattice system can be written in the form
	\begin{equation}
		\ket{\Psi (t)}_{CL} = \frac{1}{\sqrt{2}}\left(\ket{0}\ket{\psi_0} + e^{i \phi} \ket{1}\ket{\psi_1}\right),
		\label{eq:final_state}
	\end{equation}
	where $\phi \in \lbrace 0, \frac{\pi}{2}, \pi\rbrace$ is a relative phase factor and 
	$\ket{\psi_0}$ and $\ket{\psi_1}$ are the conditional states of the lattice when the state of the coin is $\ket{0}$ and $\ket{1}$ respectively. 
	The relative phase factor $\phi$ is mainly decided by the value of $\delta$ chosen for the initial coin state, value of $\theta_1$ and $\theta_2$ chosen and the number of time steps taken in the quantum walk. 
	Moreover, the states $\ket{\psi_0}$ and $\ket{\psi_1}$ are either completely real or completely imaginary.
	It should be noted that the only reason we use $\ket{\psi_i}$ to denote the conditional states of the lattice and not $\ket{\alpha_i}$ is because these states are not \textit{exactly} coherent states.

	The state $\ket{\psi_0}$ can be evaluated by applying the projector $\dyad{0} \otimes \mathds{1}_L$ to the state $\ket{\Psi (t)}_{CL}$ and then taking a partial trace over the state of the coin. In a similar fashion, we can evaluate $\ket{\psi_1}$ by applying the projector $\dyad{1} \otimes \mathds{1}_L$. 
	
	In Fig.~\ref{fig:alpha10}, we plot the evolution of the conditional states of the lattice ($\ket{\psi_0}$ and $\ket{\psi_1}$), hereafter denoted simply as states of the lattice (lattice states), for $\alpha_0 = 10$ and for different values of time steps $t$.  We observe that the initial coherent state of the lattice (Fig.~\ref{fig:step0}) splits into two parts, each one corresponding to the two orthogonal coin states. Also, the resultant lattice states $\ket{\psi_0}$ and $\ket{\psi_1}$ after $20$ time steps retain the form of a coherent state with some additional deformation as the quantum walk evolves, which is evident from Fig.~\ref{fig:step20a}. Since the resultant lattice states are very close to coherent states, we define and identify them with coherent amplitudes $\Bar{\alpha}_1$ and $\Bar{\alpha}_2$ corresponding to $\ket{\psi_0}$ and $\ket{\psi_1}$ respectively. 
	
	For more detailed explanation of the resultant HE states after an arbitrary number of times steps and a different set of values of $\theta_1$ and $\theta_2$, we refer to Appendix~\ref{appendixB}.

	We now detail a method to calculate the estimated coherent amplitudes $\Bar{\alpha}_1$ and $\Bar{\alpha}_2$ of the resultant lattice states. It should be noted that $\Bar{\alpha}_1$ and $\Bar{\alpha}_2$ depend on the number of time steps taken in the quantum walk and the coherent amplitude $\alpha_0$ of the initial coherent state. 
	
	In order to detail a method to evaluate $\Bar{\alpha}_1$ and $\Bar{\alpha}_2$ of the resultant lattice states, we first describe a method to evaluate the unknown coherent amplitude of any coherent state~\cite{scully1997quantum} which may also be deformed as described above. We consider that the amplitude, $C_j$, corresponding to the lattice state $\ket{j}$ at each lattice point $j$ is accessible (as is our case). To begin with let us denote by $\alpha^*$ as the parameter which characterizes a coherent state or a deformed coherent state after the quantum walk. One particular method to evaluate the unknown parameter $\alpha^*$ is to look at the coefficients $C_j$ (and $C_{j + 1}$) of the lattice state corresponding to each lattice point $j$ (and $j + 1$), and then evaluate
	
	\begin{equation}
		\alpha^* = \frac{\sqrt{(j + 1)} C_{j + 1}}{C_j}.
		\label{eq:alpha_value}
	\end{equation}

	In an ideal situation, when the resultant state matches exactly with a coherent state then the coefficients $C_j = e^{-|\alpha|^2/2}\frac{\alpha^j}{\sqrt{j}}$, where $\alpha$ is the coherent amplitude. In this limiting case it is then observed that in Eq.~\eqref{eq:alpha_value} $\alpha^* = \alpha$. However, in our case, the resultant state after the quantum walk is a deformed coherent state such that the coefficients are not exactly equal to those mentioned above. However, Eq.~\eqref{eq:alpha_value} yields a reliable approximation to the coherent amplitude, which we can then use to characterize the resultant lattice states.

	%\chg{Using this method the required parameters $\Bar{\alpha}_1$ and $\Bar{\alpha}_2$ can be evaluated as detailed below.}
	%In another approach, the probabilities to observe $j$ photons, given by $P_j = e^{|\alpha|^2 \frac{|\alpha|^{2j}}{j!}}$ are evaluated, followed by performing a curve-fitting to determine $\alpha^*$. This approach can also be used to determine $\Bar{\alpha}_1$ and $\Bar{\alpha}_2$ as detailed in the appendix. Additionally, we have also found that both approaches yield similar results in determining $\Bar{\alpha}_1$ and $\Bar{\alpha}_2$ (as shown in the appendix).}

For an ideal coherent state, the value of $\alpha^*$, as calculated according to Eq.~\eqref{eq:alpha_value} is the same irrespective of the coefficients $C_j$ and $C_{j + 1}$ chosen. However, for the final states $\ket{\psi_0}$ and $\ket{\psi_1}$, this is not the case because of a small dispersion of probability on neighboring lattice sites. This implies that the coherent amplitude, as calculated according to Eq.~\eqref{eq:alpha_value}, will be slightly different for different lattice points $j$. Taking this effect into account, we define $\alpha^j_i$ as the coherent amplitude of state $\ket{\psi_i}$ at the lattice point $j$. 

Next, we identify the lattice point $j^*_i$ where the amplitude corresponding to the lattice state, $\ket{\psi_i}$ is maximum. As it can be seen in Fig.~\ref{fig:step20a} the amplitude peaks at the lattice sites $j^*_0 = 120$ and $j^*_1 = 80$ for $\ket{\psi_0}$ and $\ket{\psi_1}$ respectively after $t = 20$ time steps. 
Then, we only consider a subset of lattice points around the lattice point $j^*$ in order to calculate $\alpha^j_i$ using Eq.~\eqref{eq:alpha_value}.  
Finally, we average over all these $\alpha^j_i$ to obtain $\Bar{\alpha}_i$, which we then use to characterize the coherent state part of the resultant HE state. As an example, we consider a set of $40$ lattice points around $j^*_0$ and $j^*_1$ (as seen in Fig.~\ref{fig:step20b}) to obtain the estimates of the coherent amplitude $\Bar{\alpha}_1 = 11.0131$ and $\Bar{\alpha}_2 = 8.9839$.

The reason for choosing a small finite subset of lattice points to evaluate $\Bar{\alpha}_i$ is because the dispersion of probability on neighbouring lattice points becomes prominent for $j_i\gg j^*_i$ or $j_i \ll j^*_i$. 
This leads to a large deviation in the value of $\alpha^j_i$ obtained from Eq.~\eqref{eq:alpha_value}. 
As a consequence, if we consider all values of $\alpha_i$, then the resultant $\Bar{\alpha}_i$ is found to be non-optimal, in the sense that the resultant state gives a much lower fidelity (see Sec.~\ref{sec:fidel}). 
In Sec.~\ref{sec:fidel} we provide a much more detailed discussion on why we choose $40$ lattice points around $j^*_0$ and $j^*_1$.

We perform a similar analysis for $t = 40$ time steps and plot the corresponding results in Fig.~\ref{fig:step60a} and \ref{fig:step60b}. Again, we consider a set of $40$ lattice points around $j^*_0 = 140$ and $j^*_1 = 60$, from which we obtain $\Bar{\alpha}_1 = 11.8973$ and $\Bar{\alpha}_2 = 7.7682$.

Let us define $\psi_{qw} = \lbrace \ket{\psi_0}, \ket{\psi_1}\rbrace$ be the set of lattice states generated after performing the quantum walk corresponding to coin states $\ket{0}$ and $\ket{1}$ respectively. Similarly, let $\psi_{th} = \lbrace \ket{\Bar{\alpha}_1}, \ket{\Bar{\alpha}_2}\rbrace$
be the set of theoretical coherent states with coherent amplitudes $\Bar{\alpha}_1$ and $\Bar{\alpha}_2$ respectively. As can be seen in Fig.~\ref{fig:fid}, the generated states $\psi_{qw}$ are in good agreement with theoretical coherent states $\psi_{th}$.

We also note that it is also possible to create a symmetric form of a HE state as given in Eq.~\eqref{eq:symm_he_state}. To do so, we note that the coherent amplitudes $\Bar{\alpha}_1$ and $\Bar{\alpha}_2$ can be written as $\Bar{\alpha}_1 = g \Bar{\alpha}_2$, where $g \in \mathbb{R}$. By carefully choosing a displacement operator $D_L(\alpha)$, we can shift the lattice states to obtain a symmetric HE state. Specifically, we have
\begin{equation}
	\begin{aligned}
		\ket*{\tilde{\Psi} (t)}_{CL} &=  \left[\mathds{1}_C \otimes D_L\left(-\frac{\Bar{\alpha}_2 + g\Bar{\alpha}_2}{2}\right) \right]\ket*{\Psi (t)}_{CL}\\
		&= \frac{1}{\sqrt{2}}\left(\ket{0}\ket{\psi'_0} + \ket{1}\ket{\psi'_1}\right),
	\end{aligned}
\end{equation}
where $\ket{\psi'_0}$ has coherent amplitude $\Bar{\alpha} = \frac{g \Bar{\alpha}_2 - \Bar{\alpha}_2}{2}$ and $\ket{\psi'_1}$ has coherent amplitude $-\Bar{\alpha}$. This state is symmetric and can be written in the form of Eq.~\eqref{eq:symm_he_state}. This operation simply shifts the coherent state by the appropriate magnitude.
This operation is applied once the HE state (as given in Eq.~\eqref{eq:final_state}) has been generated. Since this operation can be done for any resultant HE state, we instead focus our efforts on generating and characterizing the states in Eq.~\eqref{eq:final_state} for different time steps of the quantum walk. The only reason (to the best of our knowledge) for performing the displacement operation is to bring the resultant state in a symmetric form, which has been shown to have applications in quantum information~\cite{LJ13, ANV15, OTJ20, OTL21, BJ22, HM22, KJ13, Sheng2013, YLim2016}.

%%%%%%%%%%%%%%%%%%%%%%%%%%%%%%%%%%%%%%%%%%%%%%%%%%%%%%%%%%%%%%%%%%%%%%%%%%%%%%%%%%%%%%%%
\section{Fidelity of the resultant state}
\label{sec:fidel}

\begin{figure}
	\centering
	\subfigure[]{\includegraphics[width=0.23\textwidth]{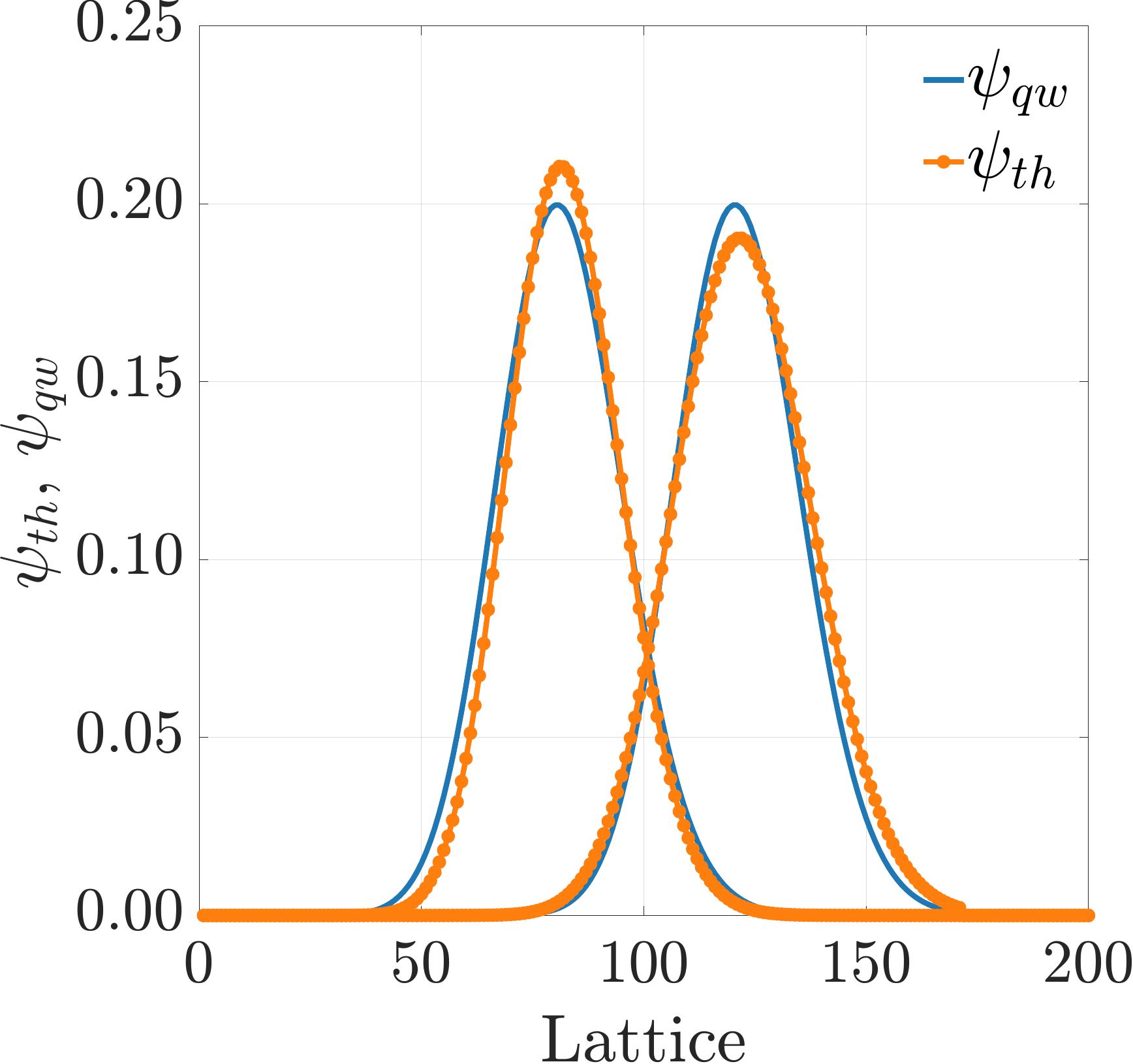}
		\label{fig:step20c}}
	\subfigure[]{\includegraphics[width=0.23\textwidth]{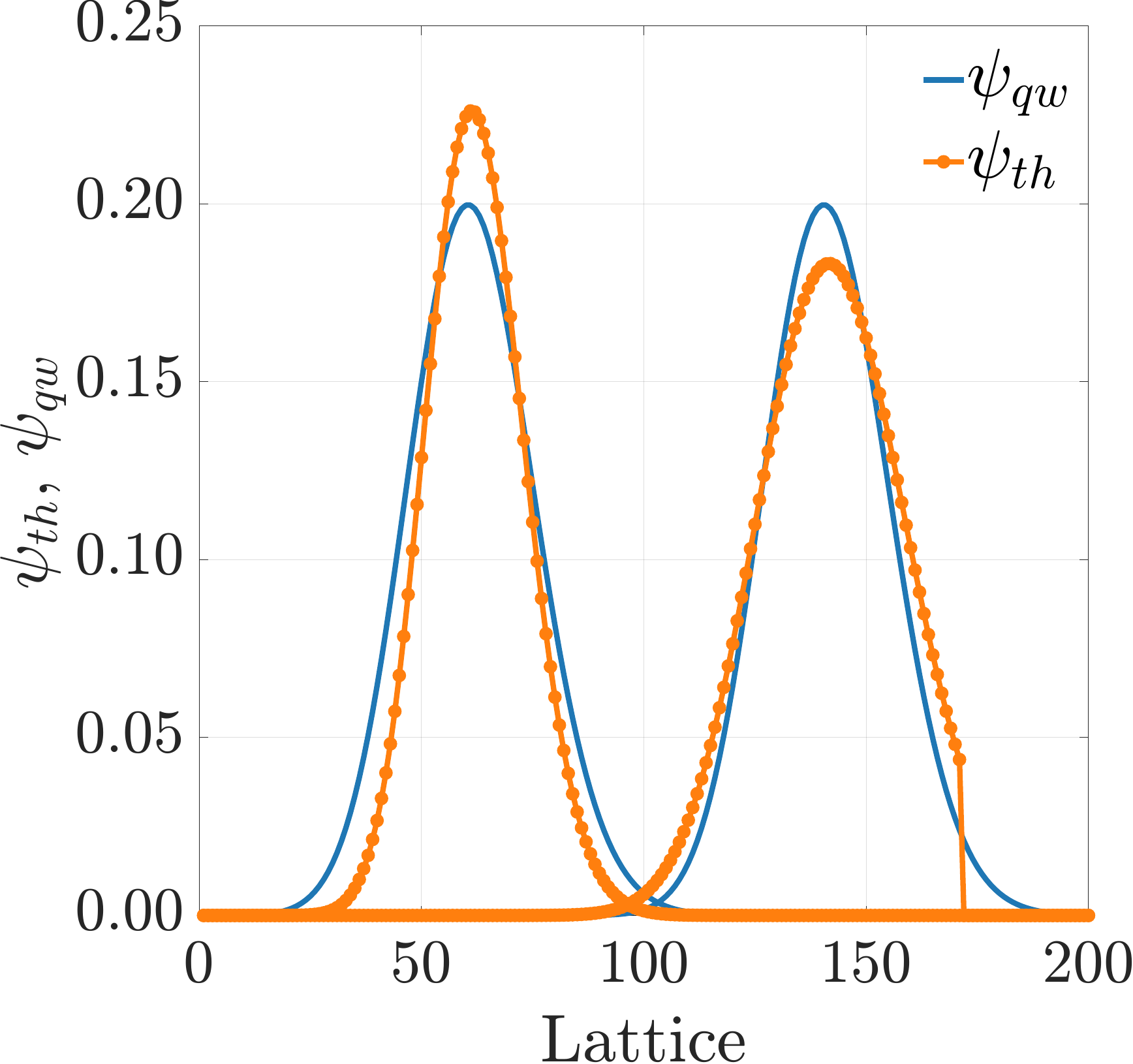}
		\label{fig:step60c}}
	\caption{The generated ($\psi_{qw}$) and the targeted coherent states ($\psi_{th}$) with coherent amplitudes $\Bar{\alpha}_1$ and $\Bar{\alpha}_2$ on the lattice for time steps (a) $t = 20$, (b) $t = 40$. We obtain a fidelity of $99.23 \%$ and $97.05 \%$ after $t = 20$ and $t = 40$ time steps respectively. Here, we obtain $\Bar{\alpha}_1 = 11.0131$ and $\Bar{\alpha}_2 = 8.9839$. All other parameters are taken to be the same as in Fig.~\ref{fig:alpha10}.}
	\label{fig:fid}
\end{figure}

Using our proposed methodology, it is possible to deterministically generate bipartite entangled states that are extremely close to HE states using only simple unitary operations. Note that the states on the lattices corresponding to the two coin states are not exactly Gaussian, and this deviation from a Gaussian envelope increases with the number of time steps (as evident from Fig.~\ref{fig:step20a} and Fig.~\ref{fig:step60a}). In this section, we quantify the `closeness' of our generated states to HE states using trace fidelity. 

The trace fidelity of a density matrix $\rho$ with respect to another density matrix $\sigma$ is defined as the probability with which $\rho$ is indistinguishable from $\sigma$ in any experimental test. Mathematically, it can be written as~\cite{NC10, J94}
\begin{equation} \label{eq:fidelity}
	\mathcal{F}(\rho, \sigma) :=  \left[\text{tr} \left(\sqrt{\sqrt{\rho}\sigma \sqrt{\rho}}\right)\right]^2.
\end{equation}
The above definition reduces to
\begin{equation}
	\mathcal{F}(\ket{\psi}, \ket{\phi}) = \abs{\ip{\psi}{\phi}}^2,
\end{equation}
if $\sigma = \dyad{\psi}$ and $\rho = \dyad{\phi}$ are pure states. 

We now compute the fidelity of the HE state generated via our method. For our purposes, we take
\begin{equation}
	\begin{aligned}
		\ket{\psi} &= \frac{1}{\sqrt{2}}\left(\ket{0}\ket{\psi_0} + \ket{1}\ket{\psi_1}\right),\\
		\ket{\phi} &= \frac{1}{\sqrt{2}}\left(\ket{0}\ket{\Bar{\alpha}_1} + \ket{1}\ket{\Bar{\alpha}_2}\right),
	\end{aligned}
\end{equation}
to compute the fidelity, where $\ket{\psi}$ is the state generated after performing the quantum walk and $\ket{\phi}$ is the theoretical HE state that our scheme is targeting. As a result, we obtain
\begin{equation}
	\mathcal{F}(\ket{\psi}, \ket{\phi}) = \abs{\ip*{\psi_0}{\Bar{\alpha}_1} + \ip*{\psi_1}{\Bar{\alpha}_2}}^2,
\end{equation} 
where $\ket*{\Bar{\alpha}_i}$ are the targeted theoretical coherent states. Secondly, we note that we evaluate the fidelity for states of the form in Eq.~\eqref{eq:final_state}, which may not be in the symmetric form. This is due to the fact that the displacement operation is simply a relabeling of the lattice sites and can be accomplished locally (on the lattice) since the system is translation invariant.

\begin{figure}
	\centering
	\subfigure[]{\includegraphics[width=0.235\textwidth]{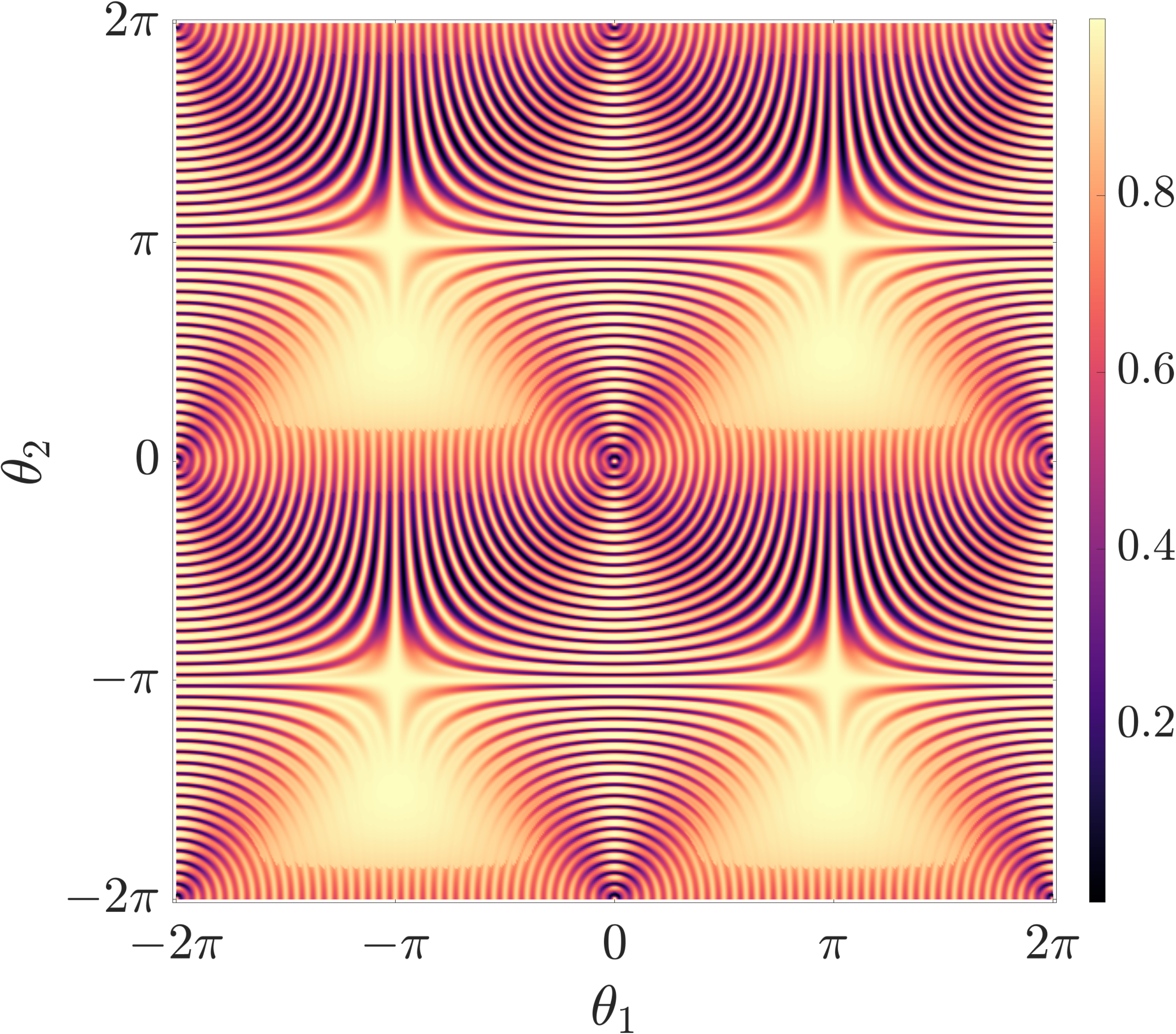}}
	\subfigure[]{\includegraphics[width=0.235\textwidth]{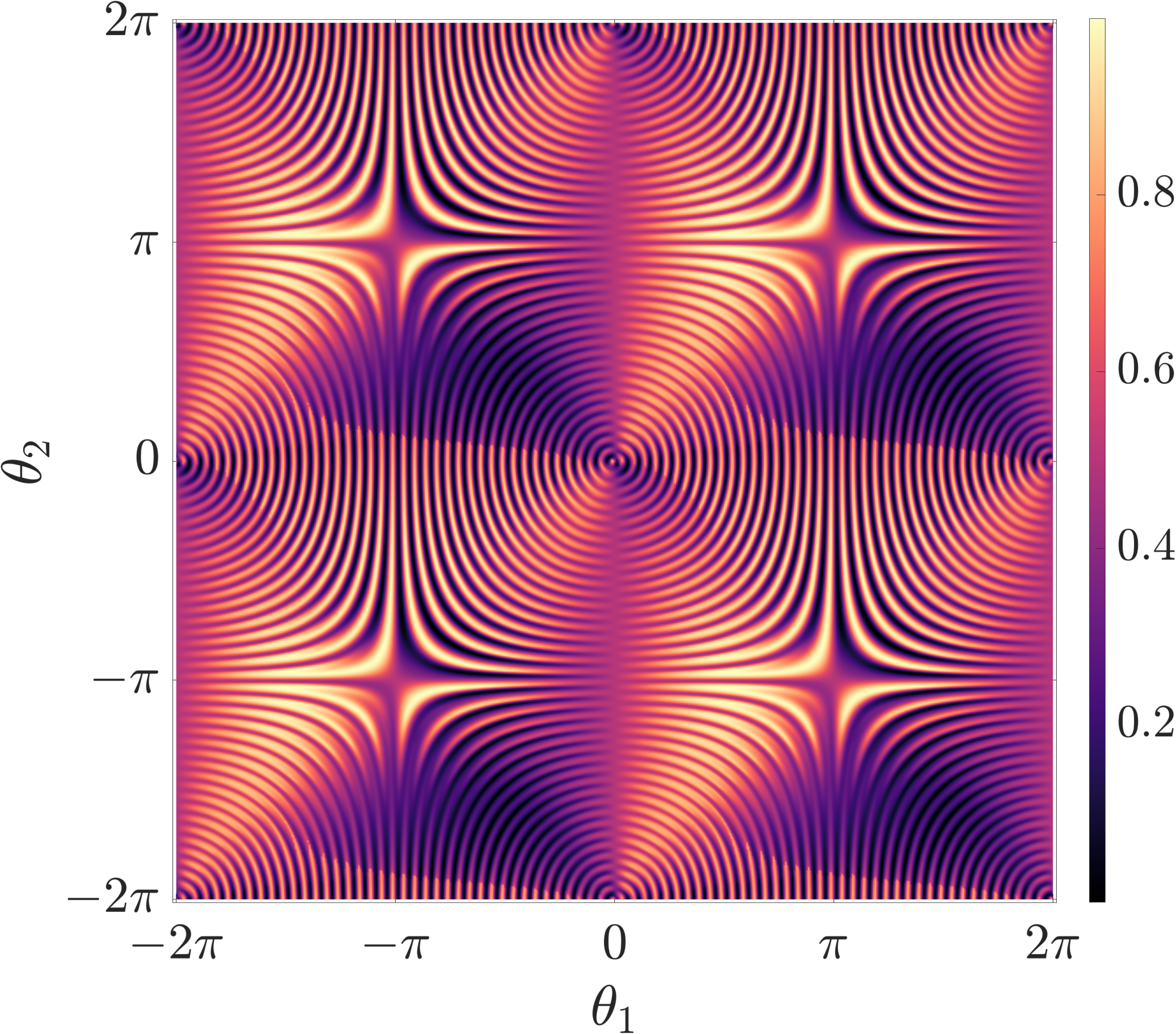}}
	\caption{A contour plot of the fidelities of the generated HE states after $t = 40$ time steps, as a function of the rotation angles of the coin operator $\theta_1$ and $\theta_2$ for two different initial states of the coin with (a) $\delta = 0$ and (b) $\delta = \pi/2$.}
	\label{fig:fidelity}
\end{figure}

We first address the choice of the parameters $\theta_1$ and $\theta_2$ in Sec.~\ref{sec:he_quant_walk}, by scanning over the full set of their permissible values and the corresponding fidelity of the HE state that can be achieved. In Fig.~\ref{fig:fidelity} we present a density plot of fidelity as a function of $\theta_1$ and $\theta_2$. We take $N = 200$ and evolve the quantum walk for $t = 40$ time steps to calculate the fidelity. We consider the same initial state of the lattice as given in Eq.~\eqref{eq:intial_lattice_coin} but two different initial states of the coin with (a) $\delta = 0$ and (b) $\delta = \pi/2$. From the density plots it can be observed that there exist several sets of values of $\theta_1$ and $\theta_2$ that yield a good fidelity of the generated HE state. For example, one set of values $\left(\theta_1, \theta_2\right) = \left( \pi, -\pi/2 \right)$ corresponding to $\delta = 0$ yields an exceptionally good fidelity of the generated HE state. 

%Firstly, it should be noted that using our scheme, the final composite state, $\ket{\Psi (t)}_{CL}$, of the lattice and the coin, after $t$ time steps, is given by Eq.~\eqref{eq:final_state}, while the theoretical state, $\sigma$, which we want to target, is given by Eq.~\eqref{eq:he_state}. 

To elucidate further aspects, we choose $\delta = 0$, $\theta_1 = \theta_2 = 0$ which also yields a HE state with good fidelity. For a lattice of size $200$, initial coherent amplitude $\alpha_0 = 10$ we evolve the system for $t = 20$ time steps and we obtain $\Bar{\alpha}_1 = 11.0131$ and $\Bar{\alpha}_2 = 8.9839$, which leads to a fidelity of $99.23 \%$. For $t = 40$ time steps (while keeping all the other parameters the same), we obtain $\Bar{\alpha}_1 = 11.8973$ and $\Bar{\alpha}_2 = 7.7682$, which leads to a fidelity of $97.05 \%$.

In Fig.~\ref{fig:step20c} and $\ref{fig:step60c}$ we plot the generated lattice states and the targeted coherent states by evaluating $\Bar{\alpha}_1$ and $\Bar{\alpha}_2$ for $t = 20$ and $t = 40$ time steps. In both cases, we find that the generated states are extremely close to the targeted coherent states with fidelity $ \approx 97\%$.

It should be noted that the value of $\delta$, characterizing the initial state of the coin, does not affect the fidelity after $t$ time steps.

\begin{figure}
	\centering
	\subfigure[]{\includegraphics[width=0.23\textwidth]{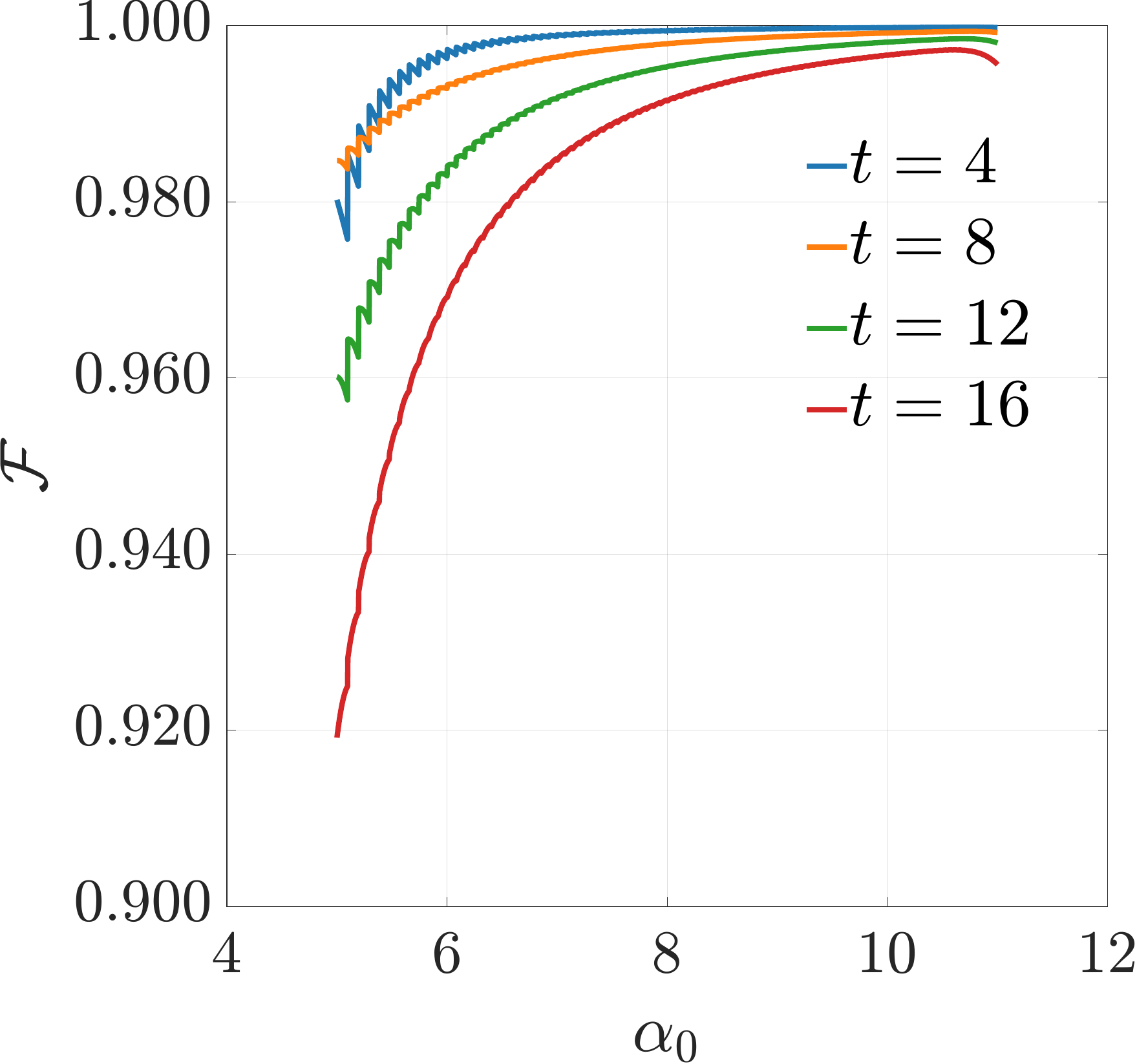}
		\label{fig:vsalpha}}
	\subfigure[]{\includegraphics[width=0.23\textwidth]{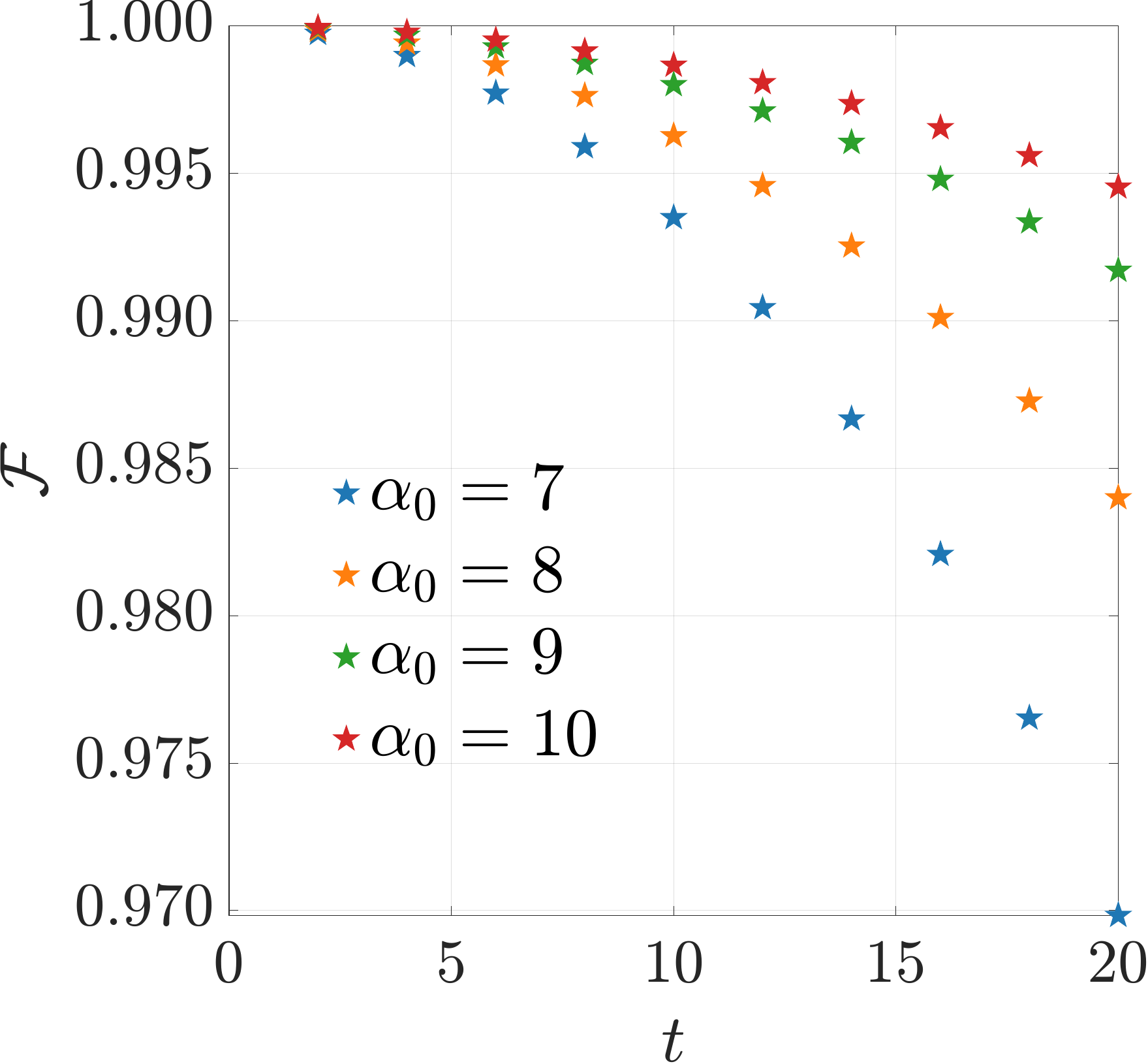}
		\label{fig:vssteps}}
	\caption{The plot of fidelity (a) with different values of initial coherent amplitude $\alpha_0$ for a fixed number of time steps, (b) with different time steps and fixed values of 
		$\alpha_0$. All other parameters are taken to be the same as in Fig.~\ref{fig:alpha10}.}
	\label{fig:fidalphatime}
\end{figure}

In Fig.~\ref{fig:fidalphatime}, we plot the fidelity of the resultant HE states for various fixed initial parameters. It can be seen that the fidelity decreases with more number of time steps. However, it can be improved by initializing the lattice with a higher coherent amplitude $\alpha_0$.

\begin{figure}
	\centering
	\subfigure[]{\includegraphics[height=0.215\textwidth]{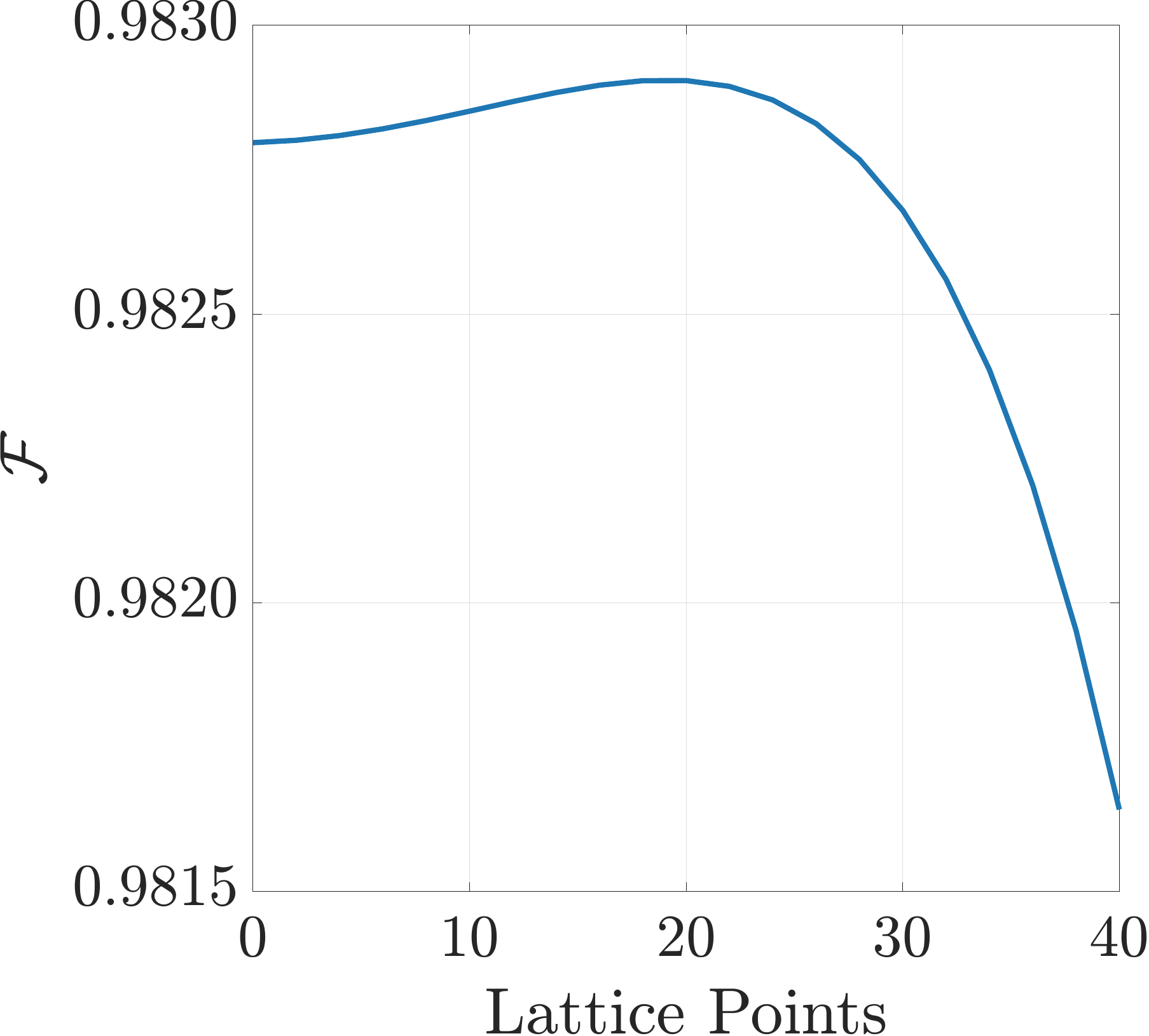}}
	\subfigure[]{\includegraphics[height=0.215\textwidth]{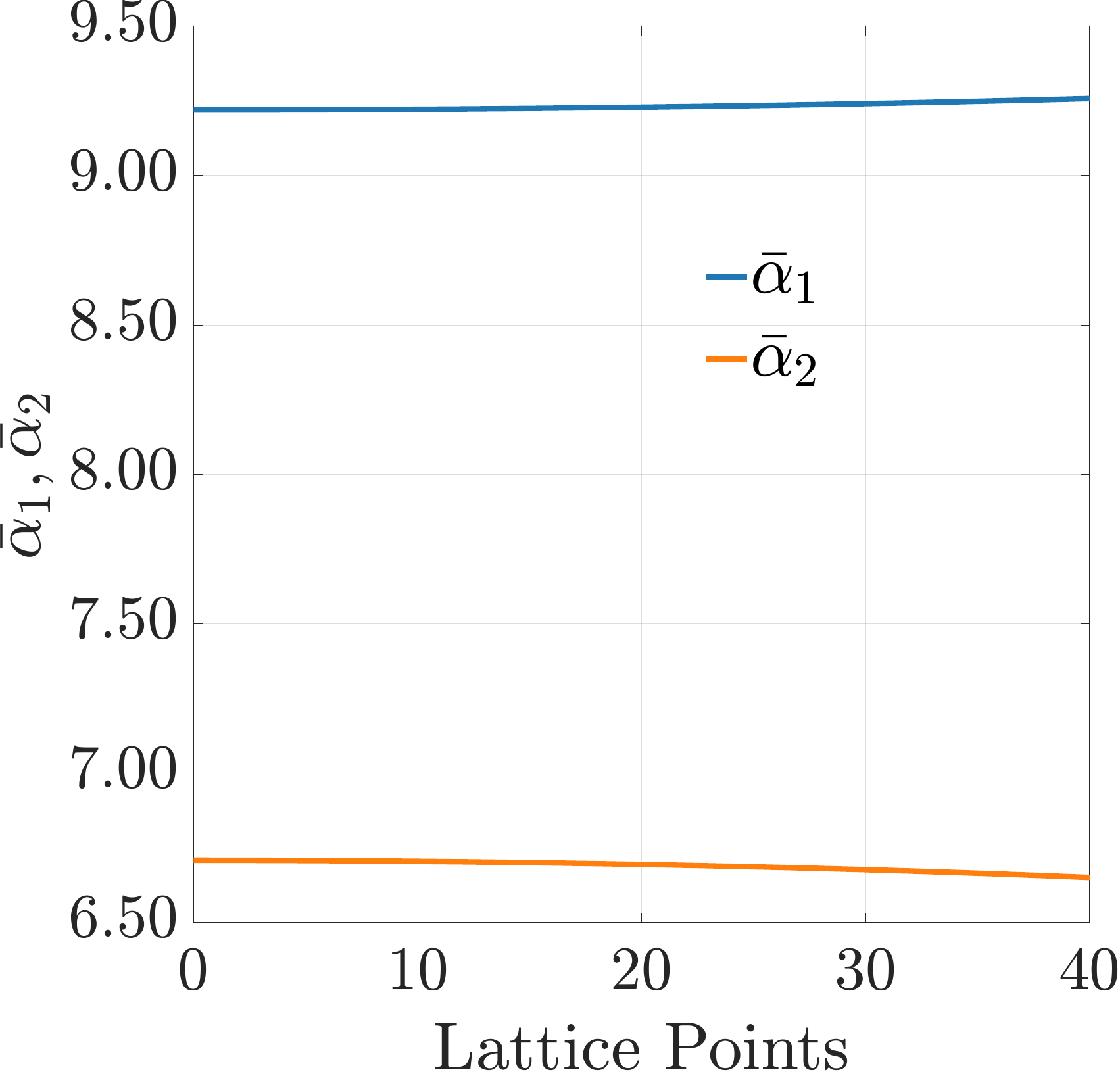}}
	\subfigure[]{\includegraphics[height=0.215\textwidth]{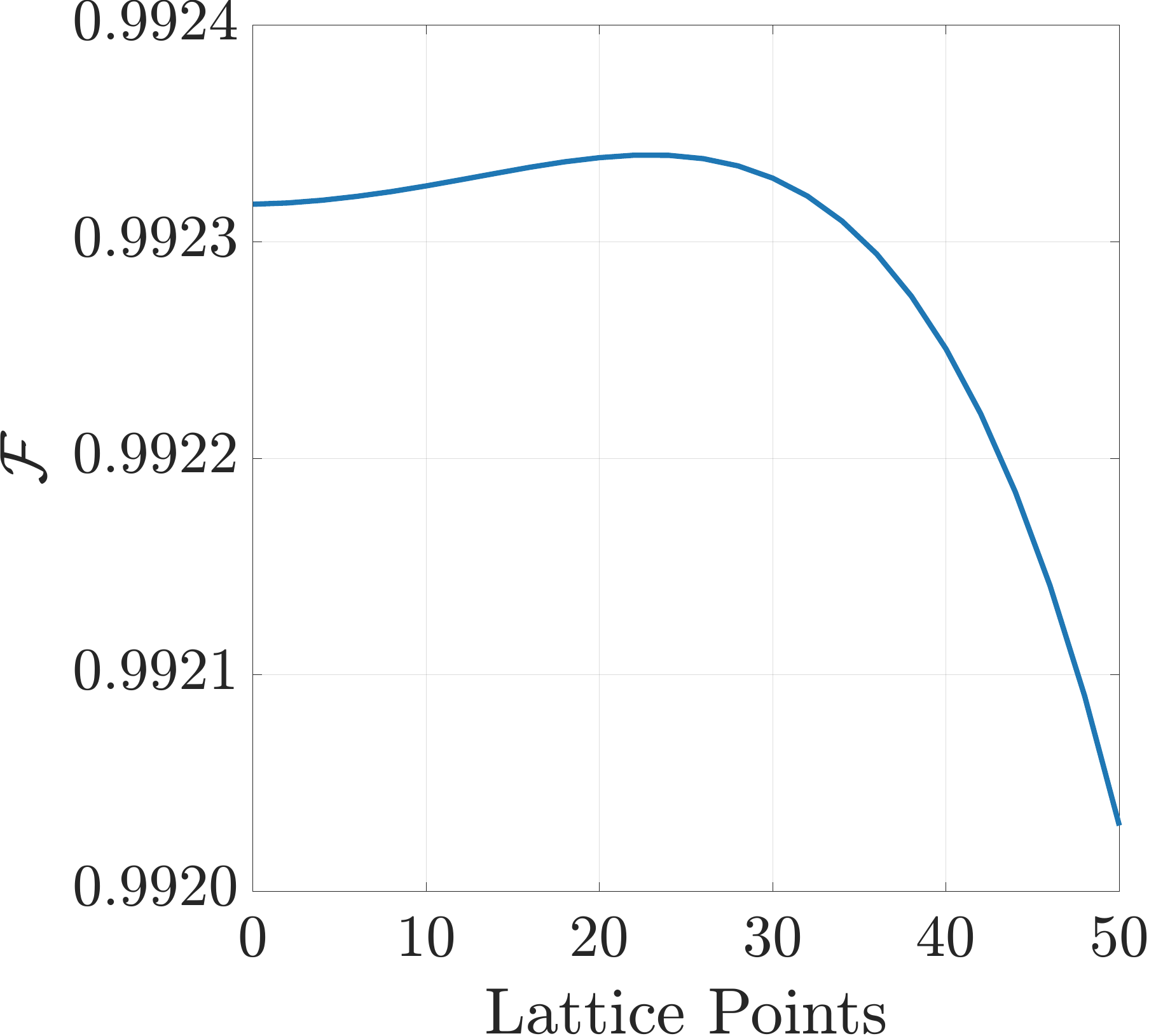}}
	\subfigure[]{\includegraphics[height=0.215\textwidth]{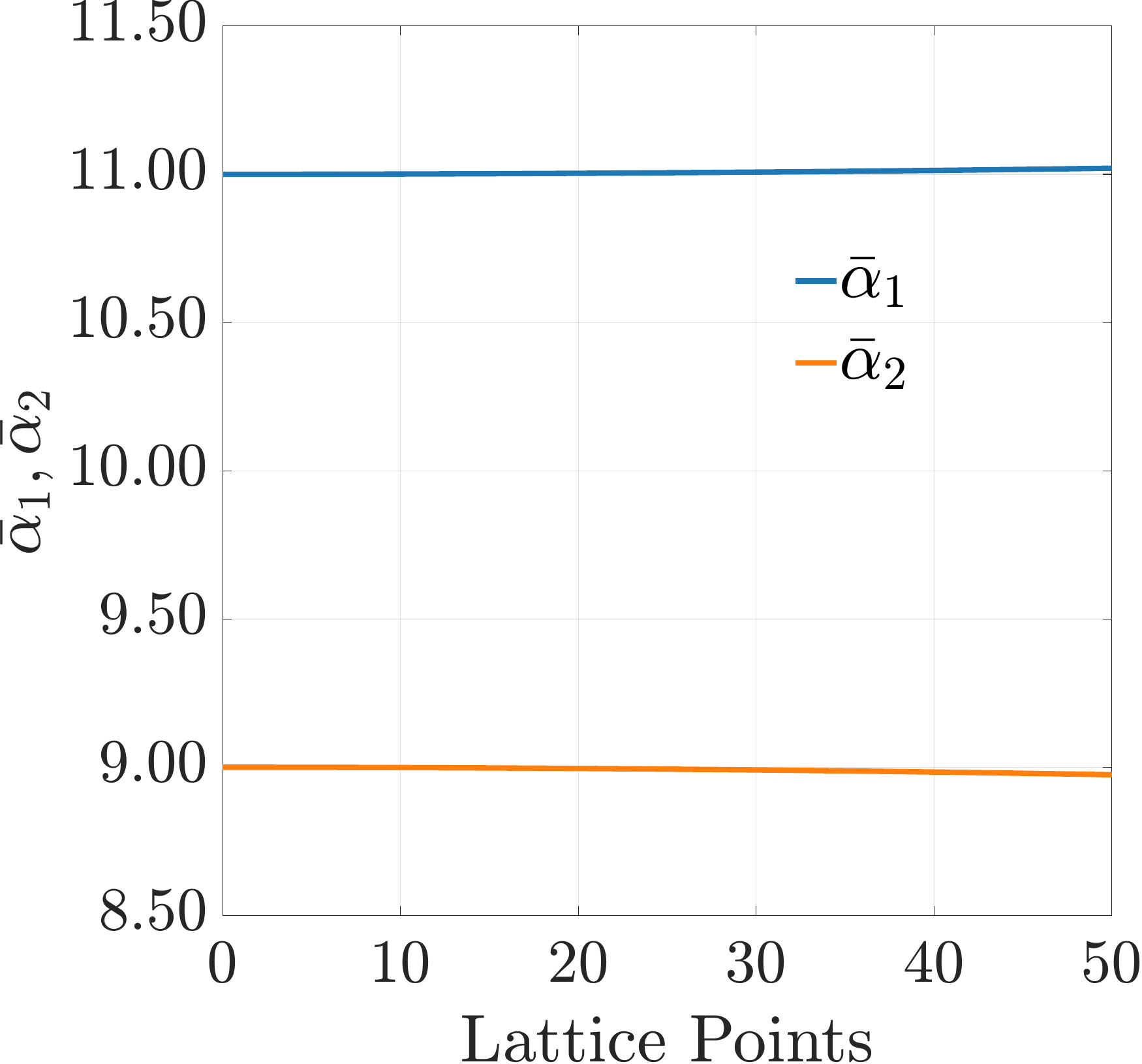}}
	\caption{The plot of fidelity and the average coherent amplitudes against the size of the subset of lattice points chosen to evaluate $\Bar{\alpha}_i$ after $t = 20$ time steps. In (a) and (b), we choose $\alpha_0 = 8$, while in (c) and (d), we choose $\alpha_0 = 10$, respectively. All other parameters are taken to be the same as in Fig.~\ref{fig:alpha10}.}
	\label{fig:fidalphalattice}
\end{figure}

In Fig.~\ref{fig:fidalphalattice}, we plot the relation between the fidelity of the resultant state with the coherent state (having coherent amplitude $\Bar{\alpha}_i$) when a fixed number of lattice points are used to compute $\Bar{\alpha}_i$ as mentioned in Sec.~\ref{sec:he_quant_walk}. 
We see that there exists an optimal number of lattice points around $j^*$ that should be chosen to evaluate $\Bar{\alpha}_i$ in order to maximize the fidelity. 
In Sec.~\ref{sec:he_quant_walk} we chose $40$ lattice points on either side of $j^*_0$ and $j^*_1$ for the case when $\alpha_0 = 10$. 
We particularly chose this number because, while it is not the optimal number of lattice points (as can be seen in Fig.~\ref{fig:fidalphalattice}), it still results in an extremely good fidelity, and goes on to show that using our technique, we are able to generate extremely good HE states.

In comparison, the fidelity of generating an HE state in the symmetric form following the results of Ref.~\cite{JZK14} is high ($\approx 99 \%$) only if the coherent amplitude of the initial state is around $\alpha_0 \approx 2.8$ and the resultant (symmetric) state has a very small coherent amplitude ($\alpha_1 = -\alpha_2\approx 0.2$). For all other cases, the authors report a decrease in the fidelity of the resultant HE state. For example, to obtain $\alpha_1 = -\alpha_2 = 0.5$ from any value of $\alpha_0$, the resultant fidelity is around $80 \%$, which is very low. Moreover, the fidelity decreases for very high values of $\alpha_1 = -\alpha_2$.

Generally, an experimental implementation of a $1$-D quantum walk can only access a finite number of lattice points. For this particular reason, the initial state of the lattice is chosen such that there are an appropriate number of experimentally accessible lattice points on either side of the walker for a fixed number of time steps. In our case, since we are identifying energy eigenstates as the lattice, then it is clear that if the walker starts off at an energy eigenstate very close to the ground state, then there are not enough lattice points to hop to on one side of the lattice for a large number of time steps. It is for this particular reason that we initialize the walker such that its coherent amplitude is not close to unity and the walker has access to ample energy eigenstates on either side. Moreover, this is also one of the reasons why our scheme does not fare well when we choose $\alpha_0$ very small. Another possible reason is the algorithm we present to evaluate $\Bar{\alpha}_i$. It becomes inapplicable if we choose a very small value of $\alpha_0$. As an example, for $\alpha_0 = 1$, there is only a single energy level ($j = 0$) for computing $\alpha_1$. In this case, our algorithm, which relies on using several energy levels to compute $\Bar{\alpha}_1$ cannot be applied. Therefore, our algorithm also fails to characterize the coherent amplitude of the resultant state for small $\alpha_0$.

%%%%%%%%%%%%%%%%%%%%%%%%%%%%%%%%%%%%%%%%%%%%%%%%%%%%%%%%%%%%%%%%%%%%%%%%%%%%%%%%%%%%%%%%
\section{Conclusion}
\label{sec:conc}

In this paper, we provide an efficient and deterministic scheme to generate high fidelity HE states which play an important role in several quantum information processing applications. 
Unlike previous proposals, our scheme, based on $1$-D SSQW, is deterministic and utilizes only simple translation operations.
We show that by identifying the lattice as energy eigenstates of a single particle in a harmonic oscillator potential well and initially preparing it in a coherent state with a carefully chosen coherent amplitude $\alpha_0$ and coin parameters $\theta_1$, $\vb{n}_1$, $\theta_2$ and $\vb{n}_2$, it is possible to generate a large class of HE states efficiently. It should be noted that our scheme always generates the same HE state for a given initial coherent amplitude $\alpha_0$, number of time steps $t$ and fixed coin parameters.

For the purpose of demonstration, we specifically chose the largest value of the coherent amplitude, $\alpha_0 = 10$, which could be handled by our machine before encountering numerically zero values (because of $\frac{1}{\sqrt{j!}}$). 
Next, using $t = 20$ time steps, we generate HE states with resultant coherent amplitudes $\Bar{\alpha}_1 = 11.0131$ and $\Bar{\alpha}_2 = 8.9839$. 
The separation between the two resultant states, as described by $|\Bar{\alpha}_1 - \Bar{\alpha}_2|$, is found to increase with the number of steps in the quantum walk.

Our scheme generates HE states with a fidelity of $99.23 \%$ for $t = 20$ time steps. 
We observe that the fidelity of preparation is higher when the number of time steps taken is small and decreases when the number of time steps is increased. 
However, it is noteworthy that even for higher time steps, specifically $t = 60$, the obtained fidelity is still above $98 \%$, which is significantly high. 
For smaller time steps, like $t = 12$, and keeping all other parameters the same as before, it is possible to obtain a fidelity of $99.98 \%$ (for more details, we refer to Appendix~\ref{appendixB}). However, in this case, the separation between the resultant coherent states on the lattice is not very large.

We note that the reported fidelities are independent of any noise (or errors) that may arise because of a specific physical architecture. 
However, due to extensive experimental research in quantum walks, control over the coin-lattice system has significantly improved over the years~\cite{QW2022a,QW2023a,DiColandrea23}, which may allow an efficient implementation of our scheme without much additional noise.

While our scheme works efficiently for higher values of $\alpha_0$, we note that it does not provide adequate results for smaller values of the same. 
As an example, our scheme fails to prepare a HE state with good enough fidelity when $\alpha_0\leq 3$. 
The major reason for this behaviour can be attributed to the fact that for very low values of $\alpha_0$, there are not enough lattice points (energy eigenstates) for the walker to hop to on the left of the initial starting position.
This leads to failure in characterizing the resultant lattice states for very low values of $\alpha_0$.
This is in stark contrast with earlier experimental schemes where it was possible to generate and characterize HE states having extremely good fidelity with small values of the final coherent amplitude. 
We point out that while the earlier approach of Ref.~\cite{JZK14} might be useful to generate HE states having a low coherent amplitude, our approach can be used to generate the same with a higher coherent amplitude.
In this sense, both the approaches can complement each other. Moreover, there may also exist other efficient methods other than our algorithm to characterize the resultant lattice states. 
However, our method already provides more than $99\%$ fidelity.

It now remains to be seen how our theoretical scheme can be experimentally implemented on some appropriately chosen physical system. 
A possible test-bed could be that of Ref.~\cite{McCormick2019} where the authors experimentally prepare Fock states $\ket{n}$ and their superpositions with very high fidelity. Specifically, the authors report that they were able to extend existing experimental techniques~\cite{Zoller1993, Zoller1995, Wineland1996, Wineland1997, Wineland2003} to create Fock states up to $\ket{n = 100}$. Their experimental setup consists of a single $^9\text{Be}^+$ ion trapped above cryogenic linear surface-electrode trap~\cite{WCB14}, which forms a quantum harmonic oscillator. In addition to this, three levels within the electronic $^2S_{1/2}$ ground-state hyperfine manifold, $\ket{F=1,{m}_{F}=-1} = \ket{0}$, $\ket{F=2,{m}_{F}=-2} = \ket{1}$  and $\ket{F=2,{m}_{F}=0} = \ket{\text{aux}}$ were used in the preparation of the Fock state space. Here, $F$ is the total angular momentum and $m_F$ is its component along the quantization axis. Next, the coherent state of the ion can then be prepared from the $\ket{n = 0}$ ground Fock state by a spatially uniform classical driving field~\cite{Wineland1996, CN65}, which can be understood as the displacement operation. 

The different transitions between energy levels in this system can be achieved as follows. First, the ion is prepared in a state $\ket{\Psi_0} = \ket{0} \otimes \ket{0} \in \mathcal{H}_\text{elec.} \otimes \mathcal{H}_{\text{ho}}$ with a fidelity more than $0.99$ where $\mathcal{H}_\text{elec}$ and $\mathcal{H}_\text{ho}$ denotes the Hilbert space of the electronic states and the motional states (corresponding to the Fock space) of the ion, respectively. The transition between an intermediate number state $n$, $\ket{0} \otimes \ket{n} \leftrightarrow \ket{1} \otimes \ket{n \pm 1}$ are then implemented with stimulated Raman transitions~\cite{MMK95}. In this setup, the two electronic states $\ket{0}$ and $\ket{1}$ can be distinguished 
with state-selective fluorescence technique~\cite{Wineland2003}. 

These aforementioned results indicate that it is possible to prepare a lattice of Fock states of a quantum harmonic oscillator and a coherent state over them with very good fidelity using a trapped ion. Moreover, the electronic levels of the ion can then be used as a coin for the quantum walk. The translation operators in our scheme would then correspond to the stimulated Raman transitions. An experimental implementation of our scheme would have to create conditional (or state-dependent) Raman transitions to implement the quantum walk. We note that conditional Raman transitions have been earlier implemented in Ref.~\cite{Schmitz2009} although in a different context, and can therefore be specifically designed by experimentalists to better suit our scheme.

We also note that there can exist a number of physical architectures over which our results can be implemented. 
However, the ``coin-lattice” system must necessarily satisfy a few conditions for a proper physical implementation.
Apart from the fact that it should be possible to define a coherent state on the lattice (like energy eigenstates of a harmonic oscillator).
Specifically, such a coin-lattice system must ensure (i) high number of lattice points ($\gtrapprox 100$), (ii) good control over the coin and lattice operations, (iii) the ability to initialise a coherent state on the lattice and (iv) if possible, allows spatial separation between the coin and lattice systems.

Property (iv), while not being a necessary ingredient in our scheme, will prove to be useful in several information processing tasks that require entanglement to be distributed among separated parties. 
However, we do not yet know whether it is experimentally attainable in quantum walks. 
We expect that the scheme presented herein will motivate interactions with experimental groups to search for such an optimal physical system that can simulate our results.

%%%%%%%%%%%%%%%%%%%%%%%%%%%%%%%%%%%%%%%%%%%%%%%%%%%%%%%%%%%%%%%%%%%%%%%%%%%%%%%%%%%%%%%%
\section*{Acknowledgements}
J.~S. acknowledges the support from the National Science and Technological Council (NSTC) of Taiwan through grant no. NSTC 112-2628-M-006-007-MY4 and NSTC 112-2811-M-006-033-MY4.
V.~M. acknowledges the support from the National Science and Technology Council (NSTC) of Taiwan through MOST 111-2636-M-007-009-, NSTC 112-2636-M-007-008- and Physics Division, National Center for Theoretical Sciences (Grant No. 112-2124-M-002-003-). 
J.~S. and V.~M. thank Prof. Yi-Ping Huang for carefully reading an earlier version of the manuscript and providing insightful comments.


%%%%%%%%%%%%%%%%%%%%%%%%%%%%%%%%%%%%%%%%%%%%%%%%%%%%%%%%%%%%%%%%%%%%%%%%%%%%%%%%%%%%%%%%
\appendix

%%%%%%%%%%%%%%%%%%%%%%%%%%%%%%%%%%%%%%%%%%%%%%%%%%%%%%%%%%%%%%%%%%%%%%%%%%%%%%%%%%%%%%%%
\section{One-dimensional discrete-time quantum walk}
\label{appendixA}

A one-dimensional discrete-time quantum walk ($1$-D DTQW) is defined over a lattice with real space dimension $1$ and a coin having a Hilbert space dimension $2$. It consists of a conditional coin-dependent shift operator $T$ and a coin flip operator $R(\theta)$ for a real parameter $\theta$. We can represent these operators in the position basis of the lattice $\{\ket{n}\} \in \mathcal{H}_{L}$ and spin basis $\{\ket{0}, \ket{1} \} \in \mathcal{H}_C$ of the coin. The composite Hilbert space of the coin-lattice system can be written as $\mathcal{H} = \mathcal{H}_C \otimes \mathcal{H}_L$. Then the operator $U(\theta) = T R(\theta)$ governs the time evolution of the walker for a unit time on the lattice. Here 
\begin{gather}
	T = \sum_j \dyad{0} \otimes \dyad{j+1}{j}  + \dyad{1} \otimes \dyad{j-1}{j}, \\
	R(\theta) = e^{-i \theta \sigma_y/2} \otimes \mathds{1}_{N},
\end{gather}
and $-2\pi\le\theta<2\pi$ is a real parameter and $\sigma_y$ is the Pauli matrix along the $y$-axis. Here, $\mathds{1}_N$ represents the identity operation on the lattice. The unitary operator which governs the time evolution of the walker for a unit time step reads
\begin{equation}
	U(\theta) = T R(\theta).
	\label{eq:unitaryevolution}
\end{equation}

For our purposes, we only consider a homogeneous system where the translation and the coin operator do not depend on the lattice site. Given an initial state $ \ket{\psi(0)} $, the state of the walker after $t$ time steps is written as
\begin{equation}
	\begin{aligned}
		\ket{\psi(t)} &= (U)^t \ket{\psi(0)} \\
		&= \sum_j \psi_{0, j} (t) \ket{0} \otimes \ket{j} + \psi_{1, j} (t) \ket{1} \otimes \ket{j},
	\end{aligned}  
\end{equation}
and the probability of finding the walker at the $j$th lattice site after $t$ time steps is given by
\begin{align}
	P(j,t) &= \abs{\bra{0} \otimes \ip{j}{\psi(t)}}^2 + \abs{\bra{1} \otimes \ip{j}{\psi(t)}}^2 \nonumber \\ 			
	&= \abs{\psi_{0, j} (t)}^2 + \abs{\psi_{1, j} (t)}^2.
	\label{eq:probdist}
\end{align}
where $\psi_{0, j} (t)$ and $\psi_{1, j} (t)$ are the amplitudes of the lattice state at time $t$ corresponding to the two coin states $\ket{0}$ and $\ket{1}$ respectively.

\begin{figure*}
	\centering
	\subfigure[]{\includegraphics[width=0.22\textwidth]{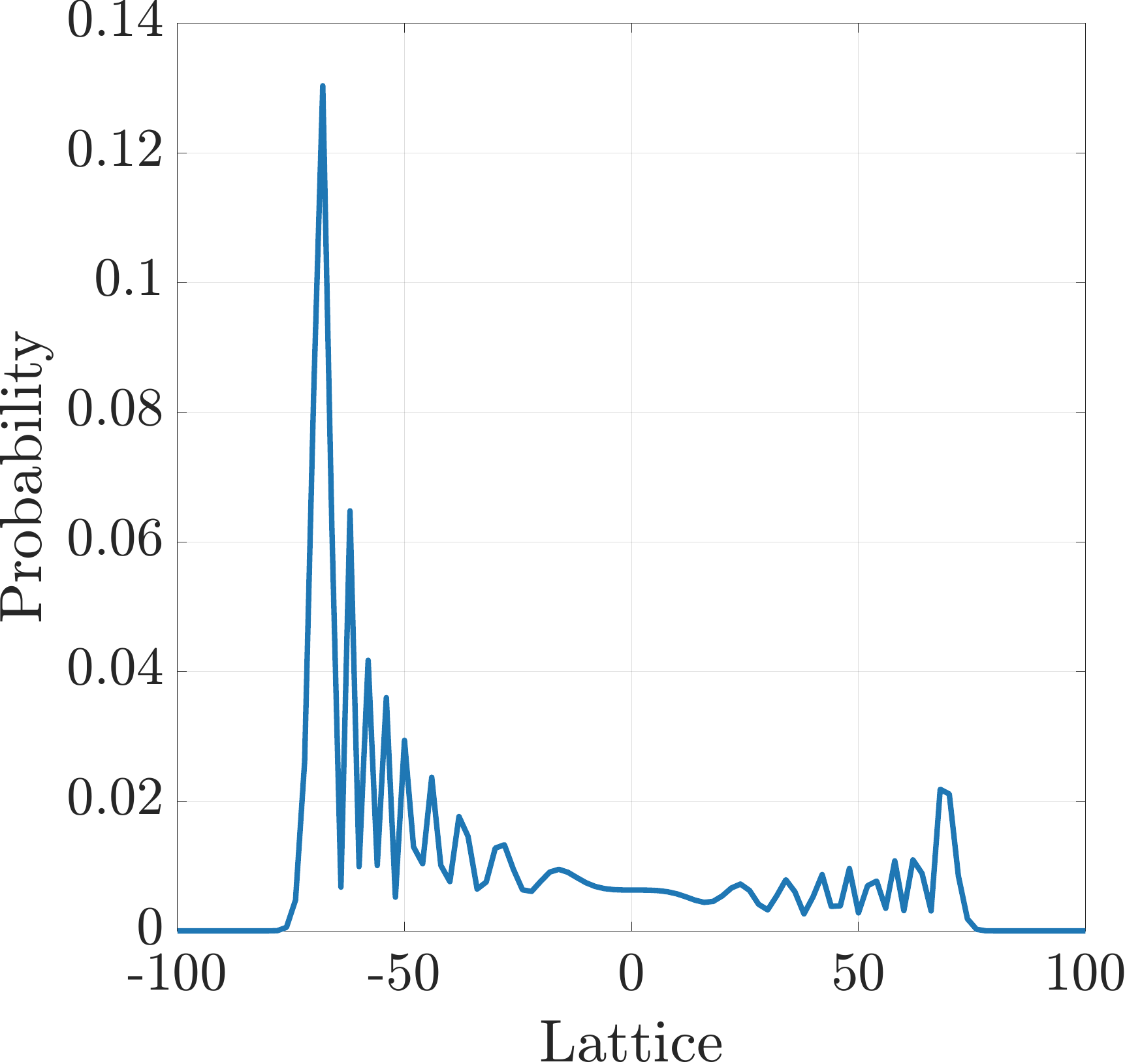}
		\label{fig:1DEvolutiondown}}
	\subfigure[]{\includegraphics[width=0.22\textwidth]{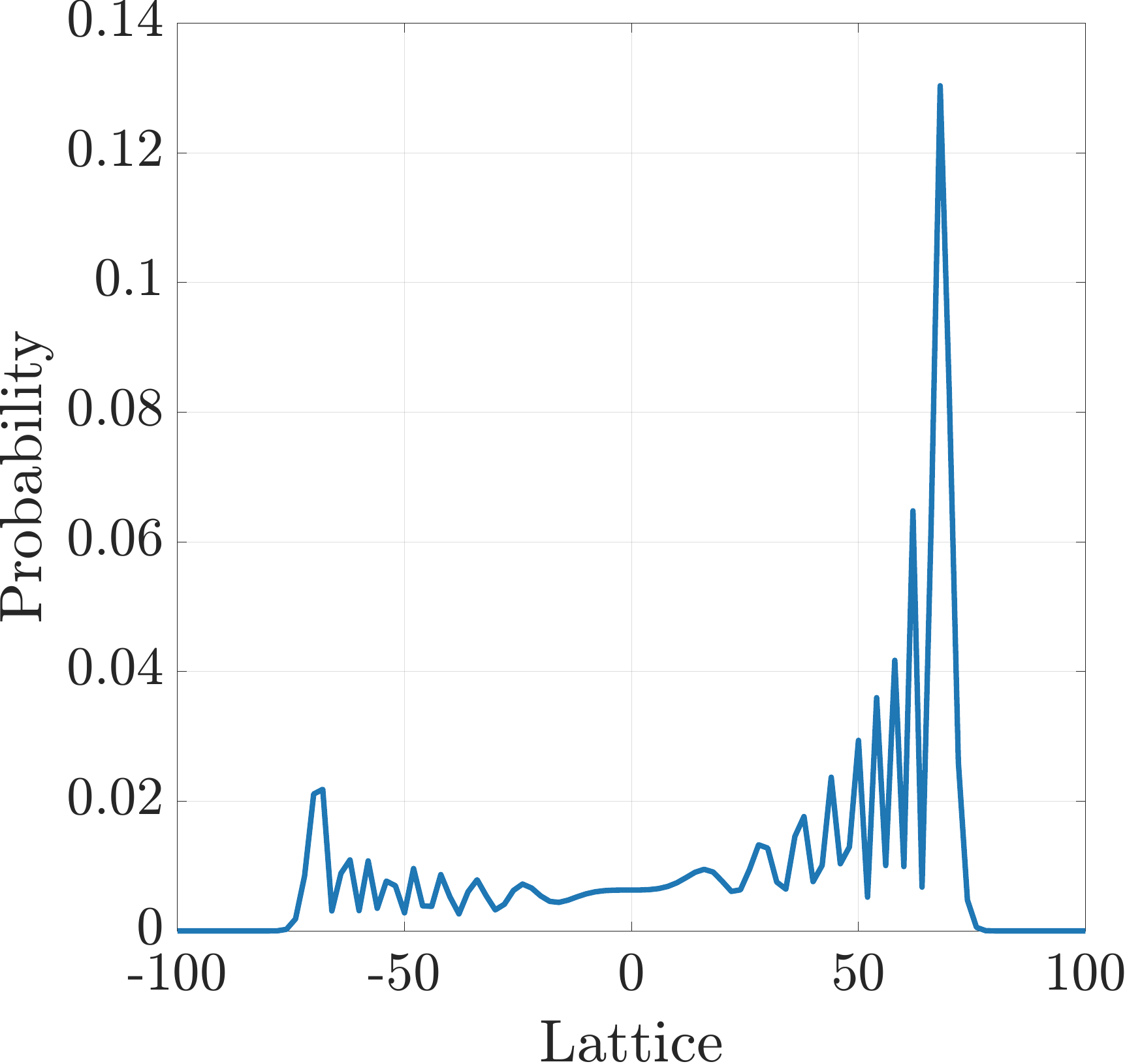}
		\label{fig:1DEvolutionup}}
	\subfigure[]{\includegraphics[width=0.22\textwidth]{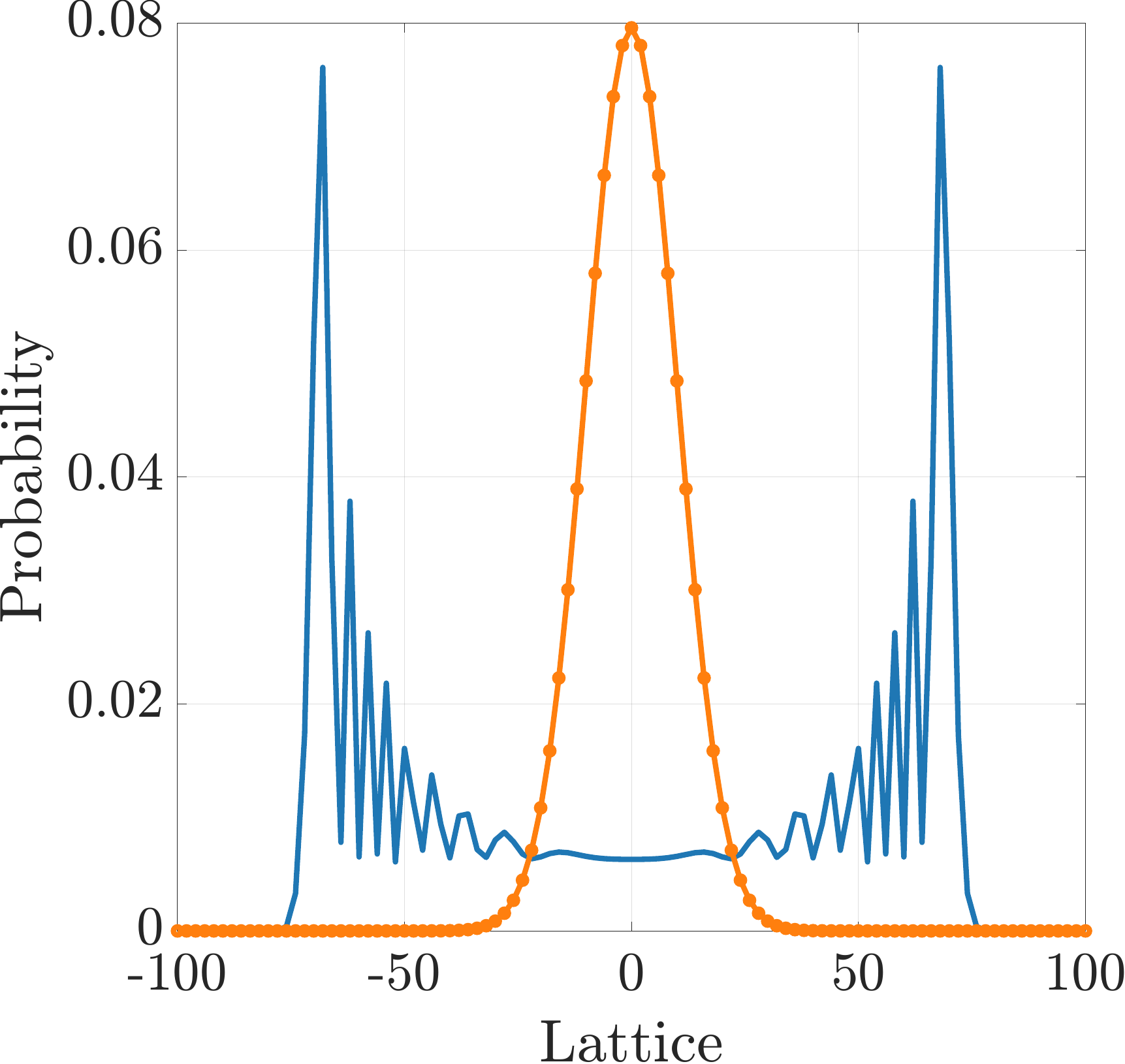}
		\label{fig:dtqw}}	
	\subfigure[]{\includegraphics[width=0.22\textwidth]{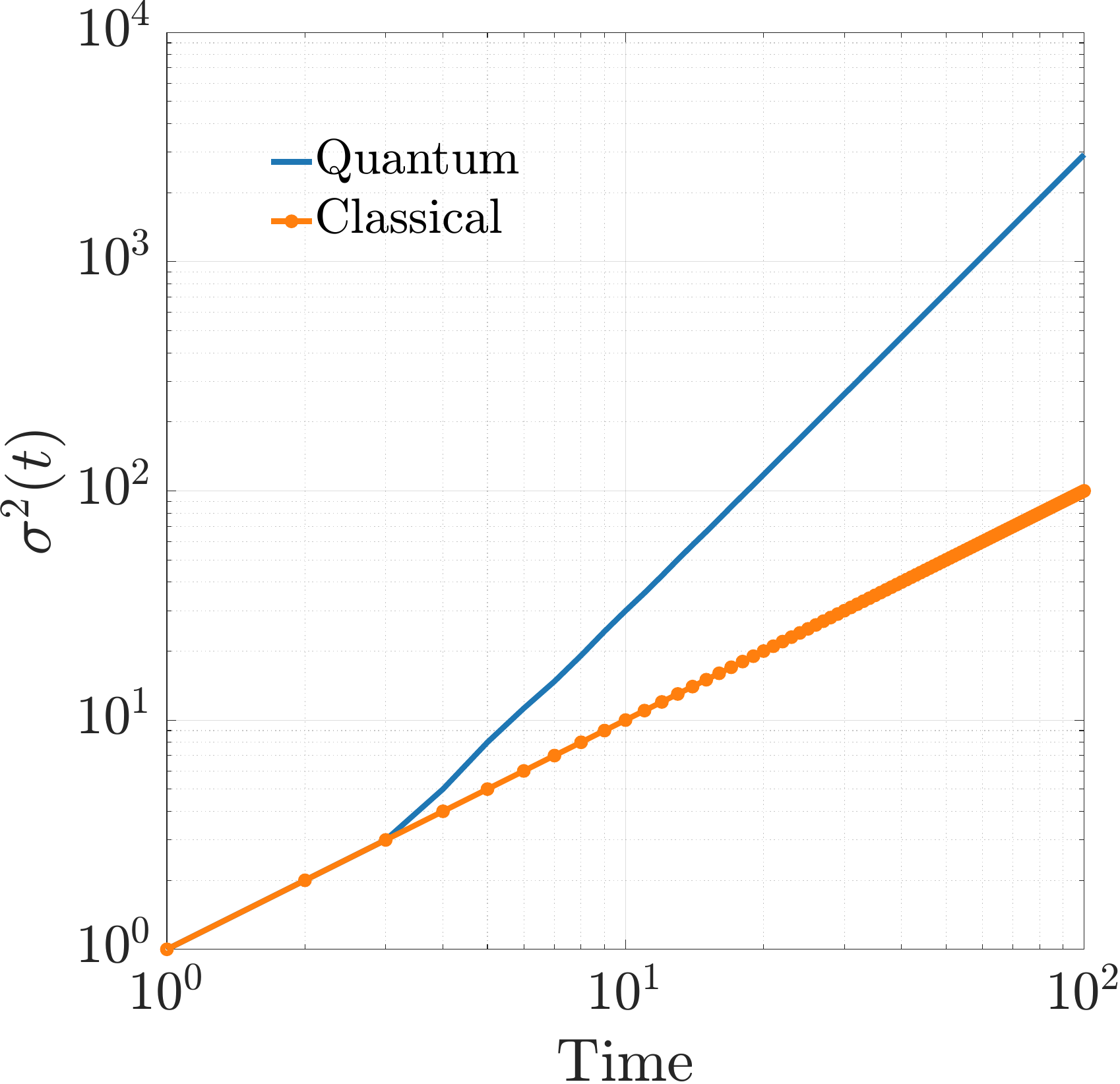}
		\label{fig:variance}}
	\caption{Probability distribution of a walker for a $1$D DTQW with $N = 201$ lattice sites after $t = 100$ time steps for different initial states. \subref{fig:1DEvolutiondown} $ \ket{\psi(0)} = \ket{1}_C \otimes \ket{0}_L $, \subref{fig:1DEvolutionup} $ \ket{\psi(0)} = \ket{0}_C \otimes \ket{0}_L $, \subref{fig:dtqw} $ \ket{\psi(0)} = (\ket{0}_C + i \ket{1}_C)/\sqrt{2} \otimes \ket{0}_L$ in contrast to classical walk (red). Note that only the points with non-zero probability are plotted. \subref{fig:variance} The variance, $\sigma^2(t)$ for quantum and classical walks as a function of time steps $t$. In this case, the axes are chosen on a logarithmic scale. We chose the coin flip operator with parameter $\theta = \pi/2$ for all the plots.} 
	\label{fig:ProbDis}
\end{figure*}

In Fig.~\ref{fig:1DEvolutiondown},~\ref{fig:1DEvolutionup},~\ref{fig:dtqw} we plot the probability distribution of the walker, as given by Eq.~\eqref{eq:probdist}, for $100$ time steps and for different initial states. In Fig.~\ref{fig:dtqw}, we compare the probability distribution for the classical and quantum walks. A contrasting behaviour between the two can be seen. In a quantum walk, the probability amplitude evolves with time steps, which leads to constructive and destructive interference effects. A constructive interference is observed at the extremities of the lattice, while a destructive interference is observed at the origin. On the other hand, no interference effects are observed in a classical random walk, and instead, we observe a Gaussian distribution. 

As can be seen in Fig.~\ref{fig:dtqw}, the variance of the probability distribution, denoted by $\sigma^2 (t)$, for a $1$-D DTQW, for some initial state is proportional to the square of the number of time steps $t$, i.e. $\sigma^2(t) \propto t^2 $, which is quadratically faster than the evolution of the variance of the probability distribution that is observed in a classical random walk. This behaviour can be seen in Fig.~\ref{fig:variance} where we plot the variance for the classical and quantum walk on a logarithmic scale. Because of this behaviour, the $1$-D DTQW and classical walk are also referred to as ballistic and diffusive processes, respectively.

%%%%%%%%%%%%%%%%%%%%%%%%%%%%%%%%%%%%%%%%%%%%%%%%%%%%%%%%%%%%%%%%%%%%%%%%%%%%%%%%%%%%%%%%

\section{Quadrature measurement on a lattice formed by energy eigenstates}
\label{sec:quad}

In this section we provide details on how to treat the lattice as energy eigenstates of a harmonic oscillator potential well.
We then describe the relationship between creation and annihilation operators of the harmonic oscillator and the lattice displacement operations. 
Moreover, we also provide a description of how to implement  quadrature-like measurements on the lattice. 

Let us consider a quantized spring-mass system with $\hat{x}$ and $\hat{p}$ as the position and momentum operators for the mass $m$ and $k$ as the spring constant.
With $\omega = \sqrt{k/m}$ as the frequency of oscillation, the dimensionless operators $X$ and $P$, generally called the quadratures of a harmonic oscillator system, can be defined as 
\begin{equation}
	X = \sqrt{\frac{m \omega}{2}} \hat{x}, \quad P = \sqrt{\frac{1}{2 m \omega}} \hat{p}.
\end{equation}

Consequently, the creation (raising) and the annihilation (lowering) operators, $\hat{a}^\dagger$ and $\hat{a}$ respectively, can then be defined in terms of these quadrature operators as 
\begin{equation}
	\hat{a}^\dagger = \frac{X-\mathsf{i}P}{\sqrt{2}}, \quad \hat{a} = \frac{X+\mathsf{i}P}{\sqrt{2}},
\end{equation}
such that $\hat{a}^\dagger \ket{j} = \sqrt{j + 1}\ket{j + 1}$ and $\hat{a} \ket{j} = \sqrt{j}\ket{j - 1}$, where $\ket j$'s ($j=0,1,..$) are the energy eigenstates of the harmonic oscillator. 

Next, we briefly describe how quadrature-like measurements can be implemented on energy eigenstates. A quadrature measurement is a measurement of the continuous outcome position ($X$) or momentum ($P$) like operators. 
The POVM corresponding to the quadrature operator $X$ is given as $\lbrace \Pi_x = \ket{x}\bra{x}\rbrace$ ($x \in \left( -\infty, \infty\right)$). Here, $\ket{x}\bra{x}$ is an operator on the real position space of the single particle. Note that this is distinct from the position of the walker $\ket{j}\bra{j}$, which (in our case) corresponds to the energy of the particle. 

Expanding the operator $\Pi_x$ in terms of the energy eigenstates, $\ket{j}$ and $\ket{m}$ we get
\begin{equation}
	\begin{aligned}
		\Pi_x &= \sum_{j, m} \ket{j}\bra{j} \ket{x}\bra{x} \ket{m}\bra{m}\\
		&= \frac{e^{-x^2}}{\sqrt{\pi}} \sum_{j, m} \frac{1}{\sqrt{2^{j + m} j! m!}} H_j(x) H_m(x) \ket{j}\bra{m},
	\end{aligned}
	\label{eq:homo_proj}
\end{equation}
where $H_j(x) = (-1)^j e^{x^2} \frac{d^j}{dx^j} e^{-x^2}$ are Hermite polynomials of order $j$ such that $\bra{x}\ket{j} = \frac{e^{-x^2/2}}{\sqrt{\sqrt{\pi} 2^j j!}} H_j(x)$. 
Corresponding quadrature distribution or the probability of obtaining an outcome $\ket{x}$ for a given state $\ket{\psi} = \sum_m C_m \ket m$ ($C_m = \bra m\ket{\psi}$) is given as
\begin{equation}
	\begin{aligned}
		&\rm{P}(x) = \text{tr}\left(\Pi_x \ket{\psi}\bra{\psi}\right)\\
		&~= \frac{e^{-x^2}}{\sqrt{\pi}} \sum_{n, m} \frac{1}{\sqrt{2^{n + m} n! m!}} H_n(x) H_m(x) \text{tr} \left(\ket{\psi}\bra{\psi} \ket{n}\bra{m}\right)\\
		&~= \frac{e^{-x^2}}{\sqrt{\pi}} \sum_{n, m} \frac{1}{\sqrt{2^{n + m} n! m!}} H_n(x) H_m(x) \sum_l \bra{l}\ket{\psi}\bra{\psi} \ket{n}\bra{m}\ket{l}\\
		&~= \frac{e^{-x^2}}{\sqrt{\pi}} \sum_{n, m} \frac{1}{\sqrt{2^{n + m} n! m!}} H_n(x) H_m(x) C_m C_n^*,
	\end{aligned}
\end{equation}

A similar analysis can also be done for the momentum-like quadrature ($P$) for which the POVM is given as $\Pi_p = \ket{p}\bra{p}$ where $\ket p$ is the momentum eigenstate. 
The corresponding distribution is given as
\begin{equation}
	\begin{aligned}
		\rm{P}(p) &= \text{tr}\left(\Pi_p \ket{\psi}\bra{\psi}\right)\\
		& = \frac{e^{-p^2}}{\sqrt{\pi}} \sum_{n, m} \frac{1}{\sqrt{2^{n + m} n! m!}} H_n(p) H_m(p) C_m C_n^*.
	\end{aligned}
\end{equation}

It should be noted that apart from optical systems quadrature measurements have been experimentally implemented for mechanical oscillators as shown in Refs.~\cite{MRF19,SQM19}.

%%%%%%%%%%%%%%%%%%%%%%%%%%%%%%%%%%%%%%%%%%%%%%%%%%%%%%%%%%%%%%%%%%%%%%%%%%%%%%%%%%%%%%%%

\section{Detailed explanation of the generation of HE states for any arbitrary time step and different coin states} 
\label{appendixB}

In the main draft, we detailed our scheme to generate hybrid entangled (HE) states and mainly focused on the scenario when the initial state of the coin was taken to be as
\begin{equation}
	\ket{\psi(t = 0)}_C = \frac{1}{\sqrt{2}}\left(\ket{0} + e^{i\delta}\ket{1}\right),
\end{equation}
with $\delta = 0$ and the coin operators characterized by $\theta_1 = \theta_2 = 0$. Here, we show that $\delta = \frac{\pi}{2}$ along with $\theta_1 = \pi$ and $\theta_2 = -\pi/2$ is also a viable option that can potentially generate a different class of HE states. Moreover, we also characterize the different types of HE states that can be generated by varying either $\delta$ or the time steps $t$ or both. 

Firstly, we note that apart from the translation effects (which shift the coherent states over the lattice), the generated coherent states after $t$ time steps show the following behaviours: 
\begin{itemize}
	
	\item[Be1:] Coefficients of both the coherent states are real and are simultaneously positive.
	
	\item[Be2:] Coefficients of both the coherent states are real and are simultaneously negative.
	
	\item[Be3:] Coefficients of both the coherent states are imaginary. One of them has positive coefficients, while the other has all negative coefficients. 
	
	\item[Be4:] Coefficients of both of the coherent states are, in general, complex.
	
\end{itemize}

Moreover, the observed behaviour can change with the number of time steps $t$ taken. For example, coherent states showing behaviour Be3 after $t$ time steps can transform into behaviour Be2 after one more step of the quantum walk.

\begin{figure*} 
	\centering
	\subfigure[]{\includegraphics[width=0.23\textwidth]{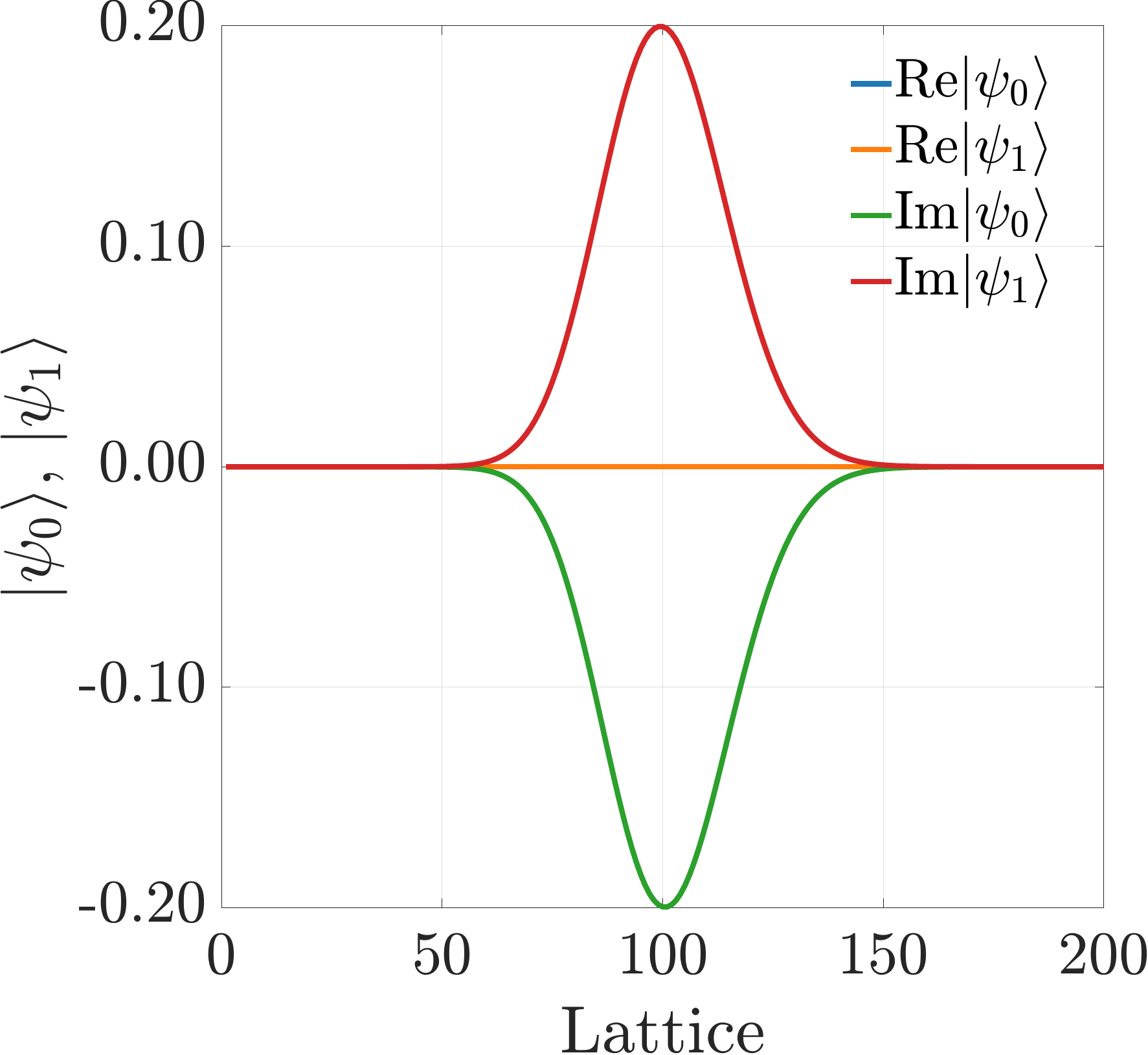}
		\label{fig:d1step1}}
	\subfigure[]{\includegraphics[width=0.23\textwidth]{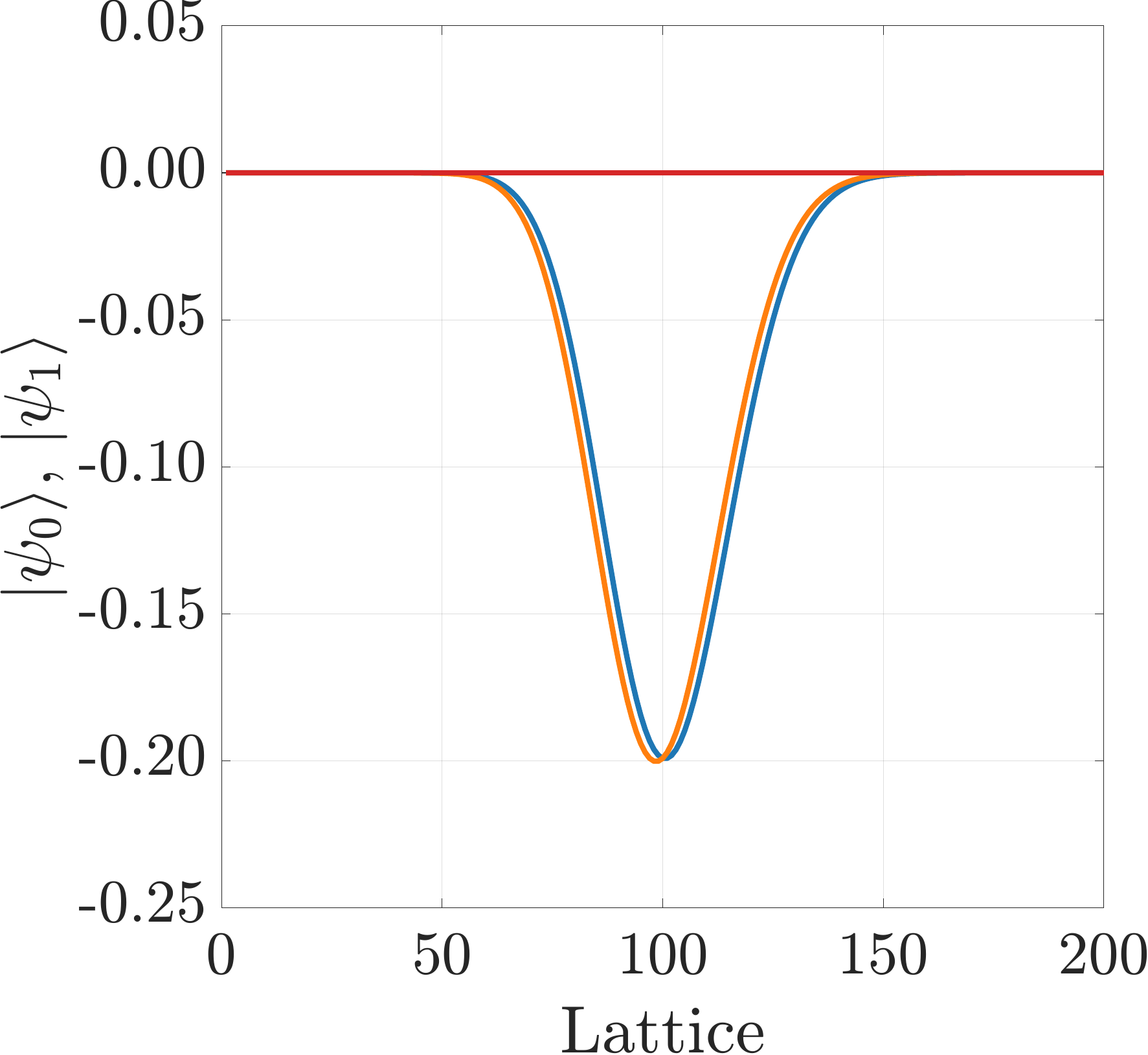}
		\label{fig:d1step2}}
	\subfigure[]{\includegraphics[width=0.23\textwidth]{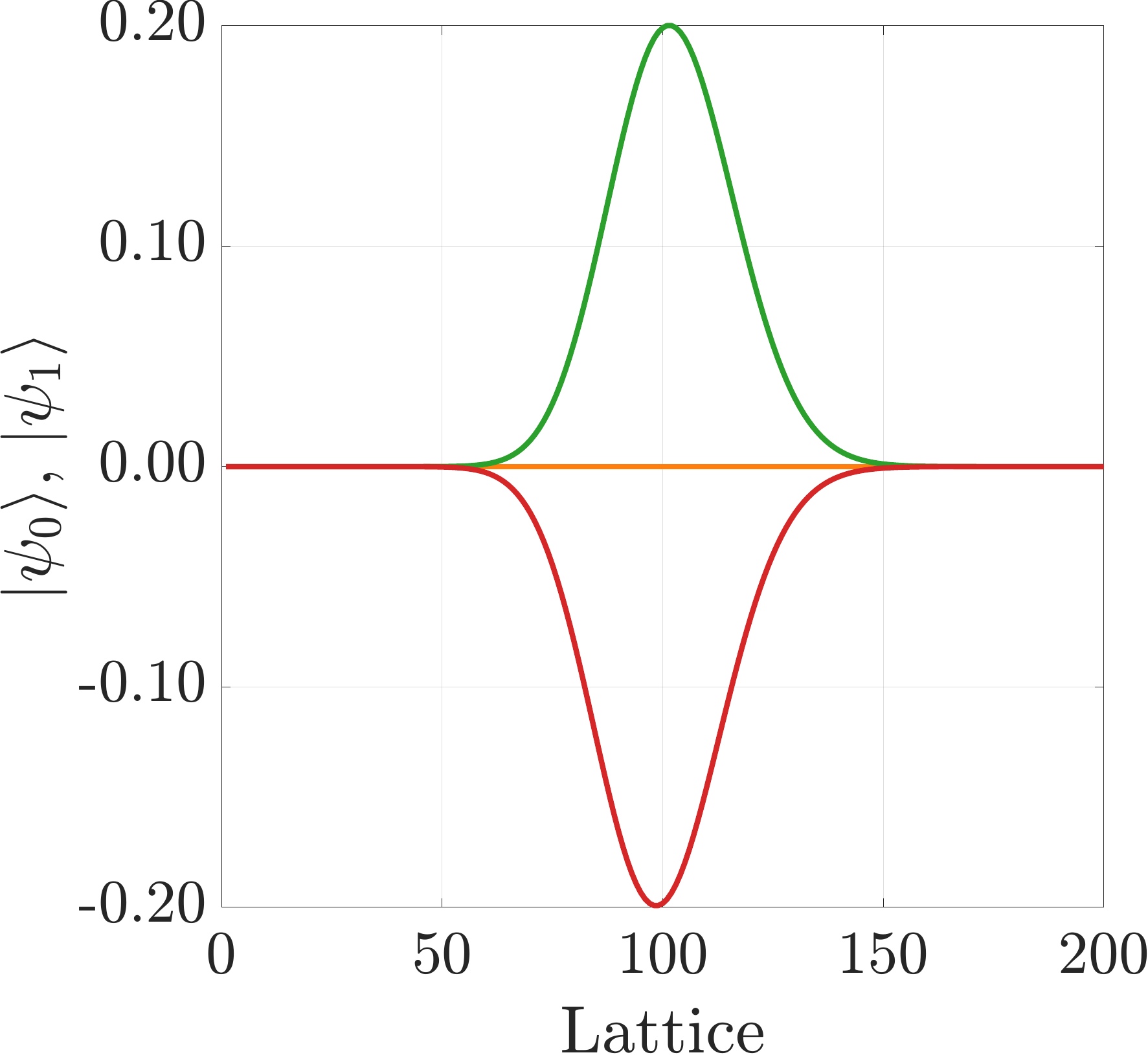}
		\label{fig:d1step3}}
	
	\subfigure[]{\includegraphics[width=0.23\textwidth]{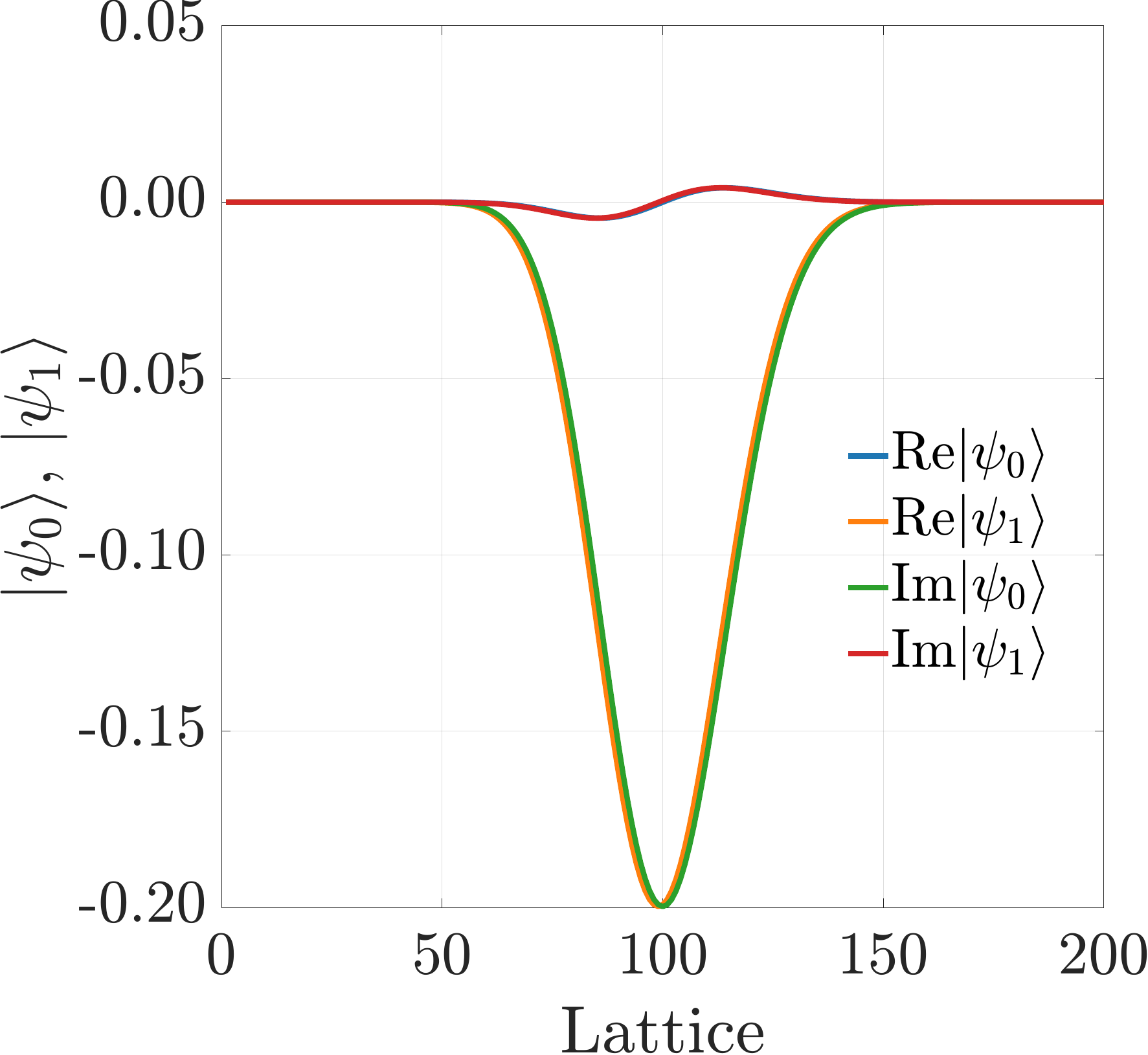}
		\label{fig:d2step1}}
	\subfigure[]{\includegraphics[width=0.23\textwidth]{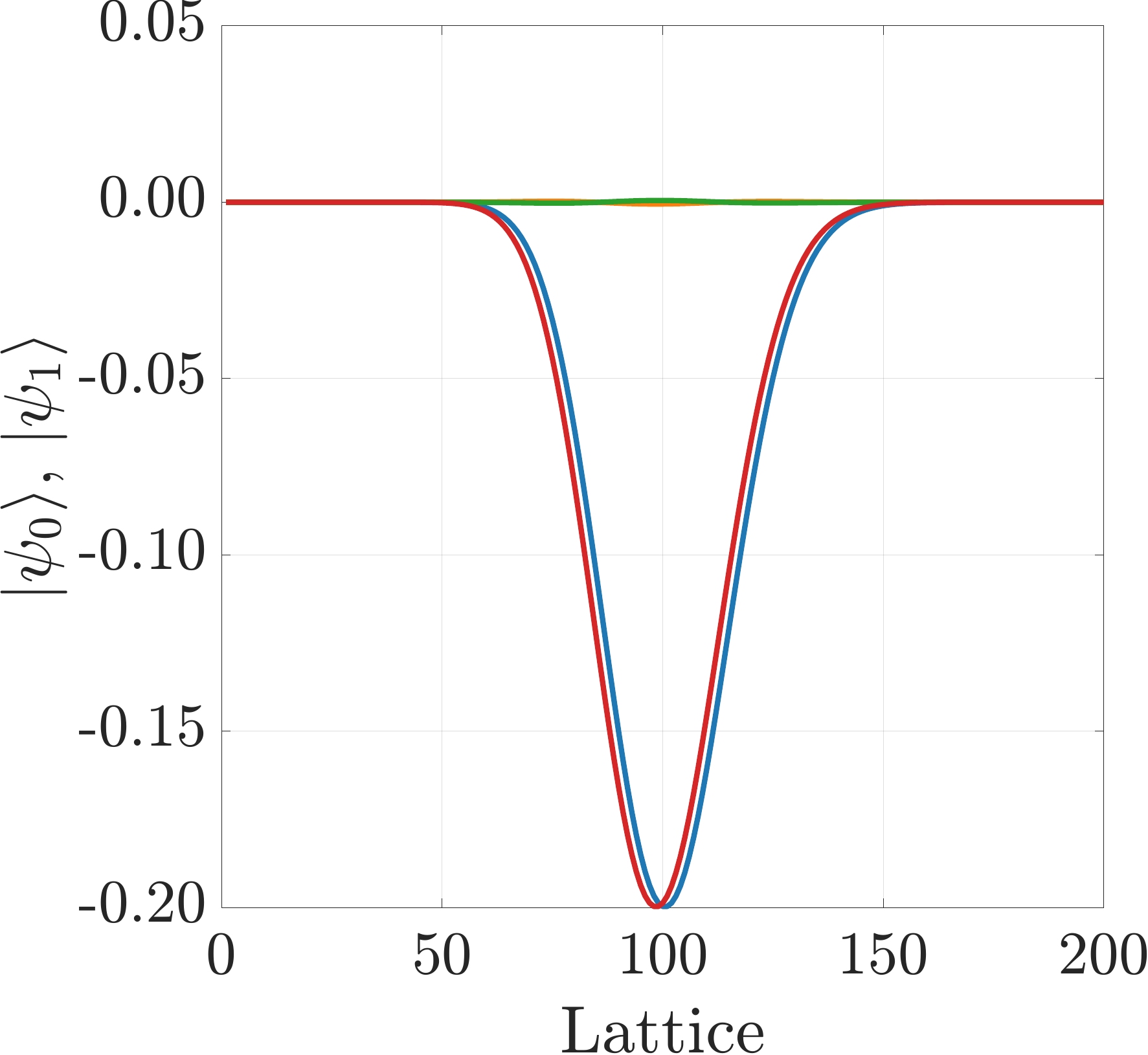}
		\label{fig:d2step2}}
	\subfigure[]{\includegraphics[width=0.23\textwidth]{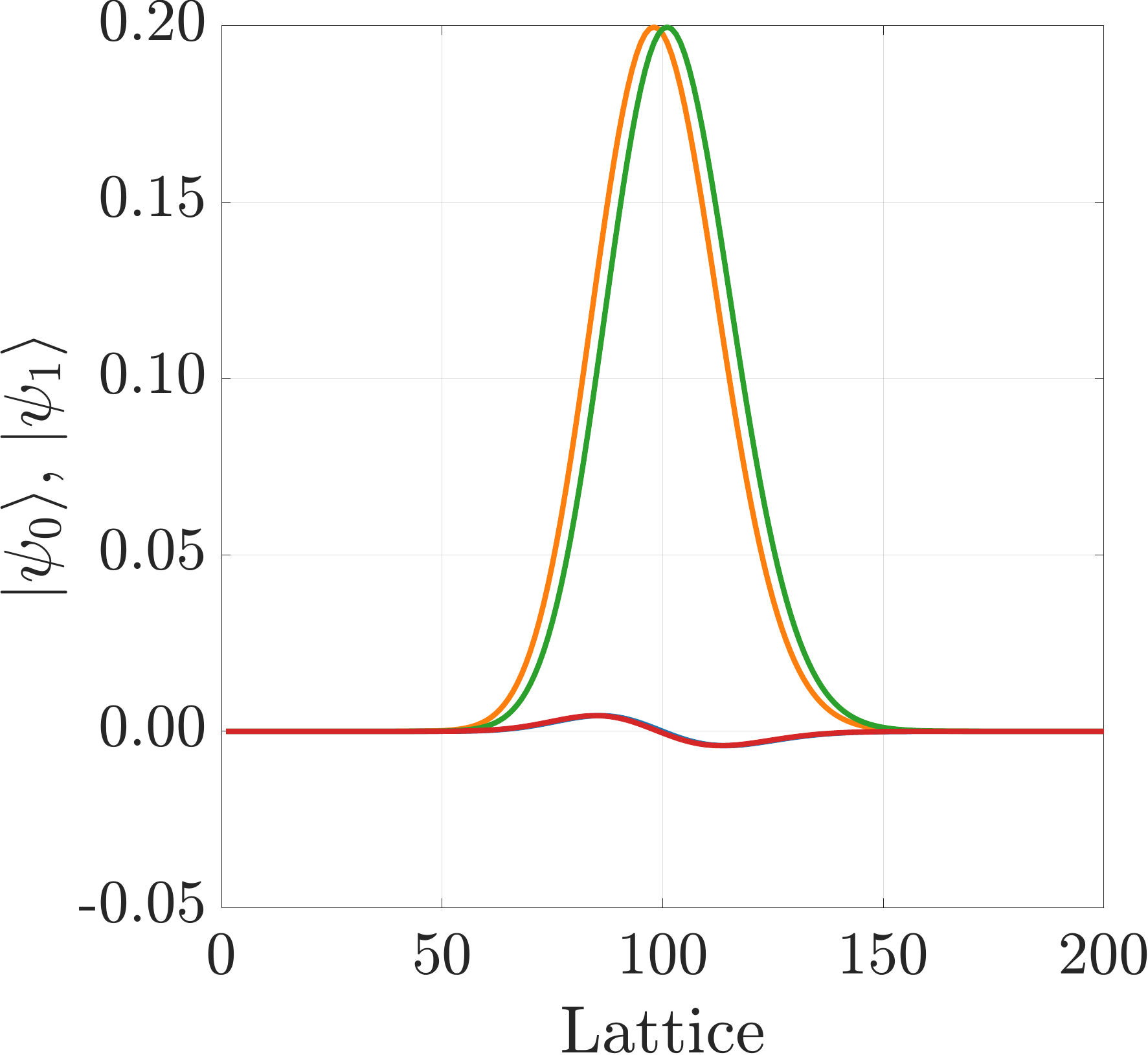}
		\label{fig:d2step3}}
	\caption{The evolution of the state of the lattice after a quantum walk with lattice size $N = 200$ after (a) $t = 1$, (b) $t = 2$ and (c) $t = 3$ time steps assuming $\delta = 0$ in the initial coin state and $\theta_1 = \pi$ and $\theta_2 = -\pi/2$. We consider $\delta = \frac{\pi}{2}$ and plot the evolution of the lattice after (d) $t = 1$, (e) $t = 2$ and (f) $t = 3$ time steps while keeping all the other parameters same.}
	\label{fig:step123}
\end{figure*}

We provide an explanation of this dynamic behaviour of the $1$-D SSQW. In fact, all the aforementioned behaviours can be elucidated by understanding the underlying dynamics for $t = 1, 2, 3$ time steps for the scenarios by choosing $\delta = 0$ and $\delta = \pi/2$ in the initial state of the coin. The pattern then repeats periodically with periodicity $t = 3$. In Fig.~\ref{fig:step123}, we plot the real and imaginary parts of the state of the lattice corresponding to the coin states $\ket{0}$ and $\ket{1}$ after different time steps and for $\delta = 0$ and $\delta = \frac{\pi}{2}$ in the initial state of the coin. As the quantum walk evolves, we obtain the state of the lattice, which can be categorized as Be1-Be4. Here, we discuss the behaviours Be2-Be4, while in the main draft, we provide a discussion on Be1.  However, it should be noted that all these behaviours yield similar fidelity of generation for the same number of time steps and initial state of the lattice.

We first begin by considering that $\delta = 0$ for the initial state of the coin.
In Fig.~\ref{fig:d1step1}, we see that after $t = 1$ time step, the states of the lattice simply acquire a relative phase factor such that one of the states has all coefficients positive while the other one has all negative. In this case, both of them only have imaginary coefficients, which is a consequence of the rotation operator $R(\theta_1)$, and we observe the behaviour Be3. The final HE state can be written as 
\begin{equation}
	\ket{\Psi (t=1)}_{CL} = \frac{i}{\sqrt{2}} \left(-\ket{0}\ket{\psi_0} + \ket{1}\ket{\psi_1}\right),
\end{equation}
where $\ket{\psi_0}$ and $\ket{\psi_1}$ are taken to be real hereafter unless otherwise mentioned.

In Fig.~\ref{fig:d1step2}, we see that after $t = 2$ time steps, both the states of the lattice are in phase with each other with a global negative factor. Moreover, the coefficients of both states are real in this case, which is a consequence of the rotation operator $R(\theta_1)$, which has now been applied twice. This corresponds to behaviour Be2, and the final state can be written as
\begin{equation}
	\ket{\Psi (t=2)}_{CL} = -\frac{1}{\sqrt{2}} \left(\ket{0}\ket{\psi_0} + \ket{1}\ket{\psi_1}\right),
\end{equation}
where the overall state now picks up a global phase factor.

In Fig.~\ref{fig:d1step3}, we see a similar behaviour like $t = 1$ time step, with the exception that the lattice state, which had all coefficients positive, now has all of them negative (and vice versa), which is a consequence of the application of $R(\theta_1)$ twice. The negative sign factor after $t = 2$ time steps is carried forward and results in the flipping of the components. This corresponds to behaviour Be3. The final HE state, in this case, can be written as 
\begin{equation}
	\ket{\Psi (t=3)}_{CL} = \frac{i}{\sqrt{2}} \left(\ket{0}\ket{\psi_0} - \ket{1}\ket{\psi_1}\right),
\end{equation}
where the overall state now picks up a relative phase factor with the global phase factor from $t = 2$ time steps.

From Fig.~\ref{fig:d2step1}-\ref{fig:d2step3}, we show the behaviour of states of the lattice when $\delta = \frac{\pi}{2}$ for the initial state of the coin. In Fig.~\ref{fig:d2step1}, we see that after $t = 1$ time step, both the states of the lattice are in phase with each other. However, both of the states have complex coefficients. In this case, the HE state can be written as,
\begin{equation}
	\ket{\Psi (t=1)}_{CL} = -\frac{1}{\sqrt{2}} \left(\ket{0}\ket{\psi_0} + \ket{1}\ket{\psi_1}\right),
	\label{eq:global_negative}
\end{equation}
where the state now picks up an overall negative global phase factor and the states $\ket{\psi_i}$ are now complex. This is an example of behaviour Be4.

In Fig.~\ref{fig:d2step2}, we see that after $t = 2$ time steps, both the states of the lattice are still in phase with each other and have negative coefficients. The state can be written as
\begin{equation}
	\ket{\Psi (t=2)}_{CL} = -\frac{1}{\sqrt{2}} \left(\ket{0}\ket{\psi_0} + i\ket{1}\ket{\psi_1}\right),
\end{equation}
where the states $\ket{\psi_i}$ are in general complex. This forms an example of behaviour Be4.

In Fig.~\ref{fig:d2step3}, all the coefficients for both the lattice states are now positive with complex coefficients. This forms an example of Be4, and the state is written as
\begin{equation}
	\ket{\Psi (t=3)}_{CL} = \frac{1}{\sqrt{2}} \left(i\ket{0}\ket{\psi_0} + \ket{1}\ket{\psi_1}\right),
\end{equation}
with the states $\ket{\psi_i}$ as complex valued.

\begin{figure*}
	\centering
	\subfigure[]{\includegraphics[width=0.195\textwidth]{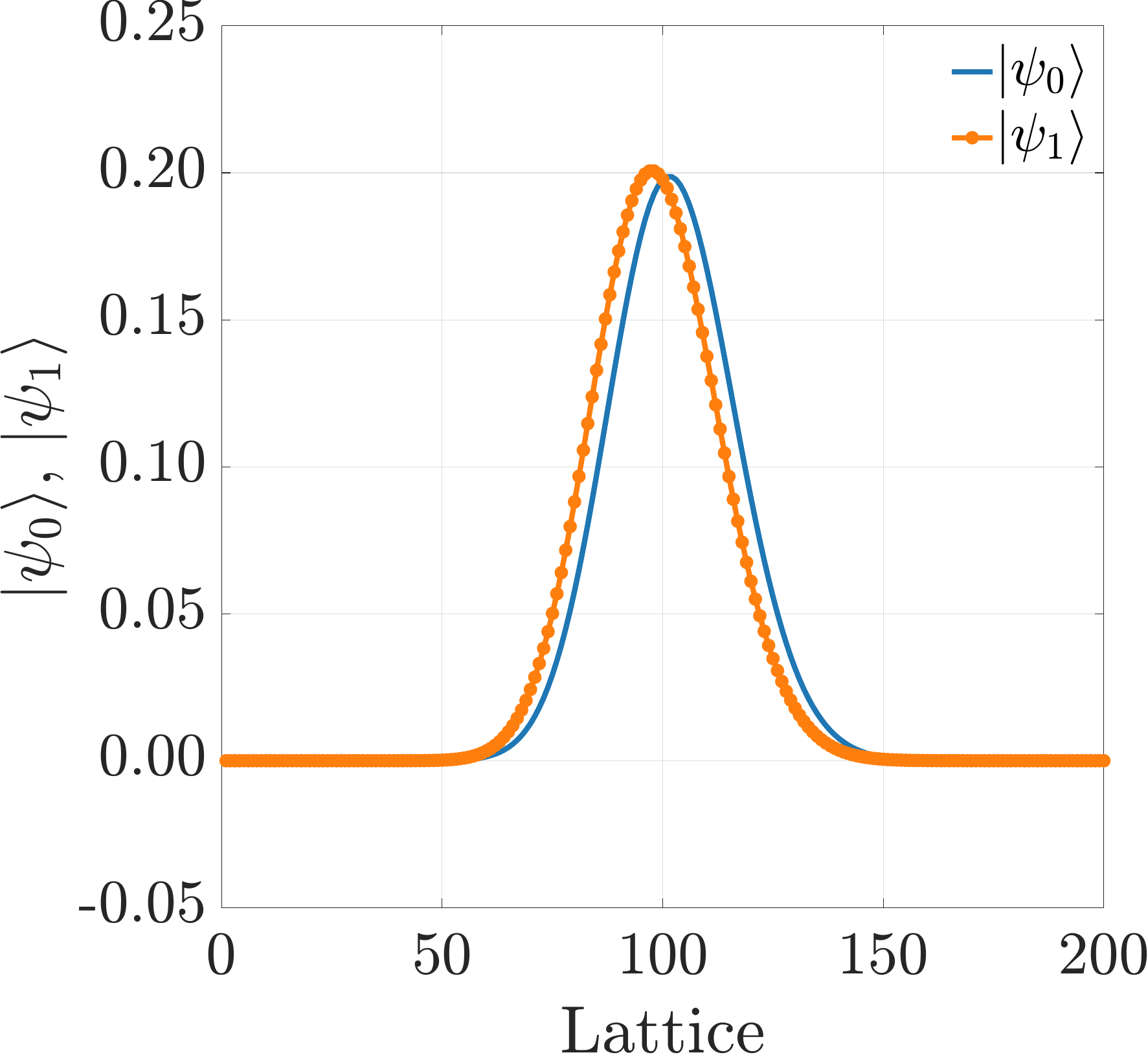}}
	\subfigure[]{\includegraphics[width=0.195\textwidth]{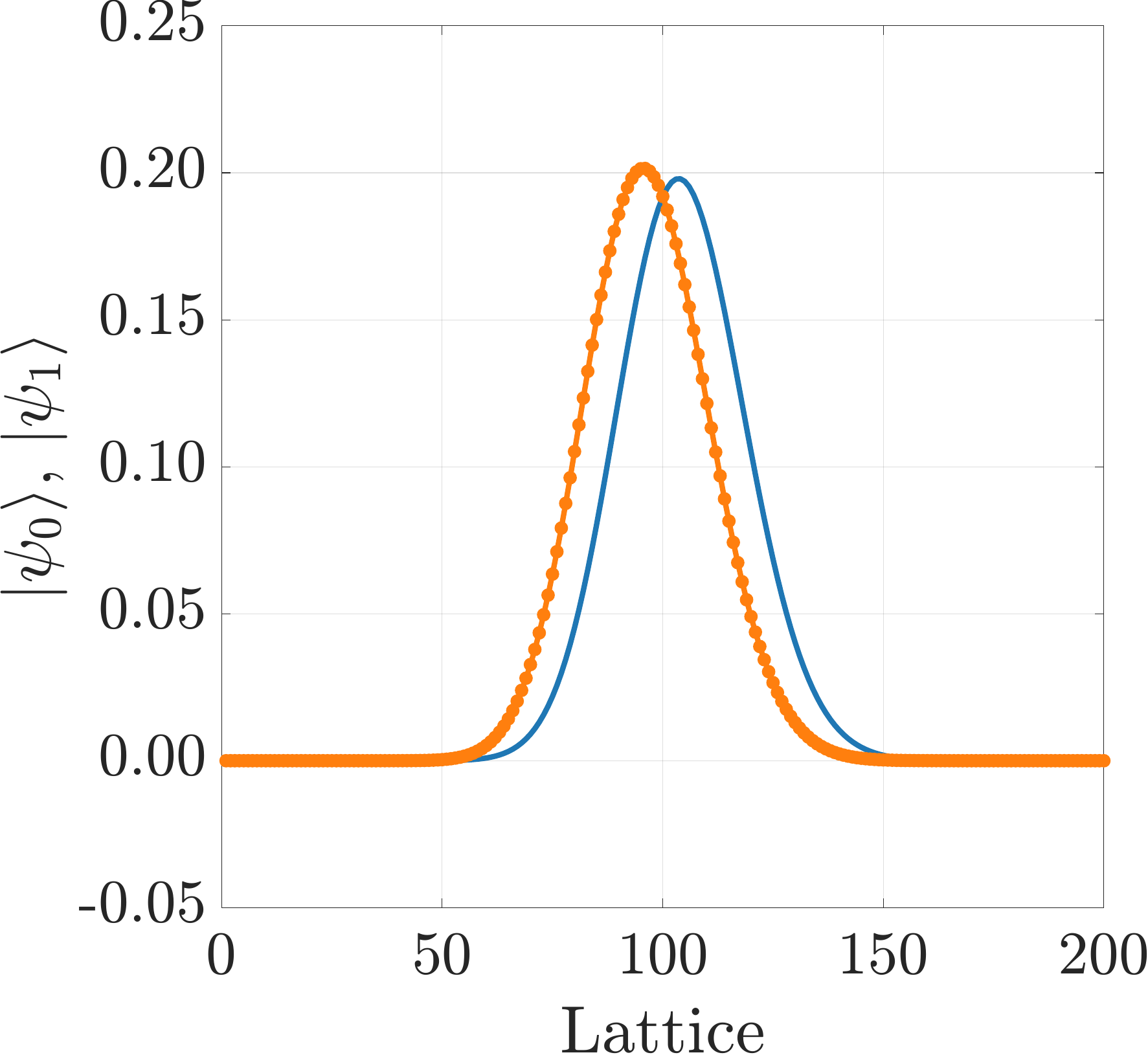}}
	\subfigure[]{\includegraphics[width=0.195\textwidth]{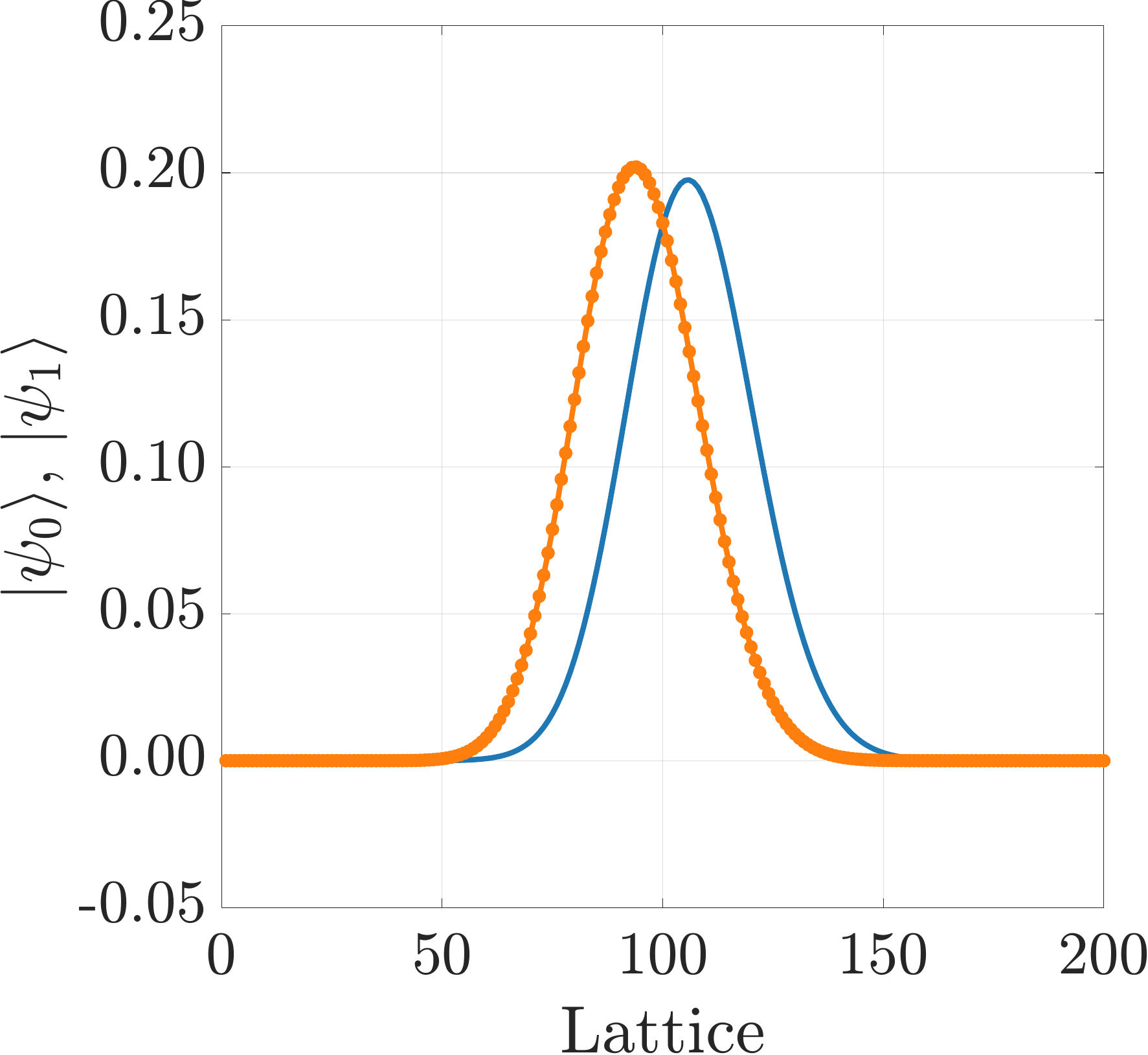}}
	\subfigure[]{\includegraphics[width=0.195\textwidth]{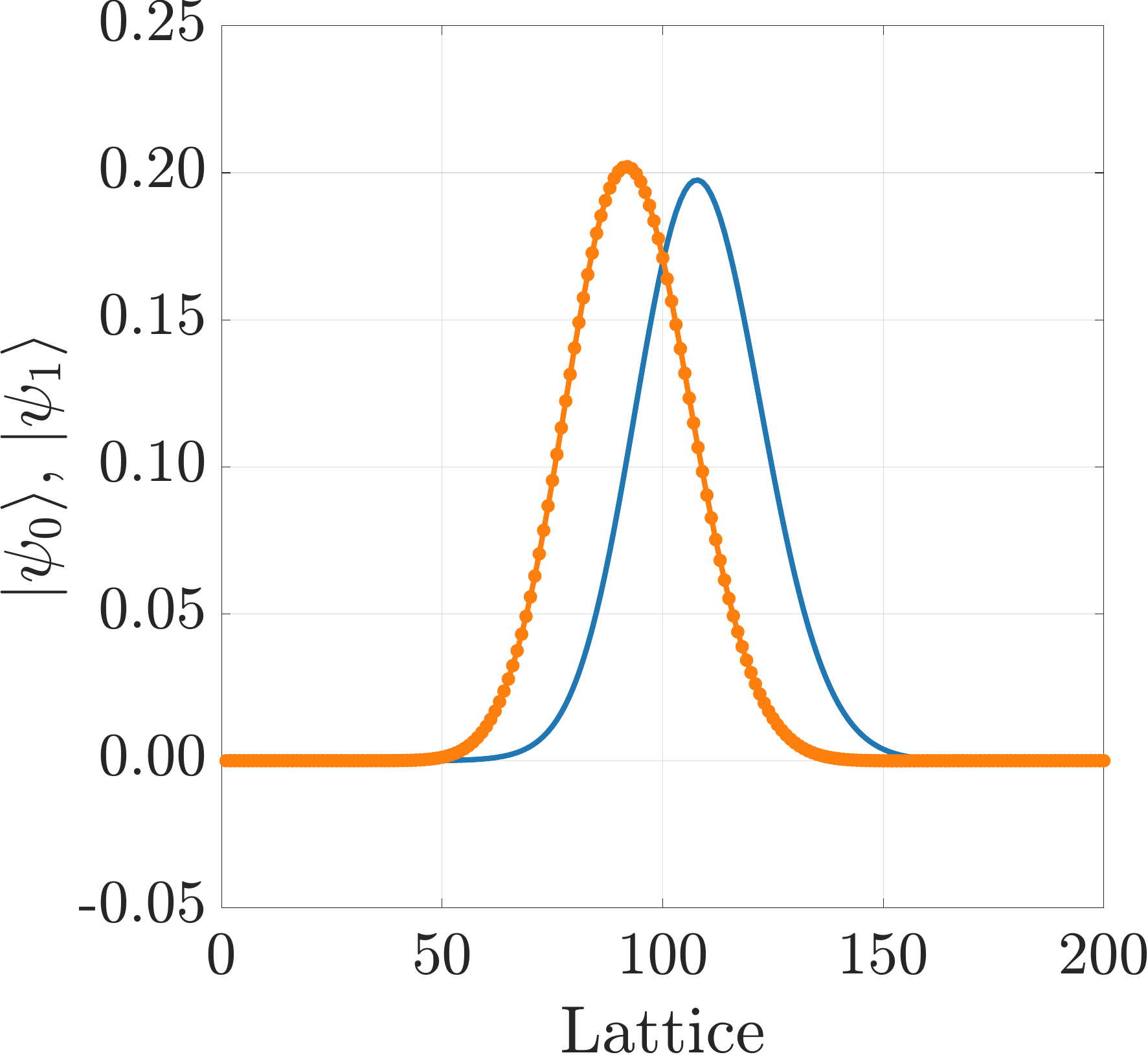}}
	\subfigure[]{\includegraphics[width=0.195\textwidth]{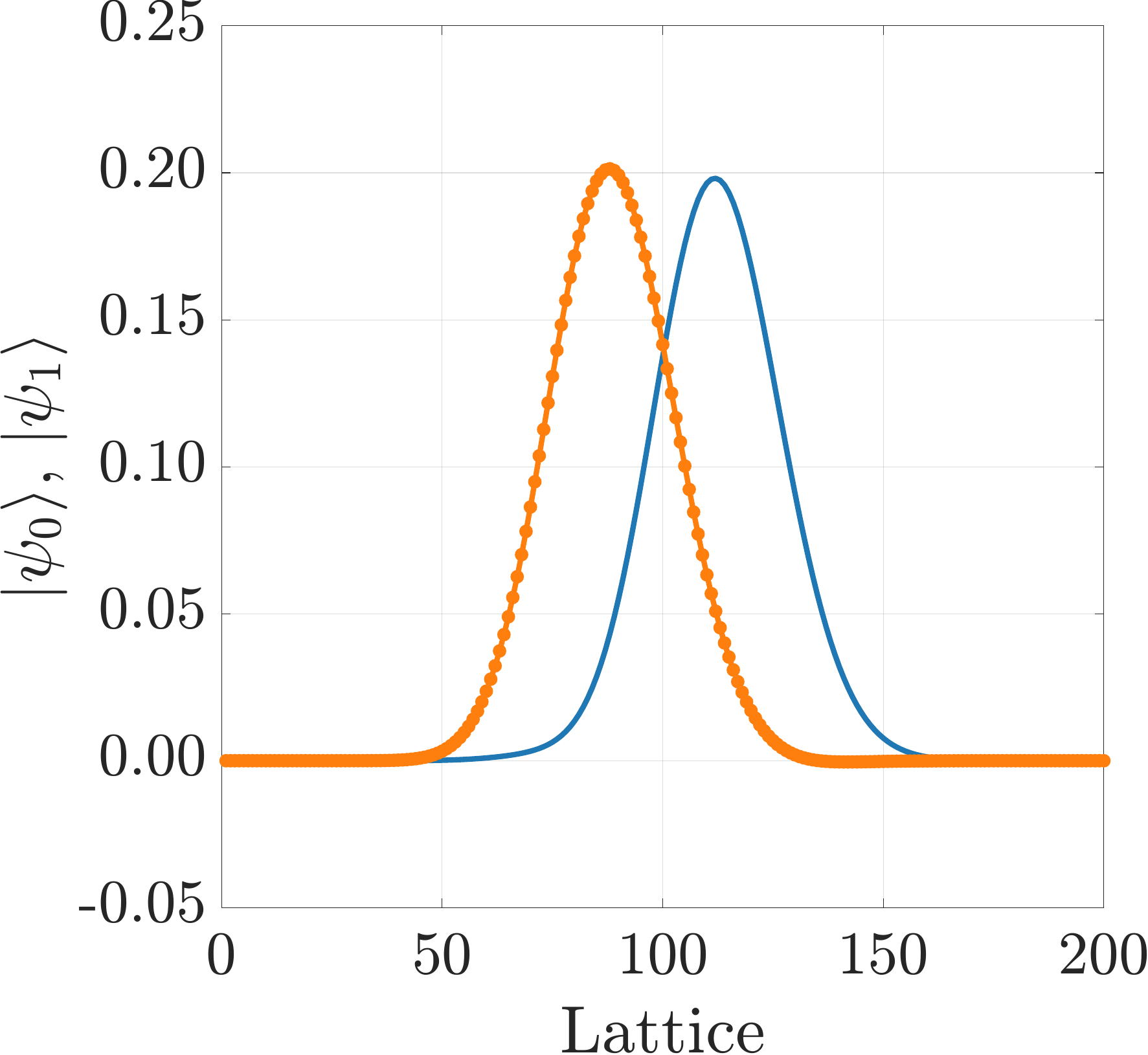}}
	
	\subfigure[]{\includegraphics[width=0.195\textwidth]{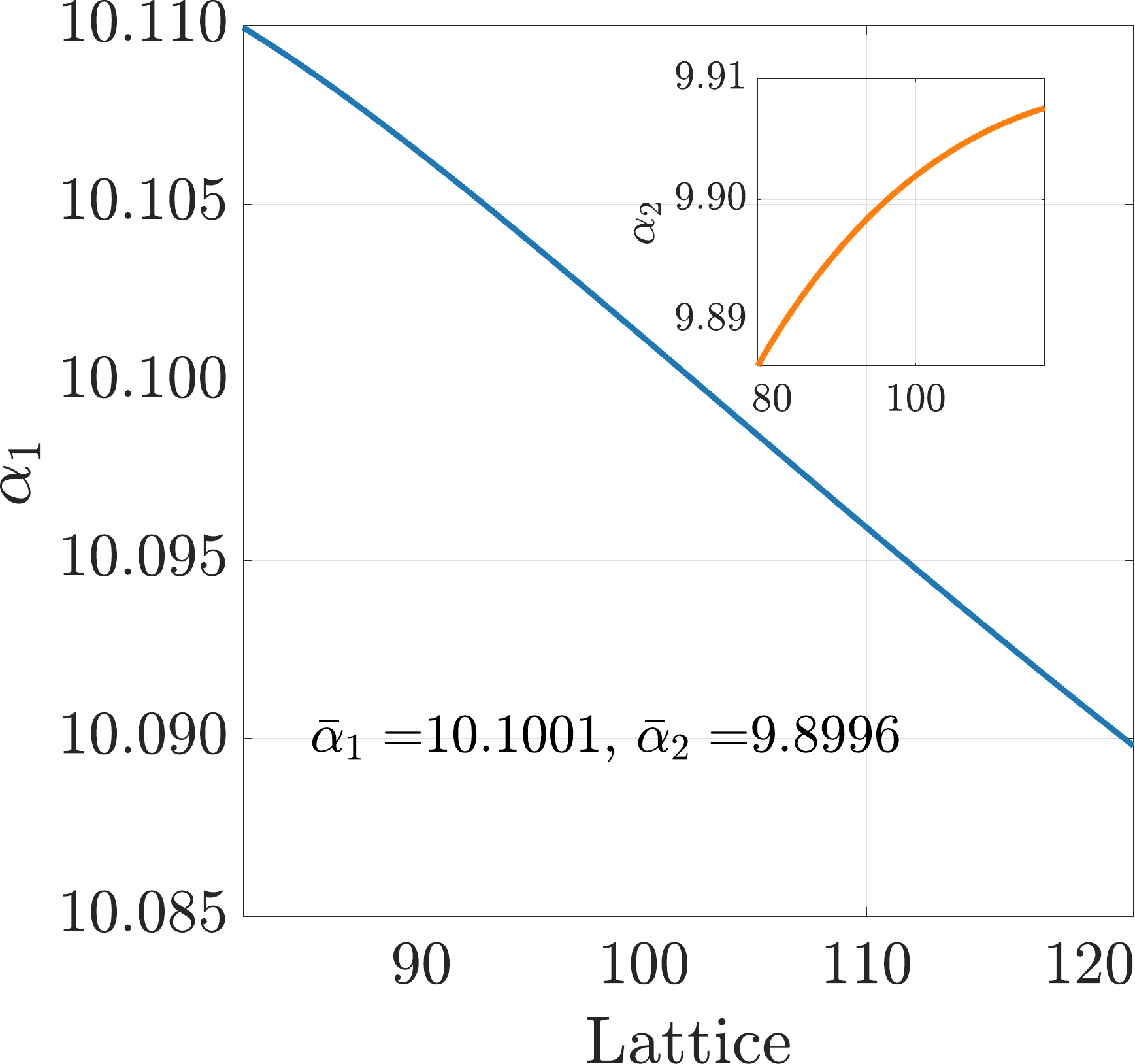}}
	\subfigure[]{\includegraphics[width=0.195\textwidth]{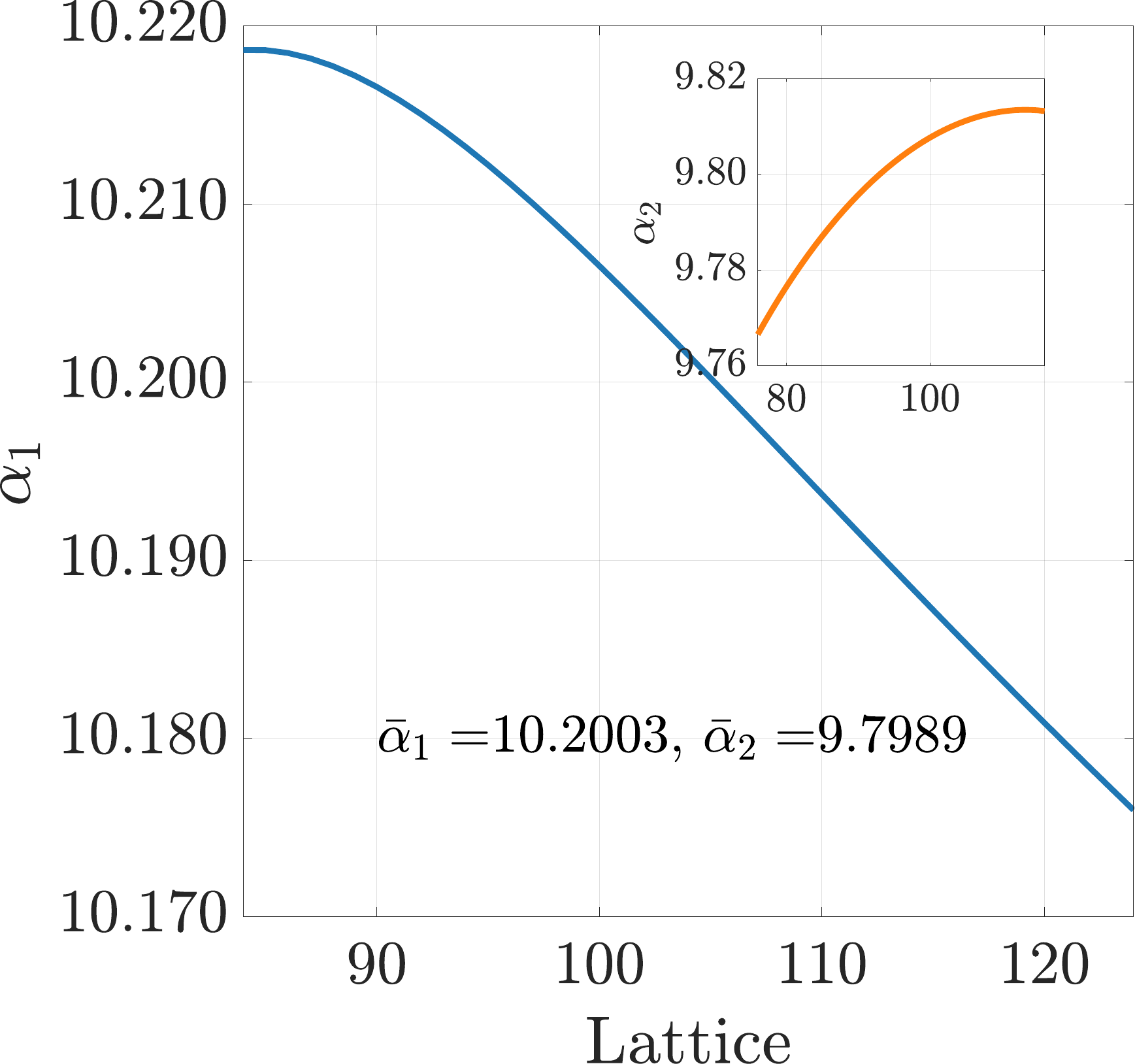}}
	\subfigure[]{\includegraphics[width=0.195\textwidth]{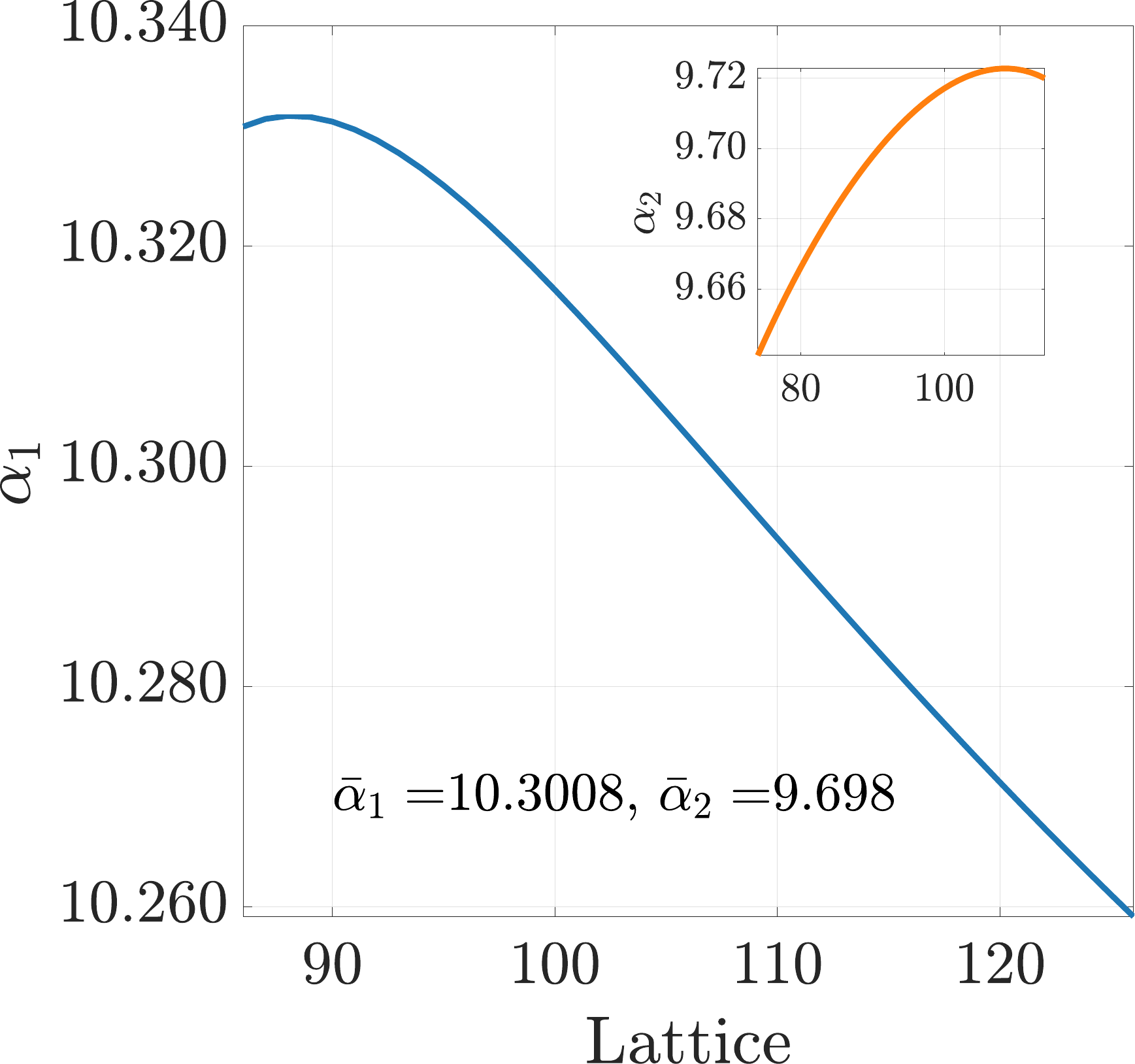}}
	\subfigure[]{\includegraphics[width=0.195\textwidth]{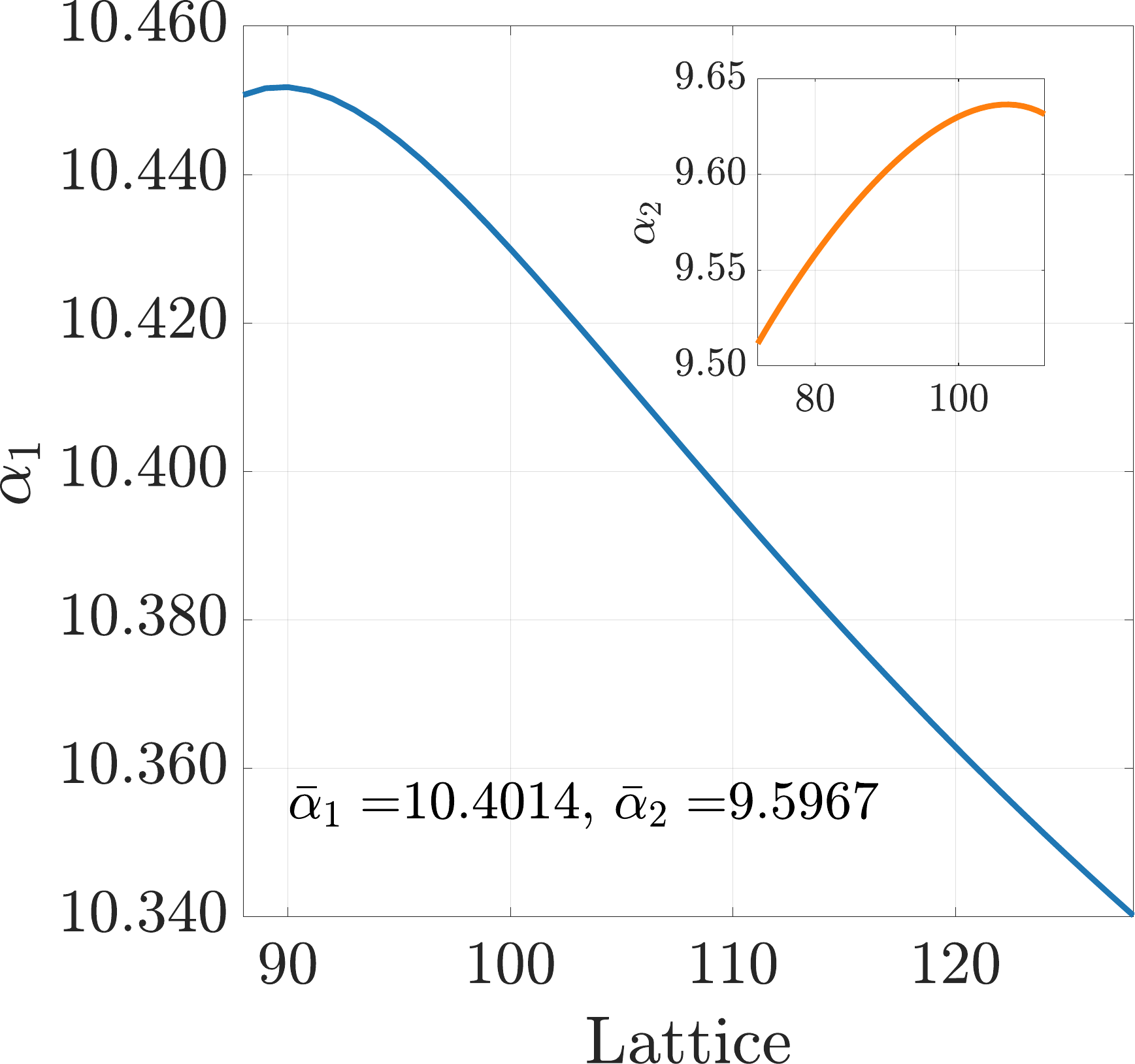}}
	\subfigure[]{\includegraphics[width=0.195\textwidth]{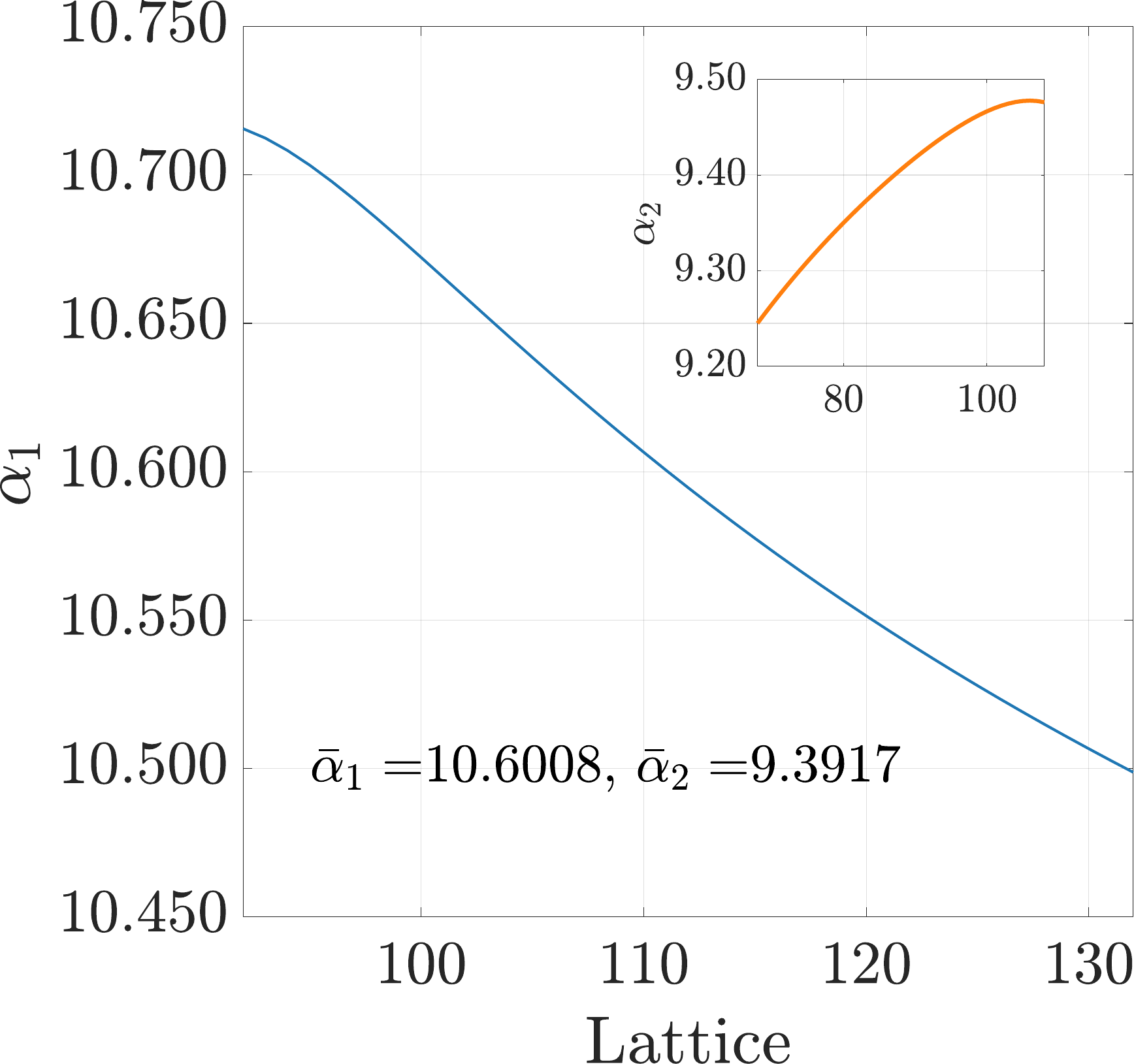}}
	
	\subfigure[]{\includegraphics[width=0.195\textwidth]{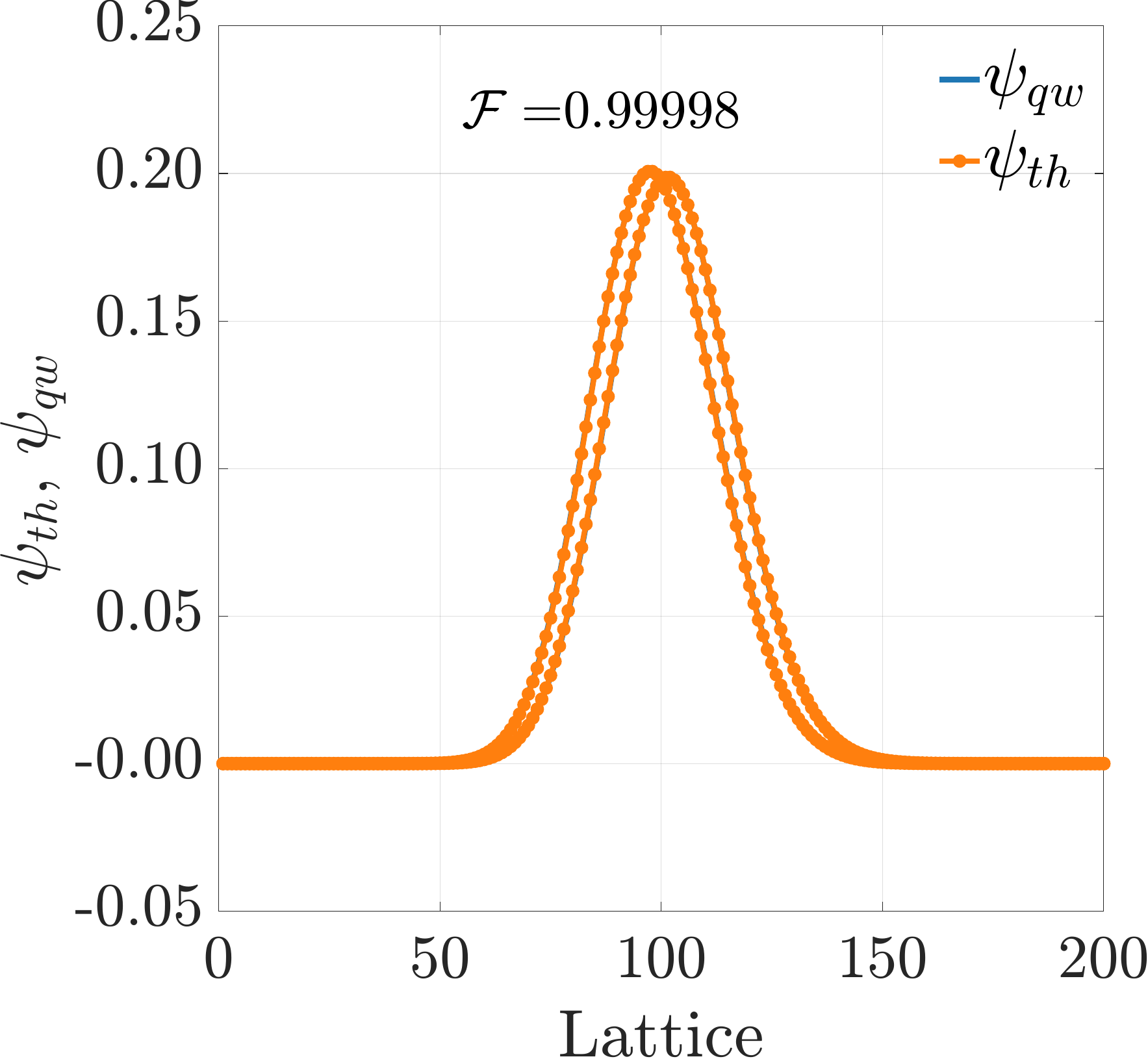}}
	\subfigure[]{\includegraphics[width=0.195\textwidth]{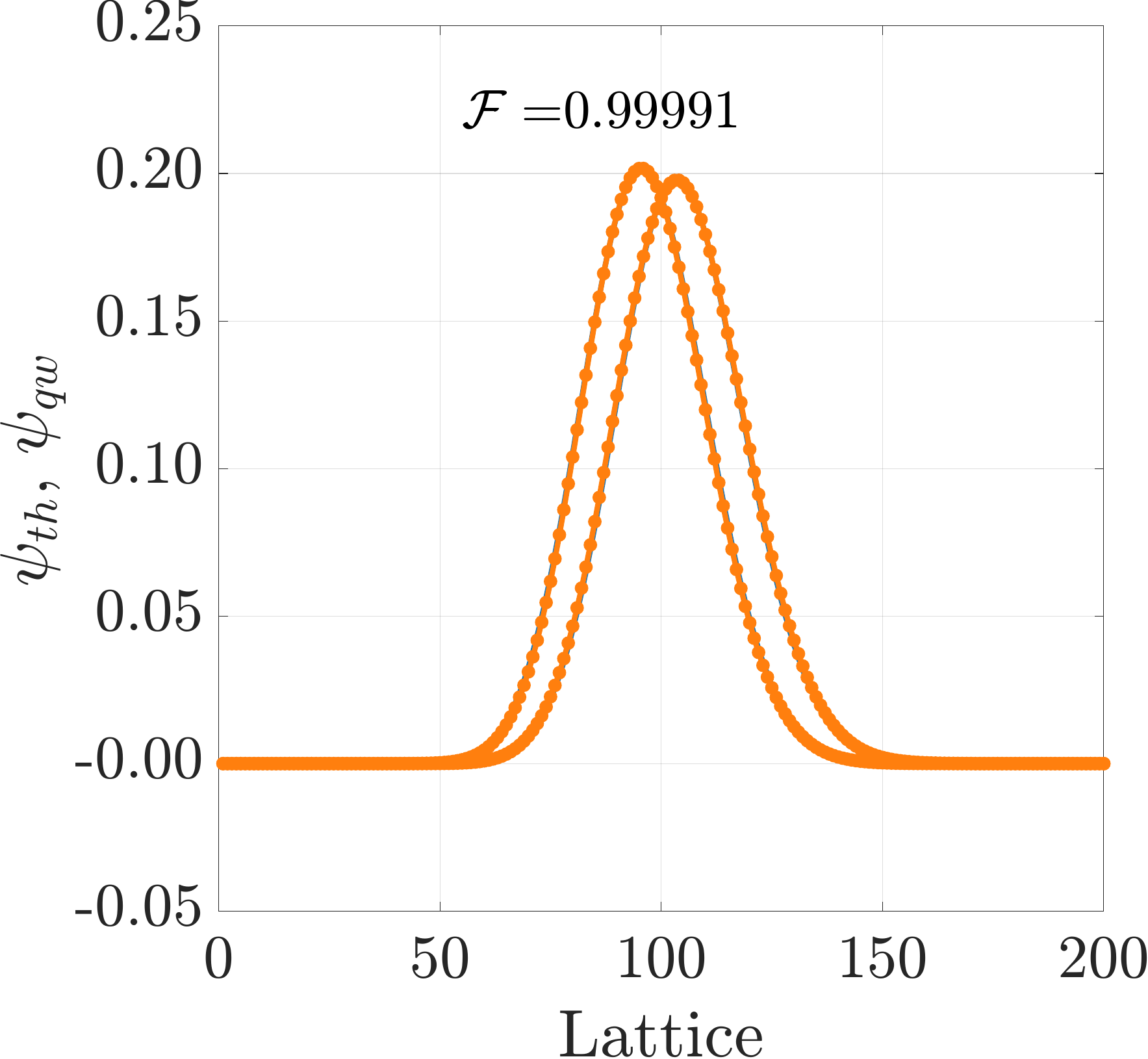}}
	\subfigure[]{\includegraphics[width=0.195\textwidth]{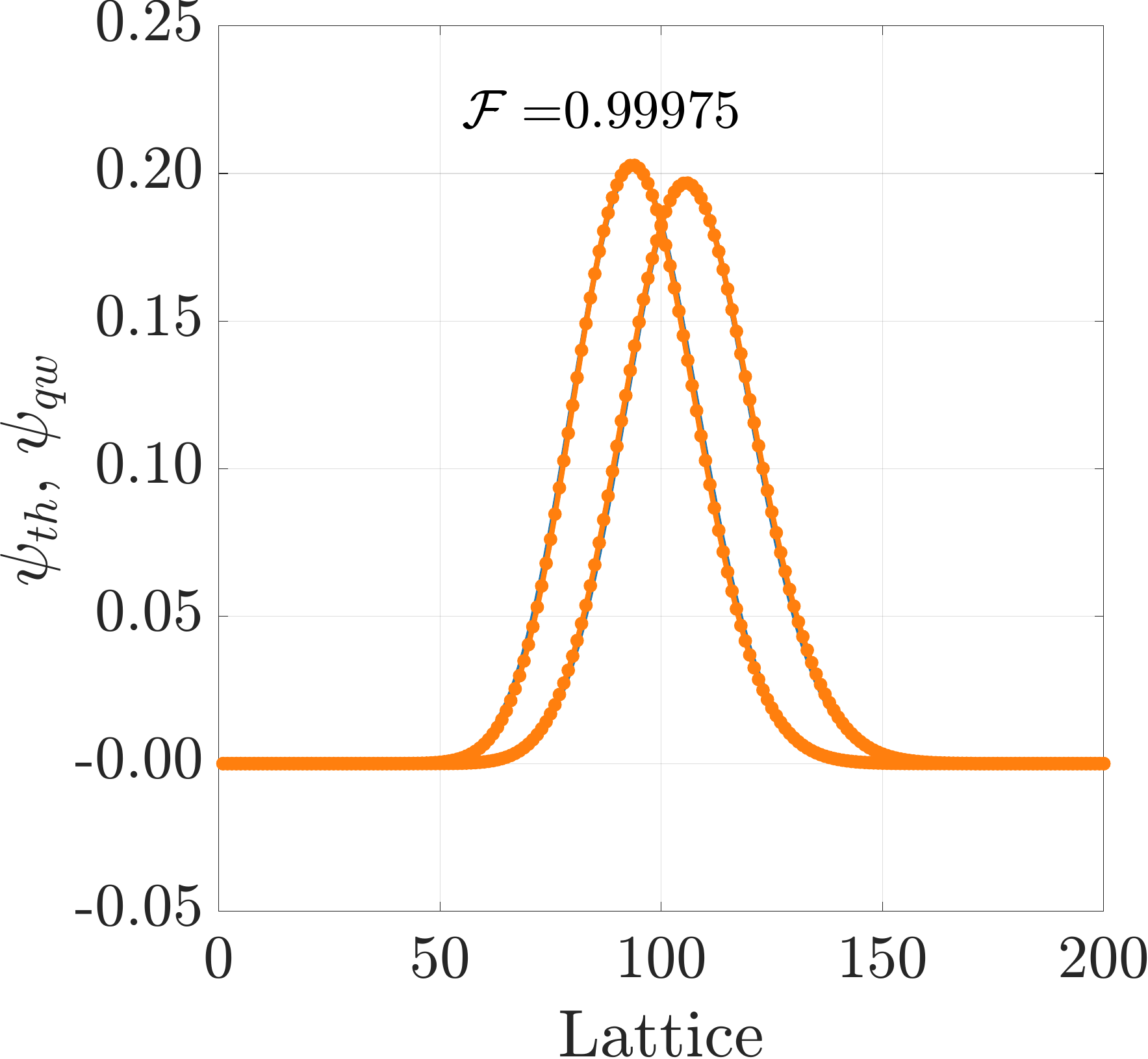}}
	\subfigure[]{\includegraphics[width=0.195\textwidth]{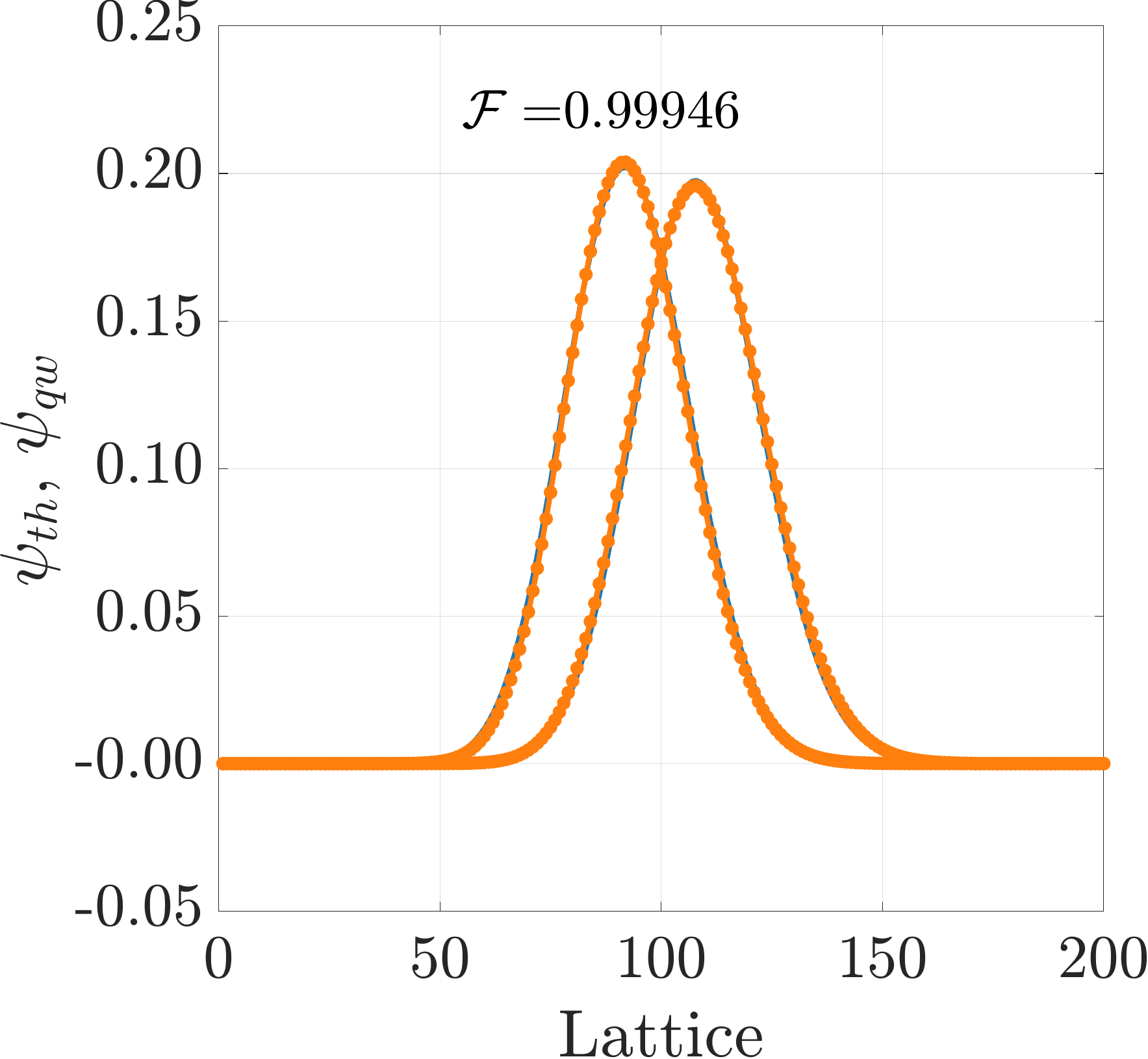}}
	\subfigure[]{\includegraphics[width=0.195\textwidth]{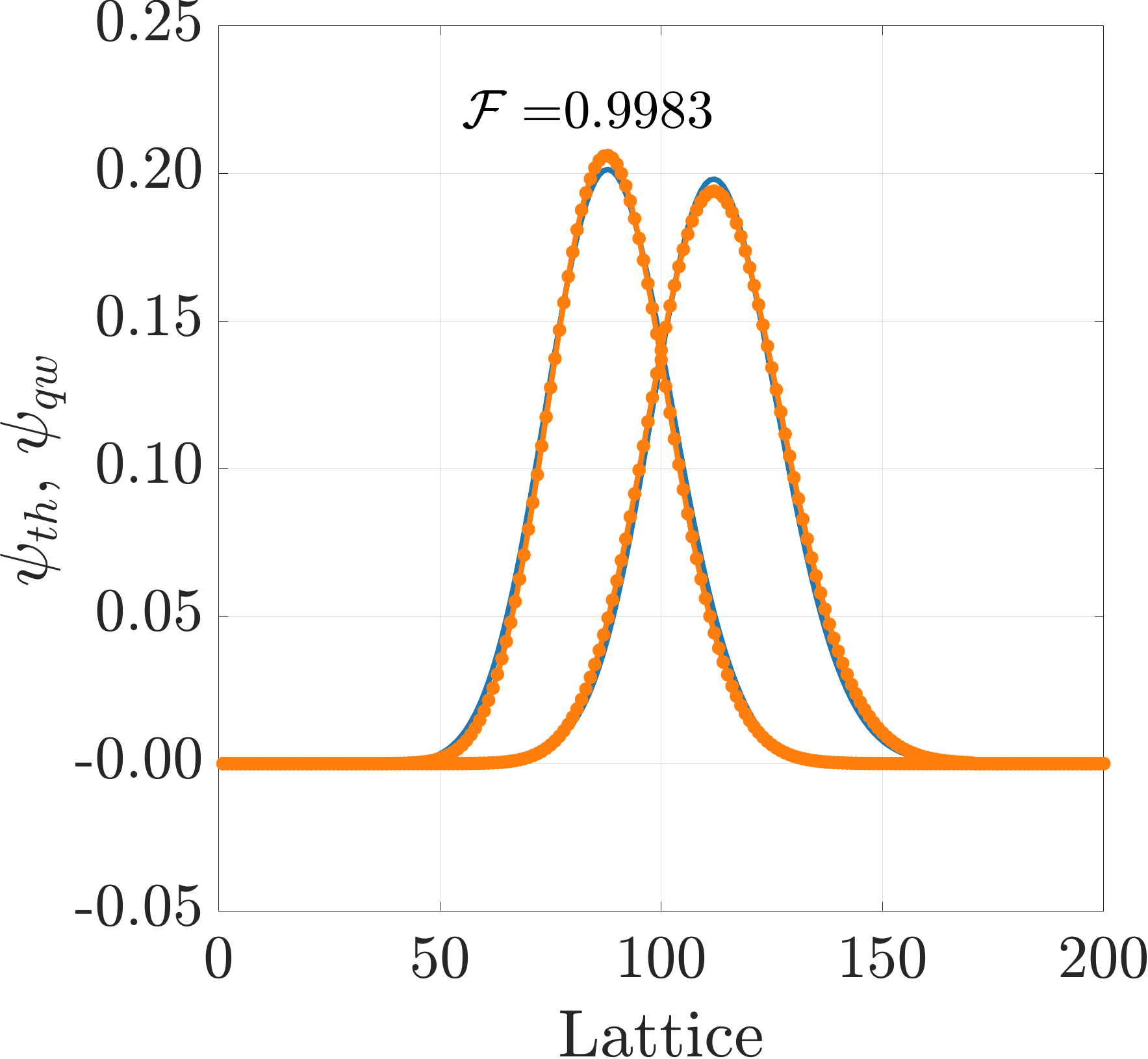}}
	
	\caption{The evolution of the state of the lattice after a quantum walk with lattice size $N = 200$ and using coin parameters $\theta_1 = \pi$, $\theta_2 = -\pi/2$ and $\delta = 0$ for (a) $t = 4$, (b) $t = 8$, (c) $t = 12$, (d) $t = 16$ and (e) $t = 24$ steps. Corresponding to the time steps in (a)-(e) we also show the variation of $\alpha_i^j$ and the corresponding fidelity in (f)-(j) and (k)-(o), respectively.}
	\label{fig:step12}
\end{figure*}

Further, in Fig.~\ref{fig:step12}, we plot the resultant lattice states for various (small) number of time steps and $\delta = 0$. As can be seen, the fidelity of the resultant states is very good; however, the separation between the coherent amplitudes of the resultant lattice states is very small. For this case, we achieve fidelity of $99.998 \%$, $99.991 \%$, $99.975 \%$, $99.946 \%$ and $99.830 \%$   when $t = 4$, $t = 8$, $t = 12$, $t = 16$ and $t = 24$ time steps.

\end{document}